\documentclass[sigplan,screen,authorversion]{acmart}

\pdfoutput=1

\usepackage{listings}
\usepackage{xspace}
\usepackage[acronym]{glossaries}
\usepackage{cleveref}
\usepackage{url}

\usepackage{microtype}

\newcommand{\idris}{Idris\xspace}
\newcommand{\IIdris}{\textsc{Idris2}\xspace}
\newcommand{\iidris}{Idris2\xspace}

\newcommand{\QC}{QuickCheck\xspace}

\AtBeginDocument{%
  }

\setcopyright{acmlicensed}
\copyrightyear{2024}
\acmYear{2024}
\acmDOI{10.1145/3678000.3678206}

\acmConference[TyDe '24]{Proceedings of the 9th ACM SIGPLAN International %
                         Workshop on Type-Driven Development}{September 6, %
                         2024}{Milan, Italy}
\acmBooktitle{Proceedings of the 9th ACM SIGPLAN International Workshop on %
              Type-Driven Development (TyDe '24), September 6, 2024, Milan, %
              Italy}

\acmISBN{979-8-4007-1103-9/24/09}

\acmSubmissionID{icfpws24tydemain-p40-p}

\newacronym{pbt}{PBT}{Property Based Testing}

\newacronym{fsm}{FSM}{Finite State Machine}

\newacronym{ism}{ISM}{Indexed State Monad}

\newacronym{dsl}{DSL}{Domain Specific Language}

\newacronym{edsl}{eDSL}{Embedded Domain Specific Language}

\newacronym{ctl}{CTL}{Computation Tree Logic}

\newacronym{ltl}{LTL}{Linear Temporal Logic}

\newacronym{tla}{TLA}{Temporal Logic of Actions}

\newacronym{prng}{PRNG}{Pseudorandom Number Generator}

\newacronym{atm}{ATM}{Automated Teller Machine}

\newacronym{arq}{ARQ}{Automatic Repeat Request}

\newacronym{qtt}{QTT}{Quantitative Type Theory}

\newacronym{lcg}{LCG}{Linear Congruence Generator}

\newacronym{gui}{GUI}{Graphical User Interface}

\usepackage{color}
\usepackage{fancyvrb}
\usepackage[utf8]{inputenc}

\fvset{fontsize=\small}

\makeatletter
\def\PY@reset{\let\PY@it=\relax\small \let\PY@bf=\relax\small%
    \let\PY@ul=\relax\small \let\PY@tc=\relax\small%
    \let\PY@bc=\relax\small \let\PY@ff=\relax\small}
\def\PY@tok#1{\csname PY@tok@#1\endcsname}
\def\PY@toks#1+{\ifx\relax#1\empty\else%
    \PY@tok{#1}\expandafter\PY@toks\fi}
\def\PY@do#1{\PY@bc{\PY@tc{\PY@ul{%
    \PY@it{\PY@bf{\PY@ff{#1}}}}}}}
\def\PY#1#2{\PY@reset\PY@toks#1+\relax+\PY@do{#2}}

\@namedef{PY@tok@w}{\def\PY@tc##1{\textcolor[rgb]{0.73,0.73,0.73}{##1}}}
\@namedef{PY@tok@c}{\let\PY@it=\textit\def\PY@tc##1{\textcolor[rgb]{0.24,0.48,0.48}{##1}}}
\@namedef{PY@tok@cp}{\def\PY@tc##1{\textcolor[rgb]{0.61,0.40,0.00}{##1}}}
\@namedef{PY@tok@k}{\let\PY@bf=\textbf\def\PY@tc##1{\textcolor[rgb]{0.00,0.50,0.00}{##1}}}
\@namedef{PY@tok@kp}{\def\PY@tc##1{\textcolor[rgb]{0.00,0.50,0.00}{##1}}}
\@namedef{PY@tok@kt}{\def\PY@tc##1{\textcolor[rgb]{0.69,0.00,0.25}{##1}}}
\@namedef{PY@tok@o}{\def\PY@tc##1{\textcolor[rgb]{0.40,0.40,0.40}{##1}}}
\@namedef{PY@tok@ow}{\let\PY@bf=\textbf\def\PY@tc##1{\textcolor[rgb]{0.67,0.13,1.00}{##1}}}
\@namedef{PY@tok@nb}{\def\PY@tc##1{\textcolor[rgb]{0.00,0.50,0.00}{##1}}}
\@namedef{PY@tok@nf}{\def\PY@tc##1{\textcolor[rgb]{0.00,0.00,1.00}{##1}}}
\@namedef{PY@tok@nc}{\let\PY@bf=\textbf\def\PY@tc##1{\textcolor[rgb]{0.00,0.00,1.00}{##1}}}
\@namedef{PY@tok@nn}{\let\PY@bf=\textbf\def\PY@tc##1{\textcolor[rgb]{0.00,0.00,1.00}{##1}}}
\@namedef{PY@tok@ne}{\let\PY@bf=\textbf\def\PY@tc##1{\textcolor[rgb]{0.80,0.25,0.22}{##1}}}
\@namedef{PY@tok@nv}{\def\PY@tc##1{\textcolor[rgb]{0.10,0.09,0.49}{##1}}}
\@namedef{PY@tok@no}{\def\PY@tc##1{\textcolor[rgb]{0.53,0.00,0.00}{##1}}}
\@namedef{PY@tok@nl}{\def\PY@tc##1{\textcolor[rgb]{0.46,0.46,0.00}{##1}}}
\@namedef{PY@tok@ni}{\let\PY@bf=\textbf\def\PY@tc##1{\textcolor[rgb]{0.44,0.44,0.44}{##1}}}
\@namedef{PY@tok@na}{\def\PY@tc##1{\textcolor[rgb]{0.41,0.47,0.13}{##1}}}
\@namedef{PY@tok@nt}{\let\PY@bf=\textbf\def\PY@tc##1{\textcolor[rgb]{0.00,0.50,0.00}{##1}}}
\@namedef{PY@tok@nd}{\def\PY@tc##1{\textcolor[rgb]{0.67,0.13,1.00}{##1}}}
\@namedef{PY@tok@s}{\def\PY@tc##1{\textcolor[rgb]{0.73,0.13,0.13}{##1}}}
\@namedef{PY@tok@sd}{\let\PY@it=\textit\def\PY@tc##1{\textcolor[rgb]{0.73,0.13,0.13}{##1}}}
\@namedef{PY@tok@si}{\let\PY@bf=\textbf\def\PY@tc##1{\textcolor[rgb]{0.64,0.35,0.47}{##1}}}
\@namedef{PY@tok@se}{\let\PY@bf=\textbf\def\PY@tc##1{\textcolor[rgb]{0.67,0.36,0.12}{##1}}}
\@namedef{PY@tok@sr}{\def\PY@tc##1{\textcolor[rgb]{0.64,0.35,0.47}{##1}}}
\@namedef{PY@tok@ss}{\def\PY@tc##1{\textcolor[rgb]{0.10,0.09,0.49}{##1}}}
\@namedef{PY@tok@sx}{\def\PY@tc##1{\textcolor[rgb]{0.00,0.50,0.00}{##1}}}
\@namedef{PY@tok@m}{\def\PY@tc##1{\textcolor[rgb]{0.40,0.40,0.40}{##1}}}
\@namedef{PY@tok@gh}{\let\PY@bf=\textbf\def\PY@tc##1{\textcolor[rgb]{0.00,0.00,0.50}{##1}}}
\@namedef{PY@tok@gu}{\let\PY@bf=\textbf\def\PY@tc##1{\textcolor[rgb]{0.50,0.00,0.50}{##1}}}
\@namedef{PY@tok@gd}{\def\PY@tc##1{\textcolor[rgb]{0.63,0.00,0.00}{##1}}}
\@namedef{PY@tok@gi}{\def\PY@tc##1{\textcolor[rgb]{0.00,0.52,0.00}{##1}}}
\@namedef{PY@tok@gr}{\def\PY@tc##1{\textcolor[rgb]{0.89,0.00,0.00}{##1}}}
\@namedef{PY@tok@ge}{\let\PY@it=\textit}
\@namedef{PY@tok@gs}{\let\PY@bf=\textbf}
\@namedef{PY@tok@ges}{\let\PY@bf=\textbf\let\PY@it=\textit}
\@namedef{PY@tok@gp}{\let\PY@bf=\textbf\def\PY@tc##1{\textcolor[rgb]{0.00,0.00,0.50}{##1}}}
\@namedef{PY@tok@go}{\def\PY@tc##1{\textcolor[rgb]{0.44,0.44,0.44}{##1}}}
\@namedef{PY@tok@gt}{\def\PY@tc##1{\textcolor[rgb]{0.00,0.27,0.87}{##1}}}
\@namedef{PY@tok@err}{\def\PY@bc##1{{\setlength{\fboxsep}{\string -\fboxrule}\fcolorbox[rgb]{1.00,0.00,0.00}{1,1,1}{\strut ##1}}}}
\@namedef{PY@tok@kc}{\let\PY@bf=\textbf\def\PY@tc##1{\textcolor[rgb]{0.00,0.50,0.00}{##1}}}
\@namedef{PY@tok@kd}{\let\PY@bf=\textbf\def\PY@tc##1{\textcolor[rgb]{0.00,0.50,0.00}{##1}}}
\@namedef{PY@tok@kn}{\let\PY@bf=\textbf\def\PY@tc##1{\textcolor[rgb]{0.00,0.50,0.00}{##1}}}
\@namedef{PY@tok@kr}{\let\PY@bf=\textbf\def\PY@tc##1{\textcolor[rgb]{0.00,0.50,0.00}{##1}}}
\@namedef{PY@tok@bp}{\def\PY@tc##1{\textcolor[rgb]{0.00,0.50,0.00}{##1}}}
\@namedef{PY@tok@fm}{\def\PY@tc##1{\textcolor[rgb]{0.00,0.00,1.00}{##1}}}
\@namedef{PY@tok@vc}{\def\PY@tc##1{\textcolor[rgb]{0.10,0.09,0.49}{##1}}}
\@namedef{PY@tok@vg}{\def\PY@tc##1{\textcolor[rgb]{0.10,0.09,0.49}{##1}}}
\@namedef{PY@tok@vi}{\def\PY@tc##1{\textcolor[rgb]{0.10,0.09,0.49}{##1}}}
\@namedef{PY@tok@vm}{\def\PY@tc##1{\textcolor[rgb]{0.10,0.09,0.49}{##1}}}
\@namedef{PY@tok@sa}{\def\PY@tc##1{\textcolor[rgb]{0.73,0.13,0.13}{##1}}}
\@namedef{PY@tok@sb}{\def\PY@tc##1{\textcolor[rgb]{0.73,0.13,0.13}{##1}}}
\@namedef{PY@tok@sc}{\def\PY@tc##1{\textcolor[rgb]{0.73,0.13,0.13}{##1}}}
\@namedef{PY@tok@dl}{\def\PY@tc##1{\textcolor[rgb]{0.73,0.13,0.13}{##1}}}
\@namedef{PY@tok@s2}{\def\PY@tc##1{\textcolor[rgb]{0.73,0.13,0.13}{##1}}}
\@namedef{PY@tok@sh}{\def\PY@tc##1{\textcolor[rgb]{0.73,0.13,0.13}{##1}}}
\@namedef{PY@tok@s1}{\def\PY@tc##1{\textcolor[rgb]{0.73,0.13,0.13}{##1}}}
\@namedef{PY@tok@mb}{\def\PY@tc##1{\textcolor[rgb]{0.40,0.40,0.40}{##1}}}
\@namedef{PY@tok@mf}{\def\PY@tc##1{\textcolor[rgb]{0.40,0.40,0.40}{##1}}}
\@namedef{PY@tok@mh}{\def\PY@tc##1{\textcolor[rgb]{0.40,0.40,0.40}{##1}}}
\@namedef{PY@tok@mi}{\def\PY@tc##1{\textcolor[rgb]{0.40,0.40,0.40}{##1}}}
\@namedef{PY@tok@il}{\def\PY@tc##1{\textcolor[rgb]{0.40,0.40,0.40}{##1}}}
\@namedef{PY@tok@mo}{\def\PY@tc##1{\textcolor[rgb]{0.40,0.40,0.40}{##1}}}
\@namedef{PY@tok@ch}{\let\PY@it=\textit\def\PY@tc##1{\textcolor[rgb]{0.24,0.48,0.48}{##1}}}
\@namedef{PY@tok@cm}{\let\PY@it=\textit\def\PY@tc##1{\textcolor[rgb]{0.24,0.48,0.48}{##1}}}
\@namedef{PY@tok@cpf}{\let\PY@it=\textit\def\PY@tc##1{\textcolor[rgb]{0.24,0.48,0.48}{##1}}}
\@namedef{PY@tok@c1}{\let\PY@it=\textit\def\PY@tc##1{\textcolor[rgb]{0.24,0.48,0.48}{##1}}}
\@namedef{PY@tok@cs}{\let\PY@it=\textit\def\PY@tc##1{\textcolor[rgb]{0.24,0.48,0.48}{##1}}}

\def\PYZbs{\char`\\}
\def\PYZus{\char`\_}
\def\PYZob{\char`\{}
\def\PYZcb{\char`\}}

\def\PYZlt{\char`\<}
\def\PYZgt{\char`\>}

\def\PYZdl{\char`\$}
\def\PYZhy{\char`\-}
\def\PYZsq{\char`\'}
\def\PYZdq{\char`\"}
\def\PYZti{\char`\~}

\makeatother

\begin{document}

\title{Type-level Property Based Testing}

\author{Thomas Ekstr{\"o}m Hansen}
\email{teh6@st-andrews.ac.uk}
\orcid{0000-0002-2472-9694}

\author{Edwin Brady}
\email{ecb10@st-andrews.ac.uk}
\orcid{0000-0002-9734-367X}

\affiliation{%
  \institution{School of Computer Science, University of St Andrews}
  \streetaddress{North Haugh}
  \city{St Andrews}
  \state{Fife}
  \country{United Kingdom}
  \postcode{KY16 9SX}
}

\begin{abstract}
  We present an automated framework for solidifying the cohesion between
  software specifications, their dependently typed models, and implementation at
  compile time. Model Checking and type checking are currently separate
  techniques for automatically verifying the correctness of programs. Using
  \acrfull{pbt},
  \acrfullpl{ism},
  and dependent types, we are able to model several interesting systems
  and network protocols, have the type checker verify that our implementation
  behaves as specified, and test that our model matches the specification's
  semantics; a step towards combining model and type checking.
\end{abstract}



\received{2024-05-27}
\received[accepted]{2024-07-10}

\maketitle

\section{Introduction/Motivation}

Stateful computer systems are ubiquitous. Embedded devices, computer
networks and banking systems all involve states and transitions between
different states. We can formally verify programs using tools like
Spin~\cite{holzmannModelCheckerSPIN1997}
or
\textsc{Uppaal}~\cite{bengtssonUPPAALToolSuite1996},
however these tools rarely scale to real-world code bases due to the State
Explosion Problem (although progress is steadily being
made)~\cite{valmariStateExplosionProblem1998,%
      clarkeProgressStateExplosion2001,%
      demriParametricAnalysisStateexplosion2006,%
      clarkeModelCheckingMy2008,%
      clarkeModelCheckingState2012%
}.
There are also several testing tools and methodologies one can employ, such as Test-Driven Development {\textemdash}
{\`a} la JUnit {\textemdash} and
Fuzz-testing~\cite{yangFindingUnderstandingBugs2011,%
  hollerFuzzingCodeFragments2012%
},
which typically target imperative programming languages.

For functional languages, \acrfull{pbt} using
QuickCheck~\cite{claessenQuickCheckLightweightTool2000}
has proven widely
successful~\cite{hughesQuickCheckTestingFun2007,%
      claessenFindingRaceConditions2009,%
      artsTestingAUTOSARSoftware2015,%
      hughesExperiencesQuickCheckTesting2016,%
      chenPropertyBasedTestingClimbing2022%
}
and been ported to other languages, both functional and
imperative~\cite{berghoferRandomTestingIsabelle2004,%
      denesQuickChickPropertyBasedTesting2014,%
      padhyeJQFCoverageguidedPropertybased2019%
}.
Test-driven methods are well understood, but the tests are only as good as the
cases which the programmer can think of. On the flip side, model-checking tools
require a secondary implementation of the program in the form of a formal model,
which opens the verification process up to errors in translation leading to a
semantic mismatch between the running code and the verified
model~\cite{artsTestingAUTOSARSoftware2015}).

Dependently Typed languages like
Agda~\cite{boveBriefOverviewAgda2009}
and
\idris~\cite{bradyIdrisGeneralpurposeDependently2013,%
      bradyIdrisQuantitativeType2021%
}
can be used for ``correct by construction'' programming, where typically an
\acrfull{edsl} is constructed to ensure the program is valid by
definition~\cite{bhattiDomainSpecificLanguages2009,%
      bradyCorrectbyConstructionConcurrencyUsing2010,%
      bradyResourceSafeSystemsProgramming2012,%
      castro-perezZooidDSLCertified2021%
}.
However, well-typed programs can go wrong. Occasionally, this is due to bugs in
the type
checker~\cite{chaliasosWelltypedProgramsCan2021}
or due to problems with how programmers use the language-provided escape
hatches~\cite{qinUnderstandingMemoryThread2020},
but there is also a third, arguably more likely, case: what if the types
themselves are subtly incorrect? One could imagine a program requiring that a
certain number remain positive but either by habit or by accident, the
programmer gives the type \texttt{Int} instead of \texttt{Nat}. This is unlikely
to be caught by the type checker, tests, or implementations, as the programmers
are likely to carry this assumption in their minds, thereby avoiding including
it in both tests and implementation error-checks. Nevertheless, the code
modelling the specification has now introduced subtly different permitted
states. How can we be sure that the \acrshort{dsl} does not accidentally permit
an incorrect state or transition?

  \subsection{Contributions}
  We make the following contributions:
  \begin{itemize}
    \item An implementation of QuickCheck for use with dependent types at
          compile time.
    \item A framework for simultaneously specifying, implementing, and testing
          a stateful model of an \acrfull{atm}, using \QC to increase confidence
          in the correctness of all 3 parts.
    \item We demonstrate the power of the framework by generalising it to
          stateful programs, both finite and infinite, and evaluate it by
          implementing an example of a network protocol.
  \end{itemize}

  In doing so, we aim to increase confidence in the correctness of the
  types we use to model stateful systems, using type level testing to help
  us understand the behaviour of state machines.

\section{QuickCheck in \IIdris}

QuickCheck is a \acrlong{pbt} tool introduced by Claessen and Hughes in
2000~\cite{claessenQuickCheckLightweightTool2000}.
Although initially written for Haskell, it has been successfully ported to many
programming languages including
Isabelle~\cite{berghoferRandomTestingIsabelle2004},
Erlang~\cite{claessenFindingRaceConditions2009},
Coq~\cite{denesQuickChickPropertyBasedTesting2014},
and
Java~\cite{padhyeJQFCoverageguidedPropertybased2019}.

We can use our \QC implementation at the type level, since types in Idris2 are
first class. Thus the test suite can be run at compile-time, acting on the
implementation itself {\textemdash} rather than a test environment {\textemdash}
and then being erased for the compiled program.

  \subsection{Regular \QC in \IIdris}\label{ssec:reg-qc}

  Most of \QC ports directly to \iidris from the original Haskell
  implementation, with some minor modifications. We start with the type for
  specifying \emph{generators} of types:

  \begin{Verbatim}[commandchars=\\\{\}]
\PY{k+kr}{data}\PY{+w}{ }\PY{k+kt}{Gen}\PY{+w}{ }\PY{o+ow}{:}\PY{+w}{ }\PY{k+kt}{Type}\PY{+w}{ }\PY{o+ow}{\PYZhy{}\PYZgt{}}\PY{+w}{ }\PY{k+kt}{Type}\PY{+w}{ }\PY{k+kr}{where}
\PY{+w}{  }\PY{n+nf}{MkGen}\PY{+w}{ }\PY{o+ow}{:}\PY{+w}{  }\PY{o+ow}{(}\PY{k+kt}{Int}\PY{+w}{ }\PY{o+ow}{\PYZhy{}\PYZgt{}}\PY{+w}{ }\PY{k+kt}{PRNGState}\PY{+w}{ }\PY{o+ow}{\PYZhy{}\PYZgt{}}\PY{+w}{ }a\PY{o+ow}{)}\PY{+w}{ }\PY{o+ow}{\PYZhy{}\PYZgt{}}\PY{+w}{ }\PY{k+kt}{Gen}\PY{+w}{ }a
  \end{Verbatim}

  \iidris has no \texttt{newtype} so we have to wrap it in a regular datatype;
  the implementation is otherwise identical. The \texttt{PRNGState} type
  represents the state type of a \acrfull{prng}, which is essential for
  allowing QuickCheck to generate example instances.
  As in Haskell, \texttt{Gen} is an instance of the \texttt{Monad} interface.
  It requires passing two \emph{independent}
  \acrshortpl{prng}~\cite{claessenQuickCheckLightweightTool2000},
  which is easily achieved if the provided \acrshort{prng} is \emph{splittable}
  {\textemdash} that two seemingly independent \acrshortpl{prng} may be derived
  from an initial
  one~\cite{burtonDistributedRandomNumber1992}.

  \begin{Verbatim}[commandchars=\\\{\}]
\PY{o+ow}{(\PYZgt{}\PYZgt{}=)}\PY{+w}{ }\PY{o+ow}{(}\PY{k+kt}{MkGen}\PY{+w}{ }g1\PY{o+ow}{)}\PY{+w}{ }c\PY{+w}{ }\PY{o+ow}{=}\PY{+w}{ }\PY{k+kt}{MkGen}\PY{+w}{ }\PY{o+ow}{(\PYZbs{}}n,\PY{+w}{ }r0\PY{+w}{ }\PY{o+ow}{=\PYZgt{}}
\PY{+w}{ }\PY{+w}{ }\PY{k+kr}{let}\PY{+w}{ }\PY{o+ow}{(}r1,\PY{+w}{ }r2\PY{o+ow}{)}\PY{+w}{ }\PY{o+ow}{=}\PY{+w}{ }split\PY{+w}{ }r0
\PY{+w}{ }\PY{+w}{ }\PY{+w}{ }\PY{+w}{ }\PY{+w}{ }\PY{+w}{ }\PY{o+ow}{(}\PY{k+kt}{MkGen}\PY{+w}{ }g2\PY{o+ow}{)}\PY{+w}{ }\PY{o+ow}{=}\PY{+w}{ }c\PY{+w}{ }\PY{o+ow}{(}g1\PY{+w}{ }n\PY{+w}{ }r1\PY{o+ow}{)}
\PY{+w}{ }\PY{+w}{ }\PY{k+kr}{in}\PY{+w}{ }g2\PY{+w}{ }n\PY{+w}{ }r2\PY{o+ow}{)}
  \end{Verbatim}

  Statistically sound splits are challenging to achieve and have been the source
  of bugs in the Haskell
  implementation~\cite{claessenSplittablePseudorandomNumber2013}.
  Schaatun~\cite{schaathunEvaluationSplittablePseudorandom2015}
  concluded that amongst many existing allegedly splittable \acrshortpl{prng},
  only the one presented by Claessen and
  Pa{\l}ka~\cite{claessenSplittablePseudorandomNumber2013}
  was sound. As such, we do not make any cryptographic promises about the
  \acrshort{prng} used in our implementation: we use a \acrlong{lcg} for its
  ease of implementation, using known good
  multipliers~\cite{steeleComputationallyEasySpectrally2021}.
  This is sufficient for demonstrating our approach.

  Another difference between \iidris and Haskell is that \texttt{(->)} is a
  binder, and defining interface instances over binders is not allowed in
  \iidris. Therefore, we manually wrap functions in a data type alongside an
  eliminator:

  \begin{Verbatim}[commandchars=\\\{\}]
\PY{k+kr}{data}\PY{+w}{ }\PY{k+kt}{Fn}\PY{+w}{ }\PY{o+ow}{:}\PY{+w}{ }\PY{k+kt}{Type}\PY{+w}{ }\PY{o+ow}{\PYZhy{}\PYZgt{}}\PY{+w}{ }\PY{k+kt}{Type}\PY{+w}{ }\PY{o+ow}{\PYZhy{}\PYZgt{}}\PY{+w}{ }\PY{k+kt}{Type}\PY{+w}{ }\PY{k+kr}{where}
\PY{+w}{  }\PY{n+nf}{MkFn}\PY{+w}{ }\PY{o+ow}{:}\PY{+w}{ }\PY{o+ow}{(}f\PY{+w}{ }\PY{o+ow}{:}\PY{+w}{ }a\PY{+w}{ }\PY{o+ow}{\PYZhy{}\PYZgt{}}\PY{+w}{ }b\PY{o+ow}{)}\PY{+w}{ }\PY{o+ow}{\PYZhy{}\PYZgt{}}\PY{+w}{ }\PY{k+kt}{Fn}\PY{+w}{ }a\PY{+w}{ }b

\PY{n+nf}{apply}\PY{+w}{ }\PY{o+ow}{:}\PY{+w}{ }\PY{k+kt}{Fn}\PY{+w}{ }a\PY{+w}{ }b\PY{+w}{ }\PY{o+ow}{\PYZhy{}\PYZgt{}}\PY{+w}{ }a\PY{+w}{ }\PY{o+ow}{\PYZhy{}\PYZgt{}}\PY{+w}{ }b
apply\PY{+w}{ }\PY{o+ow}{(}\PY{k+kt}{MkFn}\PY{+w}{ }f\PY{o+ow}{)}\PY{+w}{ }\PY{o+ow}{=}\PY{+w}{ }f
  \end{Verbatim}

  We can now implement \emph{promoting} functions of type \texttt{a -> Gen b} to
  a generator of functions from \texttt{a} to \texttt{b}, as well as implement
  the \texttt{Arbitrary} interface {\textemdash} which indicates that we can
  generate arbitrary instances of a given type {\textemdash} for functions,
  provided we know how to (1) modify a generator given some specific instance of
  the domain's type, and (2) generate arbitrary instances of the codomain's
  type:

  \begin{Verbatim}[commandchars=\\\{\}]
\PY{n+nf}{promote}\PY{+w}{ }\PY{o+ow}{:}\PY{+w}{ }\PY{o+ow}{(}a\PY{+w}{ }\PY{o+ow}{\PYZhy{}\PYZgt{}}\PY{+w}{ }\PY{k+kt}{Gen}\PY{+w}{ }b\PY{o+ow}{)}\PY{+w}{ }\PY{o+ow}{\PYZhy{}\PYZgt{}}\PY{+w}{ }\PY{k+kt}{Gen}\PY{+w}{ }\PY{o+ow}{(}\PY{k+kt}{Fn}\PY{+w}{ }a\PY{+w}{ }b\PY{o+ow}{)}
\PY{+w}{ }\PY{+w}{ }promote\PY{+w}{ }f\PY{+w}{ }\PY{o+ow}{=}\PY{+w}{ }\PY{k+kt}{MkGen}\PY{+w}{ }\PY{o+ow}{(\PYZbs{}}n,\PY{+w}{ }r\PY{+w}{ }\PY{o+ow}{=\PYZgt{}}\PY{+w}{ }\PY{k+kt}{MkFn}
\PY{+w}{ }\PY{+w}{ }\PY{+w}{ }\PY{+w}{ }\PY{o+ow}{(\PYZbs{}}x\PY{+w}{ }\PY{o+ow}{=\PYZgt{}}\PY{+w}{ }\PY{k+kr}{let}\PY{+w}{ }\PY{o+ow}{(}\PY{k+kt}{MkGen}\PY{+w}{ }gb\PY{o+ow}{)}\PY{+w}{ }\PY{o+ow}{=}\PY{+w}{ }f\PY{+w}{ }x\PY{+w}{ }\PY{k+kr}{in}\PY{+w}{ }gb\PY{+w}{ }n\PY{+w}{ }r\PY{o+ow}{))}

\PY{k+kt}{Arbitrary}\PY{+w}{ }a\PY{+w}{ }\PY{o+ow}{=\PYZgt{}}\PY{+w}{ }\PY{k+kt}{Arbitrary}\PY{+w}{ }b\PY{+w}{ }\PY{o+ow}{=\PYZgt{}}
\PY{k+kt}{Arbitrary}\PY{+w}{ }\PY{o+ow}{(}\PY{k+kt}{Fn}\PY{+w}{ }a\PY{+w}{ }b\PY{o+ow}{)}\PY{+w}{ }\PY{k+kr}{where}
\PY{+w}{ }\PY{+w}{ }arbitrary\PY{+w}{ }\PY{o+ow}{=}\PY{+w}{ }promote\PY{+w}{ }\PY{o+ow}{(}`coarbitrary`\PY{+w}{ }arbitrary\PY{o+ow}{)}
\PY{+w}{ }\PY{+w}{ }coarbitrary\PY{+w}{ }fn\PY{+w}{ }gen\PY{+w}{ }\PY{o+ow}{=}\PY{+w}{ }arbitrary\PY{+w}{ }\PY{o+ow}{\PYZgt{}\PYZgt{}=}
\PY{+w}{ }\PY{+w}{ }\PY{+w}{ }\PY{+w}{ }\PY{o+ow}{((}`coarbitrary`\PY{+w}{ }gen\PY{o+ow}{)}\PY{+w}{ }\PY{o+ow}{.}\PY{+w}{ }\PY{o+ow}{(}apply\PY{+w}{ }fn\PY{o+ow}{))}
  \end{Verbatim}

  \subsection{QuickCheck with Dependent Types}\label{ssec:dept-qc}

  Using QuickCheck with dependent types presents new challenges. Consider, for
  example, the \texttt{Vect} datatype: lists which carry their length in the
  type. QuickCheck's bread and butter is generators and the \texttt{Arbitrary}
  interface. Provided we know how to generate the type of the elements, we
  should be able to generate a vector of them. However, the following definition
  fails:

  \begin{Verbatim}[commandchars=\\\{\}]
\PY{k+kt}{Arbitrary}\PY{+w}{ }t\PY{+w}{ }\PY{o+ow}{=\PYZgt{}}\PY{+w}{ }\PY{k+kt}{Arbitrary}\PY{+w}{ }\PY{o+ow}{(}\PY{k+kt}{Vect}\PY{+w}{ }n\PY{+w}{ }t\PY{o+ow}{)}\PY{+w}{ }\PY{k+kr}{where}
\PY{+w}{ }\PY{+w}{ }arbitrary\PY{+w}{ }\PY{o+ow}{=}\PY{+w}{ }\PY{k+kr}{do}
\PY{+w}{ }\PY{+w}{ }\PY{+w}{ }\PY{+w}{ }arbN\PY{+w}{ }\PY{o+ow}{\PYZlt{}\PYZhy{}}\PY{+w}{ }arbitrary
\PY{+w}{ }\PY{+w}{ }\PY{+w}{ }\PY{+w}{ }v\PY{+w}{ }\PY{o+ow}{\PYZlt{}\PYZhy{}}\PY{+w}{ }genN\PY{+w}{ }arbN\PY{+w}{ }arbitrary
\PY{+w}{ }\PY{+w}{ }\PY{+w}{ }\PY{+w}{ }pure\PY{+w}{ }v
\PY{+w}{ }\PY{+w}{ }\PY{+w}{ }\PY{+w}{ }\PY{k+kr}{where}
\PY{+w}{      }\PY{n+nf}{genN}\PY{+w}{ }\PY{o+ow}{:}\PY{+w}{ }\PY{o+ow}{(}n\PY{+w}{ }\PY{o+ow}{:}\PY{+w}{ }\PY{k+kt}{Nat}\PY{o+ow}{)}\PY{+w}{ }\PY{o+ow}{\PYZhy{}\PYZgt{}}\PY{+w}{ }\PY{k+kt}{Gen}\PY{+w}{ }t\PY{+w}{ }\PY{o+ow}{\PYZhy{}\PYZgt{}}\PY{+w}{ }\PY{k+kt}{Gen}\PY{+w}{ }\PY{o+ow}{(}\PY{k+kt}{Vect}\PY{+w}{ }n\PY{+w}{ }t\PY{o+ow}{)}
\PY{+w}{ }\PY{+w}{ }\PY{+w}{ }\PY{+w}{ }\PY{+w}{ }\PY{+w}{ }genN\PY{+w}{ }\PY{k+kt}{Z}\PY{+w}{ }\PY{k+kr}{\PYZus{}}\PY{+w}{ }\PY{o+ow}{=}\PY{+w}{ }\PY{o+ow}{[]}
\PY{+w}{ }\PY{+w}{ }\PY{+w}{ }\PY{+w}{ }\PY{+w}{ }\PY{+w}{ }genN\PY{+w}{ }\PY{o+ow}{(}\PY{k+kt}{S}\PY{+w}{ }k\PY{o+ow}{)}\PY{+w}{ }g\PY{+w}{ }\PY{o+ow}{=}\PY{+w}{ }\PY{k+kr}{do}
\PY{+w}{ }\PY{+w}{ }\PY{+w}{ }\PY{+w}{ }\PY{+w}{ }\PY{+w}{ }\PY{+w}{ }\PY{+w}{ }x\PY{+w}{ }\PY{+w}{ }\PY{o+ow}{\PYZlt{}\PYZhy{}}\PY{+w}{ }g
\PY{+w}{ }\PY{+w}{ }\PY{+w}{ }\PY{+w}{ }\PY{+w}{ }\PY{+w}{ }\PY{+w}{ }\PY{+w}{ }xs\PY{+w}{ }\PY{o+ow}{\PYZlt{}\PYZhy{}}\PY{+w}{ }genN\PY{+w}{ }k\PY{+w}{ }g
\PY{+w}{ }\PY{+w}{ }\PY{+w}{ }\PY{+w}{ }\PY{+w}{ }\PY{+w}{ }\PY{+w}{ }\PY{+w}{ }pure\PY{+w}{ }\PY{o+ow}{(}x\PY{+w}{ }\PY{o+ow}{::}\PY{+w}{ }xs\PY{o+ow}{)}
  \end{Verbatim}

  The type checker has no way of unifying \texttt{n} with \texttt{arbN} since
  \texttt{n} is implicitly generalised and bound outwith the context of the
  interface. We are telling the type checker that \texttt{arbN} should exactly
  match \emph{some} external \texttt{n} which nobody has thought of yet;
  understandably, this fails to type check.
  One solution is to write \texttt{Arbitrary} instances only for the sizes of
  vectors we are interested in:

  \begin{Verbatim}[commandchars=\\\{\}]
\PY{k+kt}{Arbitrary}\PY{+w}{ }t\PY{+w}{ }\PY{o+ow}{=\PYZgt{}}\PY{+w}{ }\PY{k+kt}{Arbitrary}\PY{+w}{ }\PY{o+ow}{(}\PY{k+kt}{Vect}\PY{+w}{ }\PY{l+m+mi}{3}\PY{+w}{ }t\PY{o+ow}{)}\PY{+w}{ }\PY{k+kr}{where}
\PY{+w}{ }\PY{+w}{ }arbitrary\PY{+w}{ }\PY{o+ow}{=}\PY{+w}{ }\PY{k+kr}{do}
\PY{+w}{ }\PY{+w}{ }\PY{+w}{ }\PY{+w}{ }v0\PY{+w}{ }\PY{o+ow}{\PYZlt{}\PYZhy{}}\PY{+w}{ }arbitrary
\PY{+w}{ }\PY{+w}{ }\PY{+w}{ }\PY{+w}{ }v1\PY{+w}{ }\PY{o+ow}{\PYZlt{}\PYZhy{}}\PY{+w}{ }arbitrary
\PY{+w}{ }\PY{+w}{ }\PY{+w}{ }\PY{+w}{ }v2\PY{+w}{ }\PY{o+ow}{\PYZlt{}\PYZhy{}}\PY{+w}{ }arbitrary
\PY{+w}{ }\PY{+w}{ }\PY{+w}{ }\PY{+w}{ }pure\PY{+w}{ }\PY{o+ow}{[}v0,\PY{+w}{ }v1,\PY{+w}{ }v2\PY{o+ow}{]}
  \end{Verbatim}

  However, this is not very useful: we need an implementation for every
  length, and should we want to generate an arbitrary length we would have to
  reduce our arbitrary length to only the lengths which we have defined
  generators for, defeating the point of \texttt{Arbitrary} interface.
  Instead, we can use dependent pairs, where the second element's type depends
  on the value of the first, written using \texttt{(**)}. This allows us to
  indicate to the type checker that although we may not know the exact length of
  the vector at type checking time, \emph{when} we know the length, we also know
  it is related to a concrete instance of \texttt{Vect}:

  \begin{Verbatim}[commandchars=\\\{\}]
\PY{n+nf}{someVect}\PY{+w}{ }\PY{o+ow}{:}\PY{+w}{ }\PY{o+ow}{(}n\PY{+w}{ }\PY{o+ow}{:}\PY{+w}{ }\PY{k+kt}{Nat}\PY{+w}{ }\PY{o+ow}{**}\PY{+w}{ }\PY{k+kt}{Vect}\PY{+w}{ }n\PY{+w}{ }\PY{k+kt}{Nat}\PY{o+ow}{)}
someVect\PY{+w}{ }\PY{o+ow}{=}\PY{+w}{ }\PY{o+ow}{(}\PY{l+m+mi}{3}\PY{+w}{ }\PY{o+ow}{**}\PY{+w}{ }\PY{o+ow}{[}\PY{l+m+mi}{1},\PY{+w}{ }\PY{l+m+mi}{2},\PY{+w}{ }\PY{l+m+mi}{3}\PY{o+ow}{])}
  \end{Verbatim}

  \iidris has \texttt{failing} blocks which compile if and only if they contain
  a term raising an error. Passing in a string requires it to be part of the
  error message, with the compiler rejecting the block if it fails with a
  different message. We can thus confirm it is an error to provide a mismatching
  length and \texttt{Vect}:

  \begin{Verbatim}[commandchars=\\\{\}]
\PY{k+kr}{failing}\PY{+w}{ }\PY{l+s}{\PYZdq{}}\PY{l+s}{Mismatch between: 1 and 0.}\PY{l+s}{\PYZdq{}}
\PY{+w}{  }\PY{n+nf}{sizeMismatch}\PY{+w}{ }\PY{o+ow}{:}\PY{+w}{ }\PY{o+ow}{(}n\PY{+w}{ }\PY{o+ow}{:}\PY{+w}{ }\PY{k+kt}{Nat}\PY{+w}{ }\PY{o+ow}{**}\PY{+w}{ }\PY{k+kt}{Vect}\PY{+w}{ }n\PY{+w}{ }\PY{k+kt}{Nat}\PY{o+ow}{)}
\PY{+w}{ }\PY{+w}{ }sizeMismatch\PY{+w}{ }\PY{o+ow}{=}\PY{+w}{ }\PY{o+ow}{(}\PY{l+m+mi}{0}\PY{+w}{ }\PY{o+ow}{**}\PY{+w}{ }\PY{o+ow}{[}\PY{l+m+mi}{3}\PY{o+ow}{])}
  \end{Verbatim}

  And we can ask \iidris to infer the first element's value:

  \begin{Verbatim}[commandchars=\\\{\}]
\PY{n+nf}{inferLength}\PY{+w}{ }\PY{o+ow}{:}\PY{+w}{ }\PY{o+ow}{(}n\PY{+w}{ }\PY{o+ow}{:}\PY{+w}{ }\PY{k+kt}{Nat}\PY{+w}{ }\PY{o+ow}{**}\PY{+w}{ }\PY{k+kt}{Vect}\PY{+w}{ }n\PY{+w}{ }\PY{k+kt}{Nat}\PY{o+ow}{)}
inferLength\PY{+w}{ }\PY{o+ow}{=}\PY{+w}{ }\PY{o+ow}{(}\PY{k+kr}{\PYZus{}}\PY{+w}{ }\PY{o+ow}{**}\PY{+w}{ }\PY{o+ow}{[}\PY{l+m+mi}{1},\PY{+w}{ }\PY{l+m+mi}{2},\PY{+w}{ }\PY{l+m+mi}{3}\PY{o+ow}{])}
  \end{Verbatim}

  Dependent pairs allow us to define a general \texttt{Arbitrary} instance for
  \texttt{Vect}:

  \begin{Verbatim}[commandchars=\\\{\}]
\PY{k+kt}{Arbitrary}\PY{+w}{ }t\PY{+w}{ }\PY{o+ow}{=\PYZgt{}}\PY{+w}{ }\PY{k+kt}{Arbitrary}\PY{+w}{ }\PY{o+ow}{(}n\PY{+w}{ }\PY{o+ow}{:}\PY{+w}{ }\PY{k+kt}{Nat}\PY{+w}{ }\PY{o+ow}{**}\PY{+w}{ }\PY{k+kt}{Vect}\PY{+w}{ }n\PY{+w}{ }t\PY{o+ow}{)}\PY{+w}{ }\PY{k+kr}{where}
\PY{+w}{ }\PY{+w}{ }arbitrary\PY{+w}{ }\PY{o+ow}{=}\PY{+w}{ }\PY{k+kr}{do}
\PY{+w}{ }\PY{+w}{ }\PY{+w}{ }\PY{+w}{ }nElems\PY{+w}{ }\PY{o+ow}{\PYZlt{}\PYZhy{}}\PY{+w}{ }arbitrary
\PY{+w}{ }\PY{+w}{ }\PY{+w}{ }\PY{+w}{ }vect\PY{+w}{ }\PY{+w}{ }\PY{+w}{ }\PY{o+ow}{\PYZlt{}\PYZhy{}}\PY{+w}{ }genN\PY{+w}{ }nElems
\PY{+w}{ }\PY{+w}{ }\PY{+w}{ }\PY{+w}{ }pure\PY{+w}{ }\PY{o+ow}{(}\PY{k+kr}{\PYZus{}}\PY{+w}{ }\PY{o+ow}{**}\PY{+w}{ }vect\PY{o+ow}{)}
\PY{+w}{ }\PY{+w}{ }\PY{+w}{ }\PY{+w}{ }\PY{k+kr}{where}
\PY{+w}{      }\PY{n+nf}{genN}\PY{+w}{ }\PY{o+ow}{:}\PY{+w}{ }\PY{k+kt}{Arbitrary}\PY{+w}{ }a\PY{+w}{ }\PY{o+ow}{=\PYZgt{}}\PY{+w}{ }\PY{o+ow}{(}m\PY{+w}{ }\PY{o+ow}{:}\PY{+w}{ }\PY{k+kt}{Nat}\PY{o+ow}{)}\PY{+w}{ }\PY{o+ow}{\PYZhy{}\PYZgt{}}\PY{+w}{ }\PY{k+kt}{Gen}\PY{+w}{ }\PY{o+ow}{(}\PY{k+kt}{Vect}\PY{+w}{ }m\PY{+w}{ }a\PY{o+ow}{)}
\PY{+w}{ }\PY{+w}{ }\PY{+w}{ }\PY{+w}{ }\PY{+w}{ }\PY{+w}{ }genN\PY{+w}{ }\PY{k+kt}{Z}\PY{+w}{ }\PY{+w}{ }\PY{+w}{ }\PY{+w}{ }\PY{+w}{ }\PY{o+ow}{=}\PY{+w}{ }pure\PY{+w}{ }\PY{o+ow}{[]}
\PY{+w}{ }\PY{+w}{ }\PY{+w}{ }\PY{+w}{ }\PY{+w}{ }\PY{+w}{ }genN\PY{+w}{ }\PY{o+ow}{(}\PY{k+kt}{S}\PY{+w}{ }k\PY{o+ow}{)}\PY{+w}{ }\PY{o+ow}{=}\PY{+w}{ }\PY{k+kr}{do}
\PY{+w}{ }\PY{+w}{ }\PY{+w}{ }\PY{+w}{ }\PY{+w}{ }\PY{+w}{ }\PY{+w}{ }\PY{+w}{ }x\PY{+w}{ }\PY{+w}{ }\PY{o+ow}{\PYZlt{}\PYZhy{}}\PY{+w}{ }arbitrary
\PY{+w}{ }\PY{+w}{ }\PY{+w}{ }\PY{+w}{ }\PY{+w}{ }\PY{+w}{ }\PY{+w}{ }\PY{+w}{ }xs\PY{+w}{ }\PY{o+ow}{\PYZlt{}\PYZhy{}}\PY{+w}{ }genN\PY{+w}{ }k
\PY{+w}{ }\PY{+w}{ }\PY{+w}{ }\PY{+w}{ }\PY{+w}{ }\PY{+w}{ }\PY{+w}{ }\PY{+w}{ }pure\PY{+w}{ }\PY{o+ow}{(}x\PY{+w}{ }\PY{o+ow}{::}\PY{+w}{ }xs\PY{o+ow}{)}
  \end{Verbatim}

\section{Example: ATM}\label{sec:atm}
We now consider an example \acrfull{fsm} modelling the behaviour of an
\acrfull{atm}. This example is used as a showcase of how dependent types can
help with correct stateful
programming~\cite{bradyTypedrivenDevelopmentIdris2017}.
However, as we shall shortly see, while dependent types go a long way towards
helping us be confident our program is correct, they are not enough on their
own.

  \subsection{The ATM state machine}

  The \acrshort{atm} state machine consists of three states:
  \begin{itemize}
    \item \textbf{Ready} {\textemdash} The starting state of the \acrshort{atm},
          representing the machine being ready for operation.
    \item \textbf{CardInserted} {\textemdash} When a card is present in the
          \acrshort{atm}, pending authorisation.
    \item \textbf{Session} {\textemdash} An authorised session whereby the user
          can dispense an amount of money.
  \end{itemize}

\begin{figure}[h]
    \centering
    \includegraphics[width=\linewidth]{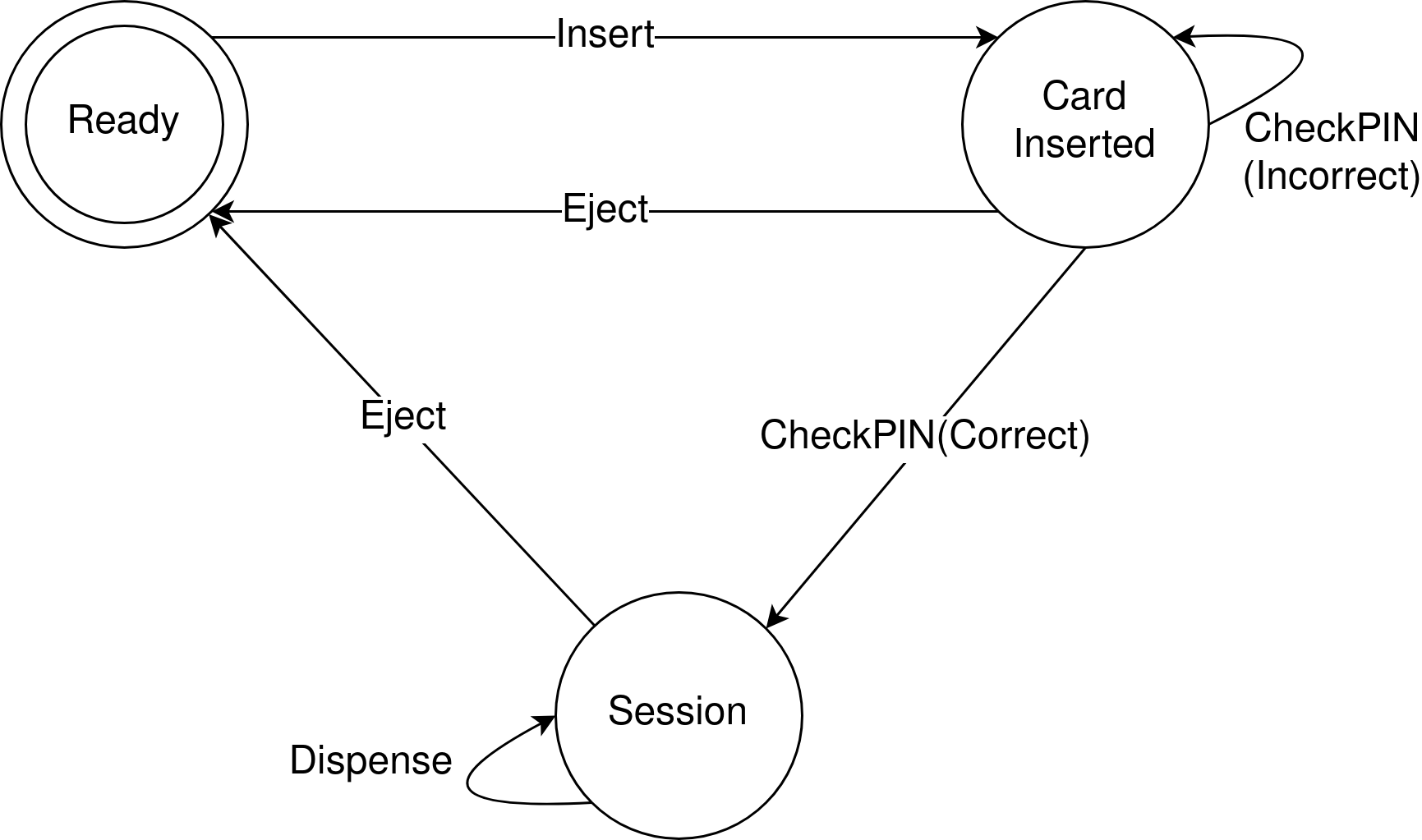}
    \caption{Diagram of the \acrshort{atm} state machine}
      \label{fig:atm}
    \Description{A state diagram with circles for each state, and labelled
                 arrows between them, with the labels containing the transition
                 names.}
  \end{figure}

  \Cref{fig:atm} illustrates the following transitions:
  \begin{itemize}
    \item \textit{Insert} {\textemdash} Inserting a bank card. This action is
          only valid when the \acrshort{atm} is in the \textbf{Ready} state and
          results in the machine changing to the \textbf{CardInserted} state.
    \item \textit{Dispense} {\textemdash} Dispensing a given amount of money.
          This is only valid when the card has been authenticated, i.e. the
          machine is in a \textbf{Session}. Since a user may want to dispense
          multiple amounts of money, \textit{Dispense} keeps the machine in its
          \textbf{Session} state.
    \item \textit{CheckPIN} {\textemdash} Verifying that the given PIN
          authenticates the card. This is only valid when the \acrshort{atm} has
          a card in it. This transition is unique in that it leads to different
          states depending on the result of checking the PIN: \textit{Incorrect}
          causes the machine to stay in the \textbf{CardInserted} state, whereas
          \textit{Correct} moves the machine to the \textbf{Sesssion} state.
    \item \textit{Eject} {\textemdash} At any point, the user may choose to
          eject their card. This takes the machine back to the \textbf{Ready}
          state.
  \end{itemize}

  \subsection{Modelling the ATM in \IIdris}

  We can model this state machine in \iidris by declaring a new data type for the
  \textit{CheckPIN} results, with constructors for each option, as well as a
  data type for the states, with a constructor per state:

  \begin{Verbatim}[commandchars=\\\{\}]
\PY{k+kr}{data}\PY{+w}{ }\PY{k+kt}{ATMState}\PY{+w}{ }\PY{o+ow}{=}\PY{+w}{ }\PY{k+kt}{Ready}\PY{+w}{ }\PY{o+ow}{|}\PY{+w}{ }\PY{k+kt}{CardInserted}\PY{+w}{ }\PY{o+ow}{|}\PY{+w}{ }\PY{k+kt}{Session}

\PY{k+kr}{data}\PY{+w}{ }\PY{k+kt}{PINok}\PY{+w}{ }\PY{o+ow}{=}\PY{+w}{ }\PY{k+kt}{Correct}\PY{+w}{ }\PY{o+ow}{|}\PY{+w}{ }\PY{k+kt}{Incorrect}
  \end{Verbatim}

  Next, we model the function describing the dependent state transition for
  \textit{CheckPIN}. This is a function from the result type, \texttt{PINok}, to
  the type of the states, \texttt{ATMState}:

  \begin{Verbatim}[commandchars=\\\{\}]
\PY{n+nf}{ChkPINfn}\PY{+w}{ }\PY{o+ow}{:}\PY{+w}{ }\PY{k+kt}{PINok}\PY{+w}{ }\PY{o+ow}{\PYZhy{}\PYZgt{}}\PY{+w}{ }\PY{k+kt}{ATMState}
ChkPINfn\PY{+w}{ }\PY{k+kt}{Correct}\PY{+w}{ }\PY{o+ow}{=}\PY{+w}{ }\PY{k+kt}{Session}
ChkPINfn\PY{+w}{ }\PY{k+kt}{Incorrect}\PY{+w}{ }\PY{o+ow}{=}\PY{+w}{ }\PY{k+kt}{CardInserted}
  \end{Verbatim}

  With the states, PIN results, and dependent transition modelled, we can now
  model the transitions themselves. As described in the \idris
  book~\cite{bradyTypedrivenDevelopmentIdris2017},
  this is where dependent types really get a chance to shine for modelling and
  programming these stateful systems: we index our operations by their result
  type, their starting state, and their state transition functions. This allows
  us to use the type declaration to state what the result type of our program
  should be, its starting state, and its end state, and having the type checker
  verify that we keep our promise and reach the end state via lawful
  transitions. Furthermore, we supply a bind operator for \texttt{do}-notation.
  The transition function, for most states, is a \texttt{const} function, as
  they only move from one state to the next. However, for \textit{CheckPIN},
  the state function is more interesting. We refer to the complete model of
  operations as an \acrfull{ism}:

  \begin{Verbatim}[commandchars=\\\{\}]
\PY{k+kr}{data}\PY{+w}{ }\PY{k+kt}{ATM}\PY{+w}{ }\PY{o+ow}{:}\PY{+w}{ }\PY{+w}{ }\PY{o+ow}{(}t\PY{+w}{ }\PY{o+ow}{:}\PY{+w}{ }\PY{k+kt}{Type}\PY{o+ow}{)}\PY{+w}{ }\PY{o+ow}{\PYZhy{}\PYZgt{}}\PY{+w}{ }\PY{k+kt}{ATMState}
\PY{+w}{ }\PY{+w}{ }\PY{+w}{ }\PY{+w}{ }\PY{+w}{ }\PY{+w}{ }\PY{+w}{ }\PY{+w}{ }\PY{+w}{ }\PY{o+ow}{\PYZhy{}\PYZgt{}}\PY{+w}{ }\PY{o+ow}{(}t\PY{+w}{ }\PY{o+ow}{\PYZhy{}\PYZgt{}}\PY{+w}{ }\PY{k+kt}{ATMState}\PY{o+ow}{)}\PY{+w}{ }\PY{o+ow}{\PYZhy{}\PYZgt{}}\PY{+w}{ }\PY{k+kt}{Type}\PY{+w}{ }\PY{k+kr}{where}
\PY{+w}{  }\PY{n+nf}{Insert}\PY{+w}{ }\PY{o+ow}{:}\PY{+w}{ }\PY{k+kt}{ATM}\PY{+w}{ }\PY{o+ow}{()}\PY{+w}{ }\PY{k+kt}{Ready}\PY{+w}{ }\PY{o+ow}{(}const\PY{+w}{ }\PY{k+kt}{CardInserted}\PY{o+ow}{)}
\PY{+w}{  }\PY{n+nf}{CheckPIN}\PY{+w}{ }\PY{o+ow}{:}\PY{+w}{ }\PY{o+ow}{(}pin\PY{+w}{ }\PY{o+ow}{:}\PY{+w}{ }\PY{k+kt}{Int}\PY{o+ow}{)}\PY{+w}{ }\PY{o+ow}{\PYZhy{}\PYZgt{}}\PY{+w}{ }\PY{k+kt}{ATM}\PY{+w}{ }\PY{k+kt}{PINok}\PY{+w}{ }\PY{k+kt}{CardInserted}
\PY{+w}{ }\PY{+w}{ }\PY{+w}{ }\PY{+w}{ }\PY{+w}{ }\PY{+w}{ }\PY{+w}{ }\PY{+w}{ }\PY{+w}{ }\PY{+w}{ }\PY{+w}{ }\PY{+w}{ }\PY{+w}{ }\PY{+w}{ }\PY{+w}{ }\PY{+w}{ }\PY{+w}{ }\PY{+w}{ }\PY{+w}{ }\PY{+w}{ }\PY{+w}{ }\PY{+w}{ }\PY{+w}{ }\PY{+w}{ }\PY{+w}{ }\PY{+w}{ }\PY{+w}{ }\PY{+w}{ }\PY{+w}{ }\PY{+w}{ }ChkPINfn
\PY{+w}{  }\PY{n+nf}{Dispense}\PY{+w}{ }\PY{o+ow}{:}\PY{+w}{ }\PY{o+ow}{(}amt\PY{+w}{ }\PY{o+ow}{:}\PY{+w}{ }\PY{k+kt}{Nat}\PY{o+ow}{)}\PY{+w}{ }\PY{o+ow}{\PYZhy{}\PYZgt{}}\PY{+w}{ }\PY{k+kt}{ATM}\PY{+w}{ }\PY{o+ow}{()}\PY{+w}{ }\PY{k+kt}{Session}
\PY{+w}{ }\PY{+w}{ }\PY{+w}{ }\PY{+w}{ }\PY{+w}{ }\PY{+w}{ }\PY{+w}{ }\PY{+w}{ }\PY{+w}{ }\PY{+w}{ }\PY{+w}{ }\PY{+w}{ }\PY{+w}{ }\PY{+w}{ }\PY{+w}{ }\PY{+w}{ }\PY{+w}{ }\PY{+w}{ }\PY{+w}{ }\PY{+w}{ }\PY{+w}{ }\PY{+w}{ }\PY{+w}{ }\PY{+w}{ }\PY{+w}{ }\PY{+w}{ }\PY{+w}{ }\PY{+w}{ }\PY{+w}{ }\PY{+w}{ }\PY{o+ow}{(}const\PY{+w}{ }\PY{k+kt}{Session}\PY{o+ow}{)}
\PY{+w}{  }\PY{n+nf}{Eject}\PY{+w}{ }\PY{o+ow}{:}\PY{+w}{ }\PY{k+kt}{ATM}\PY{+w}{ }\PY{o+ow}{()}\PY{+w}{ }st\PY{+w}{ }\PY{o+ow}{(}const\PY{+w}{ }\PY{k+kt}{Ready}\PY{o+ow}{)}
\PY{+w}{  }\PY{n+nf}{Pure}\PY{+w}{ }\PY{o+ow}{:}\PY{+w}{ }\PY{o+ow}{(}x\PY{+w}{ }\PY{o+ow}{:}\PY{+w}{ }t\PY{o+ow}{)}\PY{+w}{ }\PY{o+ow}{\PYZhy{}\PYZgt{}}\PY{+w}{ }\PY{k+kt}{ATM}\PY{+w}{ }t\PY{+w}{ }\PY{o+ow}{(}stFn\PY{+w}{ }x\PY{o+ow}{)}\PY{+w}{ }stFn
\PY{+w}{ }\PY{+w}{ }\PY{o+ow}{(\PYZgt{}\PYZgt{}=)}\PY{+w}{ }\PY{o+ow}{:}\PY{+w}{ }\PY{k+kt}{ATM}\PY{+w}{ }a\PY{+w}{ }s1\PY{+w}{ }s2f\PY{+w}{ }\PY{o+ow}{\PYZhy{}\PYZgt{}}\PY{+w}{ }\PY{o+ow}{((}x\PY{+w}{ }\PY{o+ow}{:}\PY{+w}{ }a\PY{o+ow}{)}\PY{+w}{ }\PY{o+ow}{\PYZhy{}\PYZgt{}}\PY{+w}{ }\PY{k+kt}{ATM}\PY{+w}{ }b\PY{+w}{ }\PY{o+ow}{(}s2f\PY{+w}{ }x\PY{o+ow}{)}
\PY{+w}{ }\PY{+w}{ }\PY{+w}{ }\PY{+w}{ }\PY{+w}{ }\PY{+w}{ }\PY{+w}{ }\PY{+w}{ }\PY{+w}{ }\PY{+w}{ }\PY{+w}{ }\PY{+w}{ }s3f\PY{o+ow}{)}\PY{+w}{ }\PY{o+ow}{\PYZhy{}\PYZgt{}}\PY{+w}{ }\PY{k+kt}{ATM}\PY{+w}{ }b\PY{+w}{ }s1\PY{+w}{ }s3f
  \end{Verbatim}

  We can now use the model to write stateful
  programs which are guaranteed to conform to the model:

  \begin{Verbatim}[commandchars=\\\{\}]
\PY{n+nf}{testProg}\PY{+w}{ }\PY{o+ow}{:}\PY{+w}{ }\PY{k+kt}{ATM}\PY{+w}{ }\PY{o+ow}{()}\PY{+w}{ }\PY{k+kt}{Ready}\PY{+w}{ }\PY{o+ow}{(}const\PY{+w}{ }\PY{k+kt}{Ready}\PY{o+ow}{)}
testProg\PY{+w}{ }\PY{o+ow}{=}\PY{+w}{ }\PY{k+kr}{do}
\PY{+w}{ }\PY{+w}{ }\PY{k+kt}{Insert}
\PY{+w}{ }\PY{+w}{ }\PY{k+kt}{Correct}\PY{+w}{ }\PY{o+ow}{\PYZlt{}\PYZhy{}}\PY{+w}{ }\PY{k+kt}{CheckPIN}\PY{+w}{ }\PY{l+m+mi}{1234}
\PY{+w}{ }\PY{+w}{ }\PY{+w}{ }\PY{+w}{ }\PY{o+ow}{|}\PY{+w}{ }\PY{k+kt}{Incorrect}\PY{+w}{ }\PY{o+ow}{=\PYZgt{}}\PY{+w}{ }\PY{o+ow}{?}handle\PYZus{}incorrect
\PY{+w}{ }\PY{+w}{ }\PY{k+kt}{Dispense}\PY{+w}{ }\PY{l+m+mi}{42}
\PY{+w}{ }\PY{+w}{ }\PY{k+kt}{Eject}
  \end{Verbatim}

  Here we use \iidris's pattern matching bind syntax to continue with the
  \texttt{do} block if \texttt{CheckPIN} was happy, and a hole to leave the
  unhappy case for later implementation.

  Programs which attempt to misbehave are rejected:

  \begin{Verbatim}[commandchars=\\\{\}]
\PY{k+kr}{failing}\PY{+w}{ }\PY{l+s}{\PYZdq{}}\PY{l+s}{Mismatch between: Session and CardInserted.}\PY{l+s}{\PYZdq{}}
\PY{+w}{  }\PY{n+nf}{badProg}\PY{+w}{ }\PY{o+ow}{:}\PY{+w}{ }\PY{k+kt}{ATM}\PY{+w}{ }\PY{o+ow}{()}\PY{+w}{ }\PY{k+kt}{Ready}\PY{+w}{ }\PY{o+ow}{(}const\PY{+w}{ }\PY{k+kt}{Ready}\PY{o+ow}{)}
\PY{+w}{ }\PY{+w}{ }badProg\PY{+w}{ }\PY{o+ow}{=}\PY{+w}{ }\PY{k+kr}{do}\PY{+w}{ }\PY{k+kt}{Insert}\PY{+w}{ };\PY{+w}{ }\PY{k+kt}{Dispense}\PY{+w}{ }\PY{l+m+mi}{42}
  \end{Verbatim}

  This setup looks to be correct: we have our dependently typed model which
  describes the desired semantics; we can use it to program with, expressing
  state invariants which must be obeyed and which are automatically verified;
  and whilst writing the program, the type checker keeps track of the state for
  us. This is a very strong position compared to languages without such typed
  modelling capabilities. However, while it may seem correct, there is a
  mistake in this model of the ATM: the amount of PIN retries is unlimited. The
  \texttt{ChkPINfn} only takes a \texttt{PINok} result, neither it nor the
  \textbf{CardInserted}-state keeps track of how many times the user has tried
  to enter a PIN. The following program, while not terminating, is completely
  valid as far as the type checker knows:

  \begin{Verbatim}[commandchars=\\\{\}]
\PY{k+kr}{covering}
\PY{n+nf}{loopProg}\PY{+w}{ }\PY{o+ow}{:}\PY{+w}{ }\PY{k+kt}{ATM}\PY{+w}{ }\PY{o+ow}{()}\PY{+w}{ }\PY{k+kt}{Ready}\PY{+w}{ }\PY{o+ow}{(}const\PY{+w}{ }\PY{k+kt}{Ready}\PY{o+ow}{)}
loopProg\PY{+w}{ }\PY{o+ow}{=}\PY{+w}{ }\PY{k+kr}{do}
\PY{+w}{ }\PY{+w}{ }\PY{+w}{ }\PY{+w}{ }\PY{k+kt}{Insert}
\PY{+w}{ }\PY{+w}{ }\PY{+w}{ }\PY{+w}{ }\PY{k+kr}{let}\PY{+w}{ }pin\PY{+w}{ }\PY{o+ow}{=}\PY{+w}{ }\PY{l+m+mi}{4321}
\PY{+w}{ }\PY{+w}{ }\PY{+w}{ }\PY{+w}{ }loop\PY{+w}{ }pin
\PY{+w}{ }\PY{+w}{ }\PY{k+kr}{where}
\PY{+w}{    }\PY{n+nf}{loop}\PY{+w}{ }\PY{o+ow}{:}\PY{+w}{ }\PY{k+kt}{Int}\PY{+w}{ }\PY{o+ow}{\PYZhy{}\PYZgt{}}\PY{+w}{ }\PY{k+kt}{ATM}\PY{+w}{ }\PY{o+ow}{()}\PY{+w}{ }\PY{k+kt}{CardInserted}\PY{+w}{ }\PY{o+ow}{(}const\PY{+w}{ }\PY{k+kt}{Ready}\PY{o+ow}{)}
\PY{+w}{ }\PY{+w}{ }\PY{+w}{ }\PY{+w}{ }loop\PY{+w}{ }p\PY{+w}{ }\PY{o+ow}{=}\PY{+w}{ }\PY{k+kr}{do}
\PY{+w}{ }\PY{+w}{ }\PY{+w}{ }\PY{+w}{ }\PY{+w}{ }\PY{+w}{ }\PY{k+kt}{Incorrect}\PY{+w}{ }\PY{o+ow}{\PYZlt{}\PYZhy{}}\PY{+w}{ }\PY{k+kt}{CheckPIN}\PY{+w}{ }p
\PY{+w}{ }\PY{+w}{ }\PY{+w}{ }\PY{+w}{ }\PY{+w}{ }\PY{+w}{ }\PY{+w}{ }\PY{+w}{ }\PY{o+ow}{|}\PY{+w}{ }\PY{k+kt}{Correct}\PY{+w}{ }\PY{o+ow}{=\PYZgt{}}\PY{+w}{ }\PY{o+ow}{?}omitted
\PY{+w}{ }\PY{+w}{ }\PY{+w}{ }\PY{+w}{ }\PY{+w}{ }\PY{+w}{ }loop\PY{+w}{ }p
    \end{Verbatim}

  Unfortunately, \emph{this error is not caught by the type checker.}
  Even totality checking does not help: it is a terminating computation, for
  example, to iterate over all 10,000 PINs and withdrawing all the money on
  finding the correct one. This illustrates a tricky situation: as type-driven
  programmers, we are inclined to believe that expressive types mean the type
  checker will catch our mistakes. Nevertheless, subtle errors may occur in our
  modelling, and there is no way to automatically catch these unless the
  programmer tries to write non-obvious programs.
  Who type checks the types?

  \subsection{A framework for ATM simulation}

  To gain confidence in our specification, we could try
  modelling it in a formal verification tool or model
  checker, but this does not solve the root of the problem: our models
  themselves can be wrong, and so translating them into different tools gives
  more places for introducing errors, or worse, different errors in each model.
  We would instead like to generate example instances of each part of our model,
  pass these through our state transitions, and specify properties which,
  provided well-typed inputs, the model obeys. Ideally, this should be done in
  the same development environment as the model and implementation, thus
  eliminating the risk of translation mishaps. With some work, we can achieve
  this with \QC.

  In \cref{ssec:dept-qc} we saw how dependent pairs allow us to generate
  arbitrary dependent types. This means we could, hypothetically, declare the
  following \texttt{Arbitrary} instance:

  \begin{Verbatim}[commandchars=\\\{\}]
\PY{k+kt}{Arbitrary}\PY{+w}{ }\PY{o+ow}{(}resT\PY{+w}{ }\PY{o+ow}{:}\PY{+w}{ }\PY{k+kt}{Type}\PY{+w}{ }\PY{o+ow}{**}\PY{+w}{ }nsFn\PY{+w}{ }\PY{o+ow}{:}\PY{+w}{ }resT\PY{+w}{ }\PY{o+ow}{\PYZhy{}\PYZgt{}}\PY{+w}{ }\PY{k+kt}{ATMState}
\PY{+w}{ }\PY{+w}{ }\PY{+w}{ }\PY{+w}{ }\PY{+w}{ }\PY{+w}{ }\PY{+w}{ }\PY{+w}{ }\PY{+w}{ }\PY{+w}{ }\PY{+w}{ }\PY{o+ow}{**}\PY{+w}{ }\PY{k+kt}{ATM}\PY{+w}{ }resT\PY{+w}{ }st\PY{+w}{ }nsFn\PY{o+ow}{)}
  \end{Verbatim}

  If we know the result type, state function's type, and some starting state, we
  have all the necessary information to construct a concrete instance of the
  \texttt{ATM} type. However, such an instance has a couple of problems:

  \begin{enumerate}
    \item It would only generate a single operation at a time, with no obvious
          way to trace which operations were taken when. Related to this is the
          issue of how to generate instances of the binding and sequencing
          operators, which each require a specific pair of operations to work
          correctly.
    \item To advance to the next state, we need an instance of the result
          type \texttt{resT}. However, we only know the
          \emph{type} of results the operation returns, we do not know
          \emph{which} instance of that type it returned. We could make up a
          value, but then we would fix a parameter that we want to test.
  \end{enumerate}

  \subsubsection{Separating operations from programming logic}\label{ssec:atm-op-prog}

  To address the first problem, we split the operations and the sequencing into
  separate types:

  \begin{Verbatim}[commandchars=\\\{\}]
  \PY{k+kr}{data}\PY{+w}{ }\PY{k+kt}{ATMOp}\PY{+w}{ }\PY{o+ow}{:}\PY{+w}{ }\PY{+w}{ }\PY{o+ow}{(}t\PY{+w}{ }\PY{o+ow}{:}\PY{+w}{ }\PY{k+kt}{Type}\PY{o+ow}{)}\PY{+w}{ }\PY{o+ow}{\PYZhy{}\PYZgt{}}\PY{+w}{ }\PY{k+kt}{ATMState}
  \PY{+w}{ }\PY{+w}{ }\PY{+w}{ }\PY{+w}{ }\PY{+w}{ }\PY{+w}{ }\PY{+w}{ }\PY{+w}{ }\PY{+w}{ }\PY{+w}{ }\PY{+w}{ }\PY{o+ow}{\PYZhy{}\PYZgt{}}\PY{+w}{ }\PY{o+ow}{(}t\PY{+w}{ }\PY{o+ow}{\PYZhy{}\PYZgt{}}\PY{+w}{ }\PY{k+kt}{ATMState}\PY{o+ow}{)}\PY{+w}{ }\PY{o+ow}{\PYZhy{}\PYZgt{}}\PY{+w}{ }\PY{k+kt}{Type}\PY{+w}{ }\PY{k+kr}{where}
  \PY{+w}{  }\PY{n+nf}{Insert}\PY{+w}{ }\PY{o+ow}{:}\PY{+w}{ }\PY{k+kt}{ATMOp}\PY{+w}{ }\PY{o+ow}{()}\PY{+w}{ }\PY{k+kt}{Ready}\PY{+w}{ }\PY{o+ow}{(}const\PY{+w}{ }\PY{k+kt}{CardInserted}\PY{o+ow}{)}
  \PY{+w}{  }\PY{n+nf}{CheckPIN}\PY{+w}{ }\PY{o+ow}{:}\PY{+w}{ }\PY{o+ow}{(}pin\PY{+w}{ }\PY{o+ow}{:}\PY{+w}{ }\PY{k+kt}{Int}\PY{o+ow}{)}\PY{+w}{ }\PY{o+ow}{\PYZhy{}\PYZgt{}}\PY{+w}{ }\PY{k+kt}{ATMOp}\PY{+w}{ }\PY{k+kt}{PINok}\PY{+w}{ }\PY{k+kt}{CardInserted}
  \PY{+w}{ }\PY{+w}{ }\PY{+w}{ }\PY{+w}{ }\PY{+w}{ }\PY{+w}{ }\PY{+w}{ }\PY{+w}{ }\PY{+w}{ }\PY{+w}{ }\PY{+w}{ }\PY{+w}{ }\PY{+w}{ }\PY{+w}{ }\PY{+w}{ }\PY{+w}{ }\PY{+w}{ }\PY{+w}{ }\PY{+w}{ }\PY{+w}{ }\PY{+w}{ }\PY{+w}{ }\PY{+w}{ }\PY{+w}{ }\PY{+w}{ }\PY{+w}{ }\PY{+w}{ }\PY{+w}{ }\PY{+w}{ }\PY{+w}{ }ChkPINfn
  \PY{+w}{  }\PY{n+nf}{Dispense}\PY{+w}{ }\PY{o+ow}{:}\PY{+w}{ }\PY{o+ow}{(}amt\PY{+w}{ }\PY{o+ow}{:}\PY{+w}{ }\PY{k+kt}{Nat}\PY{o+ow}{)}\PY{+w}{ }\PY{o+ow}{\PYZhy{}\PYZgt{}}\PY{+w}{ }\PY{k+kt}{ATMOp}\PY{+w}{ }\PY{o+ow}{()}\PY{+w}{ }\PY{k+kt}{Session}
  \PY{+w}{ }\PY{+w}{ }\PY{+w}{ }\PY{+w}{ }\PY{+w}{ }\PY{+w}{ }\PY{+w}{ }\PY{+w}{ }\PY{+w}{ }\PY{+w}{ }\PY{+w}{ }\PY{+w}{ }\PY{+w}{ }\PY{+w}{ }\PY{+w}{ }\PY{+w}{ }\PY{+w}{ }\PY{+w}{ }\PY{+w}{ }\PY{+w}{ }\PY{+w}{ }\PY{+w}{ }\PY{+w}{ }\PY{+w}{ }\PY{+w}{ }\PY{+w}{ }\PY{+w}{ }\PY{+w}{ }\PY{+w}{ }\PY{+w}{ }\PY{o+ow}{(}const\PY{+w}{ }\PY{k+kt}{Session}\PY{o+ow}{)}
  \PY{+w}{  }\PY{n+nf}{Eject}\PY{+w}{ }\PY{o+ow}{:}\PY{+w}{ }\PY{k+kt}{ATMOp}\PY{+w}{ }\PY{o+ow}{()}\PY{+w}{ }st\PY{+w}{ }\PY{o+ow}{(}const\PY{+w}{ }\PY{k+kt}{Ready}\PY{o+ow}{)}

  \PY{k+kr}{data}\PY{+w}{ }\PY{k+kt}{ATM}\PY{+w}{ }\PY{o+ow}{:}\PY{+w}{ }\PY{+w}{ }\PY{o+ow}{(}t\PY{+w}{ }\PY{o+ow}{:}\PY{+w}{ }\PY{k+kt}{Type}\PY{o+ow}{)}\PY{+w}{ }\PY{o+ow}{\PYZhy{}\PYZgt{}}\PY{+w}{ }\PY{k+kt}{ATMState}
  \PY{+w}{ }\PY{+w}{ }\PY{+w}{ }\PY{+w}{ }\PY{+w}{ }\PY{+w}{ }\PY{+w}{ }\PY{+w}{ }\PY{+w}{ }\PY{o+ow}{\PYZhy{}\PYZgt{}}\PY{+w}{ }\PY{o+ow}{(}t\PY{+w}{ }\PY{o+ow}{\PYZhy{}\PYZgt{}}\PY{+w}{ }\PY{k+kt}{ATMState}\PY{o+ow}{)}\PY{+w}{ }\PY{o+ow}{\PYZhy{}\PYZgt{}}\PY{+w}{ }\PY{k+kt}{Type}\PY{+w}{ }\PY{k+kr}{where}
  \PY{+w}{  }\PY{n+nf}{Op}\PY{+w}{ }\PY{o+ow}{:}\PY{+w}{ }\PY{k+kt}{ATMOp}\PY{+w}{ }t\PY{+w}{ }st\PY{+w}{ }nsFn\PY{+w}{ }\PY{o+ow}{\PYZhy{}\PYZgt{}}\PY{+w}{ }\PY{k+kt}{ATM}\PY{+w}{ }t\PY{+w}{ }st\PY{+w}{ }nsFn
  \PY{+w}{  }\PY{n+nf}{Pure}\PY{+w}{ }\PY{o+ow}{:}\PY{+w}{ }\PY{o+ow}{(}x\PY{+w}{ }\PY{o+ow}{:}\PY{+w}{ }t\PY{o+ow}{)}\PY{+w}{ }\PY{o+ow}{\PYZhy{}\PYZgt{}}\PY{+w}{ }\PY{k+kt}{ATM}\PY{+w}{ }t\PY{+w}{ }\PY{o+ow}{(}nsFn\PY{+w}{ }x\PY{o+ow}{)}\PY{+w}{ }nsFn
  \PY{+w}{ }\PY{+w}{ }\PY{o+ow}{(\PYZgt{}\PYZgt{}=)}\PY{+w}{ }\PY{o+ow}{:}\PY{+w}{ }\PY{k+kt}{ATM}\PY{+w}{ }a\PY{+w}{ }s1\PY{+w}{ }s2f\PY{+w}{ }\PY{o+ow}{\PYZhy{}\PYZgt{}}\PY{+w}{ }\PY{o+ow}{((}x\PY{+w}{ }\PY{o+ow}{:}\PY{+w}{ }a\PY{o+ow}{)}\PY{+w}{ }\PY{o+ow}{\PYZhy{}\PYZgt{}}\PY{+w}{ }\PY{k+kt}{ATM}\PY{+w}{ }b\PY{+w}{ }\PY{o+ow}{(}s2f\PY{+w}{ }x\PY{o+ow}{)}
  \PY{+w}{ }\PY{+w}{ }\PY{+w}{ }\PY{+w}{ }\PY{+w}{ }\PY{+w}{ }\PY{+w}{ }\PY{+w}{ }\PY{+w}{ }\PY{+w}{ }\PY{+w}{ }\PY{+w}{ }s3f\PY{o+ow}{)}\PY{+w}{ }\PY{o+ow}{\PYZhy{}\PYZgt{}}\PY{+w}{ }\PY{k+kt}{ATM}\PY{+w}{ }b\PY{+w}{ }s1\PY{+w}{ }s3f
  \end{Verbatim}

  This separation allows us to access the next-state function directly from the
  \texttt{ATMOp} type. We already specify it as part of the type, so given a
  concrete \texttt{ATMOp} we should have access to its next-state function. The
  only caveat is that we need to be operating at the type level. \iidris erases
  runtime-irrelevant proofs and types by default, meaning we are only allowed to
  access them in parts of the program that are themselves erased, that is, those
  defined at quantity $0$ in
  \acrshort{qtt}~\cite{mcbrideGotPlentyNuttin2016,%
      atkeySyntaxSemanticsQuantitative2018,%
      bradyIdrisQuantitativeType2021%
  }.

  \begin{Verbatim}[commandchars=\\\{\}]
\PY{l+m+mi}{0}\PY{+w}{ }nextState\PY{+w}{ }\PY{o+ow}{:}\PY{+w}{ }\PY{+w}{ }\PY{o+ow}{(}st\PY{+w}{ }\PY{o+ow}{:}\PY{+w}{ }\PY{k+kt}{ATMState}\PY{o+ow}{)}\PY{+w}{ }\PY{o+ow}{\PYZhy{}\PYZgt{}}\PY{+w}{ }\PY{k+kt}{ATMOp}\PY{+w}{ }t\PY{+w}{ }st\PY{+w}{ }nsFn
\PY{+w}{ }\PY{+w}{ }\PY{+w}{ }\PY{+w}{ }\PY{+w}{ }\PY{+w}{ }\PY{+w}{ }\PY{+w}{ }\PY{+w}{ }\PY{+w}{ }\PY{+w}{ }\PY{+w}{ }\PY{o+ow}{\PYZhy{}\PYZgt{}}\PY{+w}{ }\PY{o+ow}{(}res\PY{+w}{ }\PY{o+ow}{:}\PY{+w}{ }t\PY{o+ow}{)}\PY{+w}{ }\PY{o+ow}{\PYZhy{}\PYZgt{}}\PY{+w}{ }\PY{k+kt}{ATMState}
nextState\PY{+w}{ }st\PY{+w}{ }\PY{k+kr}{\PYZus{}}\PY{+w}{ }res\PY{+w}{ }\PY{o+ow}{=}\PY{+w}{ }nsFn\PY{+w}{ }res
  \end{Verbatim}

  On its own, this function is no more useful than the \texttt{Arbitrary}
  instance from earlier, but as we shall demonstrate, extracting the state
  transition function allows us to establish clear links between the model,
  tests, and implementation.

  \subsubsection{Tracing operations and programs}\label{ssec:atm-plumbing}

  We now consider the second issue: needing to keep track of the result type
  \emph{along with} a concrete instance of the result type, in a form we can
  control and test. To do this, we store an operation along with its result in a
  record type, \texttt{OpRes}:

  \begin{Verbatim}[commandchars=\\\{\}]
\PY{k+kr}{record}\PY{+w}{ }\PY{k+kt}{OpRes}\PY{+w}{ }\PY{o+ow}{(}resT\PY{+w}{ }\PY{o+ow}{:}\PY{+w}{ }\PY{k+kt}{Type}\PY{o+ow}{)}\PY{+w}{ }\PY{o+ow}{(}currSt\PY{+w}{ }\PY{o+ow}{:}\PY{+w}{ }\PY{k+kt}{ATMState}\PY{o+ow}{)}
\PY{+w}{ }\PY{+w}{ }\PY{+w}{ }\PY{+w}{ }\PY{+w}{ }\PY{+w}{ }\PY{+w}{ }\PY{+w}{ }\PY{+w}{ }\PY{+w}{ }\PY{+w}{ }\PY{+w}{ }\PY{+w}{ }\PY{o+ow}{(}nsFn\PY{+w}{ }\PY{o+ow}{:}\PY{+w}{ }resT\PY{+w}{ }\PY{o+ow}{\PYZhy{}\PYZgt{}}\PY{+w}{ }\PY{k+kt}{ATMState}\PY{o+ow}{)}\PY{+w}{ }\PY{k+kr}{where}
\PY{+w}{ }\PY{+w}{ }\PY{k+kr}{constructor}\PY{+w}{ }\PY{k+kt}{MkOpRes}
\PY{+w}{  }\PY{n+nf}{op}\PY{+w}{ }\PY{o+ow}{:}\PY{+w}{ }\PY{k+kt}{ATMOp}\PY{+w}{ }resT\PY{+w}{ }currSt\PY{+w}{ }nsFn
\PY{+w}{  }\PY{n+nf}{res}\PY{+w}{ }\PY{o+ow}{:}\PY{+w}{ }resT
\PY{+w}{ }\PY{+w}{ }\PY{o+ow}{\PYZob{}}\PY{k+kr}{auto}\PY{+w}{ }rShow\PY{+w}{ }\PY{o+ow}{:}\PY{+w}{ }\PY{k+kt}{Show}\PY{+w}{ }resT\PY{o+ow}{\PYZcb{}}
  \end{Verbatim}

  The auto-implicit \texttt{rShow} is so that \QC can print counterexamples if
  necessary. Given \texttt{OpRes}, all that remains is to trace how a state and
  operations are chained. We first consider each individual step, where an
  \texttt{OpRes} and its resulting \texttt{ATMState} are stored in the same
  record, and then create a type which, for a given bound, stores the trace:

  \begin{Verbatim}[commandchars=\\\{\}]
\PY{k+kr}{record}\PY{+w}{ }\PY{k+kt}{TraceStep}\PY{+w}{ }\PY{k+kr}{where}
\PY{+w}{ }\PY{+w}{ }\PY{k+kr}{constructor}\PY{+w}{ }\PY{k+kt}{MkTS}
\PY{+w}{  }\PY{n+nf}{opRes}\PY{+w}{ }\PY{o+ow}{:}\PY{+w}{ }\PY{k+kt}{OpRes}\PY{+w}{ }rT\PY{+w}{ }aSt\PY{+w}{ }aStFn
\PY{+w}{  }\PY{n+nf}{resSt}\PY{+w}{ }\PY{o+ow}{:}\PY{+w}{ }\PY{k+kt}{ATMState}

\PY{k+kr}{data}\PY{+w}{ }\PY{k+kt}{ATMTrace}\PY{+w}{ }\PY{o+ow}{:}\PY{+w}{ }\PY{k+kt}{ATMState}\PY{+w}{ }\PY{o+ow}{\PYZhy{}\PYZgt{}}\PY{+w}{ }\PY{k+kt}{Nat}\PY{+w}{ }\PY{o+ow}{\PYZhy{}\PYZgt{}}\PY{+w}{ }\PY{k+kt}{Type}\PY{+w}{ }\PY{k+kr}{where}
\PY{+w}{  }\PY{n+nf}{MkATMTrace}\PY{+w}{ }\PY{o+ow}{:}\PY{+w}{ }\PY{o+ow}{(}initSt\PY{+w}{ }\PY{o+ow}{:}\PY{+w}{ }\PY{k+kt}{ATMState}\PY{o+ow}{)}\PY{+w}{ }\PY{o+ow}{\PYZhy{}\PYZgt{}}\PY{+w}{ }\PY{o+ow}{\PYZob{}}bound\PY{+w}{ }\PY{o+ow}{:}\PY{+w}{ }\PY{k+kt}{Nat}\PY{o+ow}{\PYZcb{}}
\PY{+w}{ }\PY{+w}{ }\PY{+w}{ }\PY{+w}{ }\PY{+w}{ }\PY{+w}{ }\PY{+w}{ }\PY{+w}{ }\PY{+w}{ }\PY{+w}{ }\PY{+w}{ }\PY{+w}{ }\PY{+w}{ }\PY{o+ow}{\PYZhy{}\PYZgt{}}\PY{+w}{ }\PY{o+ow}{(}trace\PY{+w}{ }\PY{o+ow}{:}\PY{+w}{ }\PY{k+kt}{Vect}\PY{+w}{ }bound\PY{+w}{ }\PY{k+kt}{TraceStep}\PY{o+ow}{)}
\PY{+w}{ }\PY{+w}{ }\PY{+w}{ }\PY{+w}{ }\PY{+w}{ }\PY{+w}{ }\PY{+w}{ }\PY{+w}{ }\PY{+w}{ }\PY{+w}{ }\PY{+w}{ }\PY{+w}{ }\PY{+w}{ }\PY{o+ow}{\PYZhy{}\PYZgt{}}\PY{+w}{ }\PY{k+kt}{ATMTrace}\PY{+w}{ }initSt\PY{+w}{ }bound
  \end{Verbatim}

  With this framework in place, we are finally ready to write meaningful
  \texttt{Arbitrary} instances for testing our \acrshort{atm} model.

  \subsection{Arbitrary \texttt{OpRes}}\label{ssec:arb-opres}

  In order to generate arbitrary traces, we first need to be able to generate
  arbitrary operation-result pairs. Generating an \texttt{OpRes} requires
  knowing what the current state is, as we need it to determine the set of valid
  operations from it.  As with generating \texttt{Vect} instances, we use
  dependent pairs to capture the chain of things we need to know about the type
  of the operation. Once we know the concrete \texttt{resT} type, we know the
  type of our \texttt{nsFn}, which means we know the type of our \texttt{OpRes}:

  \begin{Verbatim}[commandchars=\\\{\}]
\PY{o+ow}{(}resT\PY{+w}{ }\PY{o+ow}{:}\PY{+w}{ }\PY{k+kt}{Type}\PY{+w}{ }\PY{o+ow}{**}\PY{+w}{ }nsFn\PY{+w}{ }\PY{o+ow}{:}\PY{+w}{ }resT\PY{+w}{ }\PY{o+ow}{\PYZhy{}\PYZgt{}}\PY{+w}{ }\PY{k+kt}{ATMState}
\PY{+w}{ }\PY{+w}{ }\PY{o+ow}{**}\PY{+w}{ }\PY{k+kt}{OpRes}\PY{+w}{ }resT\PY{+w}{ }st\PY{+w}{ }nsFn\PY{o+ow}{)}
  \end{Verbatim}

  For the concrete implementation, we can now pattern-match on the implicitly
  given \texttt{st}. This restricts which operations are available, as only some
  are valid in a given state, allowing us to return an operation chosen randomly
  from the set of compatible operations. We can also assign weights to the
  individual elements via QuickCheck's \texttt{frequency} function, thereby
  controlling how often certain operations are picked compared to others. Our
  \texttt{Arbitrary} instance for \texttt{OpRes} thus becomes:

  \begin{Verbatim}[commandchars=\\\{\}]
\PY{o+ow}{\PYZob{}}currSt\PY{+w}{ }\PY{o+ow}{:}\PY{+w}{ }\PY{k+kt}{ATMState}\PY{o+ow}{\PYZcb{}}\PY{+w}{ }\PY{o+ow}{\PYZhy{}\PYZgt{}}
\PY{k+kt}{Arbitrary}\PY{+w}{ }\PY{o+ow}{(}resT\PY{+w}{ }\PY{o+ow}{:}\PY{+w}{ }\PY{k+kr}{\PYZus{}}\PY{+w}{ }\PY{o+ow}{**}\PY{+w}{ }nsFn\PY{+w}{ }\PY{o+ow}{:}\PY{+w}{ }resT\PY{+w}{ }\PY{o+ow}{\PYZhy{}\PYZgt{}}\PY{+w}{ }\PY{k+kt}{ATMState}
\PY{+w}{ }\PY{+w}{ }\PY{+w}{ }\PY{+w}{ }\PY{+w}{ }\PY{+w}{ }\PY{+w}{ }\PY{+w}{ }\PY{+w}{ }\PY{+w}{ }\PY{+w}{ }\PY{+w}{ }\PY{o+ow}{**}\PY{+w}{ }\PY{k+kt}{OpRes}\PY{+w}{ }resT\PY{+w}{ }currSt\PY{+w}{ }nsFn\PY{o+ow}{)}\PY{+w}{ }\PY{k+kr}{where}
\PY{+w}{ }\PY{+w}{ }arbitrary\PY{+w}{ }\PY{o+ow}{\PYZob{}}currSt\PY{o+ow}{=}\PY{k+kt}{Ready}\PY{o+ow}{\PYZcb{}}\PY{+w}{ }\PY{o+ow}{=}
\PY{+w}{ }\PY{+w}{ }\PY{+w}{ }\PY{+w}{ }pure\PY{+w}{ }\PY{o+ow}{(}\PY{k+kr}{\PYZus{}}\PY{+w}{ }\PY{o+ow}{**}\PY{+w}{ }\PY{k+kr}{\PYZus{}}\PY{+w}{ }\PY{o+ow}{**}\PY{+w}{ }\PY{k+kt}{MkOpRes}\PY{+w}{ }\PY{k+kt}{Insert}\PY{+w}{ }\PY{o+ow}{())}

\PY{+w}{ }\PY{+w}{ }arbitrary\PY{+w}{ }\PY{o+ow}{\PYZob{}}currSt\PY{o+ow}{=}\PY{k+kt}{CardInserted}\PY{o+ow}{\PYZcb{}}\PY{+w}{ }\PY{o+ow}{=}\PY{+w}{ }\PY{k+kr}{do}
\PY{+w}{ }\PY{+w}{ }\PY{+w}{ }\PY{+w}{ }\PY{k+kr}{let}\PY{+w}{ }arbPIN\PY{+w}{ }\PY{o+ow}{=}\PY{+w}{ }\PY{l+m+mi}{0}
\PY{+w}{ }\PY{+w}{ }\PY{+w}{ }\PY{+w}{ }\PY{k+kr}{let}\PY{+w}{ }correct\PY{+w}{ }\PY{o+ow}{=}\PY{+w}{ }\PY{o+ow}{(}\PY{k+kr}{\PYZus{}}\PY{+w}{ }\PY{o+ow}{**}\PY{+w}{ }\PY{k+kr}{\PYZus{}}\PY{+w}{ }\PY{o+ow}{**}\PY{+w}{ }\PY{k+kt}{MkOpRes}
\PY{+w}{ }\PY{+w}{ }\PY{+w}{ }\PY{+w}{ }\PY{+w}{ }\PY{+w}{ }\PY{+w}{ }\PY{+w}{ }\PY{+w}{ }\PY{+w}{ }\PY{+w}{ }\PY{+w}{ }\PY{+w}{ }\PY{+w}{ }\PY{+w}{ }\PY{+w}{ }\PY{+w}{ }\PY{+w}{ }\PY{+w}{ }\PY{+w}{ }\PY{o+ow}{(}\PY{k+kt}{CheckPIN}\PY{+w}{ }arbPIN\PY{o+ow}{)}
\PY{+w}{ }\PY{+w}{ }\PY{+w}{ }\PY{+w}{ }\PY{+w}{ }\PY{+w}{ }\PY{+w}{ }\PY{+w}{ }\PY{+w}{ }\PY{+w}{ }\PY{+w}{ }\PY{+w}{ }\PY{+w}{ }\PY{+w}{ }\PY{+w}{ }\PY{+w}{ }\PY{+w}{ }\PY{+w}{ }\PY{+w}{ }\PY{+w}{ }\PY{k+kt}{Correct}\PY{o+ow}{)}
\PY{+w}{ }\PY{+w}{ }\PY{+w}{ }\PY{+w}{ }\PY{k+kr}{let}\PY{+w}{ }incorrect\PY{+w}{ }\PY{o+ow}{=}\PY{+w}{ }\PY{o+ow}{(}\PY{k+kr}{\PYZus{}}\PY{+w}{ }\PY{o+ow}{**}\PY{+w}{ }\PY{k+kr}{\PYZus{}}\PY{+w}{ }\PY{o+ow}{**}\PY{+w}{ }\PY{k+kt}{MkOpRes}
\PY{+w}{ }\PY{+w}{ }\PY{+w}{ }\PY{+w}{ }\PY{+w}{ }\PY{+w}{ }\PY{+w}{ }\PY{+w}{ }\PY{+w}{ }\PY{+w}{ }\PY{+w}{ }\PY{+w}{ }\PY{+w}{ }\PY{+w}{ }\PY{+w}{ }\PY{+w}{ }\PY{+w}{ }\PY{+w}{ }\PY{+w}{ }\PY{+w}{ }\PY{+w}{ }\PY{+w}{ }\PY{o+ow}{(}\PY{k+kt}{CheckPIN}\PY{+w}{ }arbPIN\PY{o+ow}{)}
\PY{+w}{ }\PY{+w}{ }\PY{+w}{ }\PY{+w}{ }\PY{+w}{ }\PY{+w}{ }\PY{+w}{ }\PY{+w}{ }\PY{+w}{ }\PY{+w}{ }\PY{+w}{ }\PY{+w}{ }\PY{+w}{ }\PY{+w}{ }\PY{+w}{ }\PY{+w}{ }\PY{+w}{ }\PY{+w}{ }\PY{+w}{ }\PY{+w}{ }\PY{+w}{ }\PY{+w}{ }\PY{k+kt}{Incorrect}\PY{o+ow}{)}
\PY{+w}{ }\PY{+w}{ }\PY{+w}{ }\PY{+w}{ }\PY{k+kr}{let}\PY{+w}{ }eject\PY{+w}{ }\PY{o+ow}{=}\PY{+w}{ }\PY{o+ow}{(}\PY{k+kr}{\PYZus{}}\PY{+w}{ }\PY{o+ow}{**}\PY{+w}{ }\PY{k+kr}{\PYZus{}}\PY{+w}{ }\PY{o+ow}{**}\PY{+w}{ }\PY{k+kt}{MkOpRes}\PY{+w}{ }\PY{k+kt}{Eject}\PY{+w}{ }\PY{o+ow}{())}
\PY{+w}{ }\PY{+w}{ }\PY{+w}{ }\PY{+w}{ }frequency\PY{+w}{ }\PY{o+ow}{\PYZdl{}}\PY{+w}{ }\PY{o+ow}{[}\PY{+w}{ }\PY{o+ow}{(}\PY{l+m+mi}{1},\PY{+w}{ }pure\PY{+w}{ }correct\PY{o+ow}{)}
\PY{+w}{ }\PY{+w}{ }\PY{+w}{ }\PY{+w}{ }\PY{+w}{ }\PY{+w}{ }\PY{+w}{ }\PY{+w}{ }\PY{+w}{ }\PY{+w}{ }\PY{+w}{ }\PY{+w}{ }\PY{+w}{ }\PY{+w}{ }\PY{+w}{ }\PY{+w}{ },\PY{+w}{ }\PY{o+ow}{(}\PY{l+m+mi}{4},\PY{+w}{ }pure\PY{+w}{ }incorrect\PY{o+ow}{)}
\PY{+w}{ }\PY{+w}{ }\PY{+w}{ }\PY{+w}{ }\PY{+w}{ }\PY{+w}{ }\PY{+w}{ }\PY{+w}{ }\PY{+w}{ }\PY{+w}{ }\PY{+w}{ }\PY{+w}{ }\PY{+w}{ }\PY{+w}{ }\PY{+w}{ }\PY{+w}{ },\PY{+w}{ }\PY{o+ow}{(}\PY{l+m+mi}{1},\PY{+w}{ }pure\PY{+w}{ }eject\PY{o+ow}{)}\PY{+w}{ }\PY{o+ow}{]}

\PY{+w}{ }\PY{+w}{ }arbitrary\PY{+w}{ }\PY{o+ow}{\PYZob{}}currSt\PY{o+ow}{=}\PY{k+kt}{Session}\PY{o+ow}{\PYZcb{}}\PY{+w}{ }\PY{o+ow}{=}\PY{+w}{ }\PY{k+kr}{do}
\PY{+w}{ }\PY{+w}{ }\PY{+w}{ }\PY{+w}{ }arbAmount\PY{+w}{ }\PY{o+ow}{\PYZlt{}\PYZhy{}}\PY{+w}{ }arbitrary
\PY{+w}{ }\PY{+w}{ }\PY{+w}{ }\PY{+w}{ }\PY{k+kr}{let}\PY{+w}{ }dispense\PY{+w}{ }\PY{o+ow}{=}\PY{+w}{ }\PY{o+ow}{(}\PY{k+kr}{\PYZus{}}\PY{+w}{ }\PY{o+ow}{**}\PY{+w}{ }\PY{k+kr}{\PYZus{}}\PY{+w}{ }\PY{o+ow}{**}\PY{+w}{ }\PY{k+kt}{MkOpRes}
\PY{+w}{ }\PY{+w}{ }\PY{+w}{ }\PY{+w}{ }\PY{+w}{ }\PY{+w}{ }\PY{+w}{ }\PY{+w}{ }\PY{+w}{ }\PY{+w}{ }\PY{+w}{ }\PY{+w}{ }\PY{+w}{ }\PY{+w}{ }\PY{+w}{ }\PY{+w}{ }\PY{+w}{ }\PY{+w}{ }\PY{+w}{ }\PY{+w}{ }\PY{o+ow}{(}\PY{k+kt}{Dispense}\PY{+w}{ }arbAmount\PY{o+ow}{)}
\PY{+w}{ }\PY{+w}{ }\PY{+w}{ }\PY{+w}{ }\PY{+w}{ }\PY{+w}{ }\PY{+w}{ }\PY{+w}{ }\PY{+w}{ }\PY{+w}{ }\PY{+w}{ }\PY{+w}{ }\PY{+w}{ }\PY{+w}{ }\PY{+w}{ }\PY{+w}{ }\PY{+w}{ }\PY{+w}{ }\PY{+w}{ }\PY{+w}{ }\PY{o+ow}{())}
\PY{+w}{ }\PY{+w}{ }\PY{+w}{ }\PY{+w}{ }\PY{k+kr}{let}\PY{+w}{ }eject\PY{+w}{ }\PY{o+ow}{=}\PY{+w}{ }\PY{o+ow}{(}\PY{k+kr}{\PYZus{}}\PY{+w}{ }\PY{o+ow}{**}\PY{+w}{ }\PY{k+kr}{\PYZus{}}\PY{+w}{ }\PY{o+ow}{**}\PY{+w}{ }\PY{k+kt}{MkOpRes}\PY{+w}{ }\PY{k+kt}{Eject}\PY{+w}{ }\PY{o+ow}{())}
\PY{+w}{ }\PY{+w}{ }\PY{+w}{ }\PY{+w}{ }oneof\PY{+w}{ }\PY{o+ow}{\PYZdl{}}\PY{+w}{ }map\PY{+w}{ }pure\PY{+w}{ }\PY{o+ow}{[}dispense,\PY{+w}{ }eject\PY{o+ow}{]}
  \end{Verbatim}

  \subsection{Arbitrary \texttt{ATMTrace}}\label{ssec:arb-atm-trace}

  To chain steps together, we make an instance of \texttt{Arbitrary ATMTrace}.
  Recall that traces are bounded by their depth, therefore we pattern-matching
  on the depth.

  \subsubsection{If $depth = 0$:}
  the remaining trace must be empty as the depth bound has been reached.

  \begin{Verbatim}[commandchars=\\\{\}]
arbitrary\PY{+w}{ }\PY{o+ow}{\PYZob{}}depth\PY{o+ow}{=}\PY{l+m+mi}{0}\PY{o+ow}{\PYZcb{}}\PY{+w}{ }\PY{o+ow}{=}\PY{+w}{ }pure\PY{+w}{ }\PY{o+ow}{\PYZdl{}}\PY{+w}{ }\PY{k+kt}{MkATMTrace}\PY{+w}{ }iSt\PY{+w}{ }\PY{o+ow}{[]}
  \end{Verbatim}

  \subsubsection{If $depth = (S\ b)$:}
  the trace depth has not been reached and we need to generate at least one more
  trace step.

  \begin{Verbatim}[commandchars=\\\{\}]
arbitrary\PY{+w}{ }\PY{o+ow}{\PYZob{}}depth\PY{o+ow}{=}\PY{o+ow}{(}\PY{k+kt}{S}\PY{+w}{ }b\PY{o+ow}{)\PYZcb{}}\PY{+w}{ }\PY{o+ow}{=}\PY{+w}{ }\PY{k+kr}{do}
\PY{+w}{ }\PY{+w}{ }\PY{o+ow}{?}arbitrary\PYZus{}trace\PYZus{}rhs
  \end{Verbatim}

  The first step is to generate an arbitrary \texttt{OpRes}, which we can now do
  thanks to the implementation from \cref{ssec:arb-opres}. We may not know the
  \texttt{OpRes}'s result type or state function
  but we can capture this in type variables:

  \begin{Verbatim}[commandchars=\\\{\}]
arbitrary\PY{+w}{ }\PY{o+ow}{\PYZob{}}depth\PY{o+ow}{=}\PY{o+ow}{(}\PY{k+kt}{S}\PY{+w}{ }b\PY{o+ow}{)\PYZcb{}}\PY{+w}{ }\PY{o+ow}{=}\PY{+w}{ }\PY{k+kr}{do}
\PY{+w}{ }\PY{+w}{ }opRes\PY{+w}{ }\PY{o+ow}{\PYZlt{}\PYZhy{}}\PY{+w}{ }the\PY{+w}{ }\PY{o+ow}{(}\PY{k+kt}{Gen}\PY{+w}{ }\PY{o+ow}{(}resT\PY{+w}{ }\PY{o+ow}{**}\PY{+w}{ }fn\PY{+w}{ }\PY{o+ow}{**}
\PY{+w}{ }\PY{+w}{ }\PY{+w}{ }\PY{+w}{ }\PY{+w}{ }\PY{+w}{ }\PY{+w}{ }\PY{+w}{ }\PY{+w}{ }\PY{+w}{ }\PY{+w}{ }\PY{+w}{ }\PY{+w}{ }\PY{+w}{ }\PY{+w}{ }\PY{+w}{ }\PY{+w}{ }\PY{+w}{ }\PY{+w}{ }\PY{+w}{ }\PY{+w}{ }\PY{+w}{ }\PY{k+kt}{OpRes}\PY{+w}{ }resT\PY{+w}{ }iSt\PY{+w}{ }fn\PY{o+ow}{))}\PY{+w}{ }arbitrary
\PY{+w}{ }\PY{+w}{ }\PY{o+ow}{?}arbitrary\PYZus{}trace\PYZus{}rhs
  \end{Verbatim}

  We have to use `\texttt{the}' to give a precise type to \texttt{arbitrary}, as
  \iidris cannot infer it from the shape of the \texttt{opRes} variable.
  Asking the compiler for the type of the hole gives us:

  \begin{Verbatim}[commandchars=\\\{\}]
\PY{+w}{  }\PY{n+nf}{iSt}\PY{+w}{ }\PY{o+ow}{:}\PY{+w}{ }\PY{k+kt}{ATMState}
\PY{+w}{  }\PY{n+nf}{b}\PY{+w}{ }\PY{o+ow}{:}\PY{+w}{ }\PY{k+kt}{Nat}
\PY{+w}{  }\PY{n+nf}{opRes}\PY{+w}{ }\PY{o+ow}{:}\PY{+w}{ }\PY{o+ow}{(}resT\PY{+w}{ }\PY{o+ow}{:}\PY{+w}{ }\PY{k+kt}{Type}\PY{+w}{ }\PY{o+ow}{**}\PY{+w}{ }\PY{o+ow}{(}fn\PY{+w}{ }\PY{o+ow}{:}\PY{+w}{ }resT\PY{+w}{ }\PY{o+ow}{\PYZhy{}\PYZgt{}}\PY{+w}{ }\PY{k+kt}{ATMState}\PY{+w}{ }\PY{o+ow}{**}
\PY{+w}{ }\PY{+w}{ }\PY{+w}{ }\PY{+w}{ }\PY{+w}{ }\PY{+w}{ }\PY{+w}{ }\PY{+w}{ }\PY{+w}{ }\PY{+w}{ }\PY{+w}{ }\PY{+w}{ }\PY{k+kt}{OpRes}\PY{+w}{ }resT\PY{+w}{ }iSt\PY{+w}{ }fn\PY{o+ow}{))}
\PY{c+c1}{\PYZhy{}\PYZhy{}\PYZhy{}\PYZhy{}\PYZhy{}\PYZhy{}\PYZhy{}\PYZhy{}\PYZhy{}\PYZhy{}\PYZhy{}\PYZhy{}\PYZhy{}\PYZhy{}\PYZhy{}\PYZhy{}\PYZhy{}\PYZhy{}\PYZhy{}\PYZhy{}\PYZhy{}\PYZhy{}\PYZhy{}\PYZhy{}\PYZhy{}\PYZhy{}\PYZhy{}\PYZhy{}\PYZhy{}\PYZhy{}}
\PY{n+nf}{arbitrary\PYZus{}trace\PYZus{}rhs}\PY{+w}{ }\PY{o+ow}{:}\PY{+w}{ }\PY{k+kt}{Gen}\PY{+w}{ }\PY{o+ow}{(}\PY{k+kt}{ATMTrace}\PY{+w}{ }iSt\PY{+w}{ }\PY{o+ow}{(}\PY{k+kt}{S}\PY{+w}{ }b\PY{o+ow}{))}
  \end{Verbatim}

  The generated dependent pair containing our new \texttt{OpRes} is not too
  useful as a single variable. However, we can split out its components via
  pattern-matching.

  \begin{Verbatim}[commandchars=\\\{\}]
arbitrary\PY{+w}{ }\PY{o+ow}{\PYZob{}}depth\PY{o+ow}{=}\PY{o+ow}{(}\PY{k+kt}{S}\PY{+w}{ }b\PY{o+ow}{)\PYZcb{}}\PY{+w}{ }\PY{o+ow}{=}\PY{+w}{ }\PY{k+kr}{do}
\PY{+w}{ }\PY{+w}{ }\PY{o+ow}{(}resT\PY{+w}{ }\PY{o+ow}{**}\PY{+w}{ }nsFn\PY{+w}{ }\PY{o+ow}{**}\PY{+w}{ }\PY{o+ow}{(}\PY{k+kt}{MkOpRes}\PY{+w}{ }op\PY{+w}{ }res\PY{o+ow}{))}\PY{+w}{ }\PY{o+ow}{\PYZlt{}\PYZhy{}}\PY{+w}{ }the
\PY{+w}{ }\PY{+w}{ }\PY{+w}{ }\PY{+w}{ }\PY{+w}{ }\PY{+w}{ }\PY{+w}{ }\PY{+w}{ }\PY{+w}{ }\PY{+w}{ }\PY{+w}{ }\PY{+w}{ }\PY{+w}{ }\PY{+w}{ }\PY{+w}{ }\PY{+w}{ }\PY{+w}{ }\PY{+w}{ }\PY{+w}{ }\PY{+w}{ }\PY{+w}{ }\PY{+w}{ }\PY{+w}{ }\PY{+w}{ }\PY{+w}{ }\PY{+w}{ }\PY{+w}{ }\PY{+w}{ }\PY{+w}{ }\PY{+w}{ }\PY{+w}{ }\PY{+w}{ }\PY{+w}{ }\PY{+w}{ }\PY{o+ow}{(}\PY{k+kt}{Gen}\PY{+w}{ }\PY{c+cm}{\PYZob{}\PYZhy{}}\PY{c+cm}{...}\PY{c+cm}{\PYZhy{}\PYZcb{}}\PY{o+ow}{)}
\PY{+w}{ }\PY{+w}{ }\PY{o+ow}{?}arbitrary\PYZus{}trace\PYZus{}rhs
  \end{Verbatim}

  This gives us access to several critical pieces of data:

  \begin{itemize}
    \item \texttt{op} {\textemdash} The operation itself, so we can log what
          operation led where.
    \item \texttt{res} {\textemdash} The result of the operation. In order to
          continue constructing our trace, we need to know what state we moved
          to, which requires applying the next-state function to a concrete
          result; this is exactly what is given here.
    \item \texttt{nsFn} {\textemdash} The next-state function \emph{as written
          directly in the type.} It is worth re-emphasising this: we are
          guaranteed to use the same transition function as our
          model/specification, because we are extracting it from the type which
          uses it! We can access this because the entire process is happening at
          type checking time, so we can use elements which will be erased at run
          time.
  \end{itemize}

  And we can confirm this by taking a look at the updated information for our
  hole:

  \begin{Verbatim}[commandchars=\\\{\}]
\PY{+w}{  }\PY{n+nf}{iSt}\PY{+w}{ }\PY{o+ow}{:}\PY{+w}{ }\PY{k+kt}{ATMState}
\PY{+w}{  }\PY{n+nf}{b}\PY{+w}{ }\PY{o+ow}{:}\PY{+w}{ }\PY{k+kt}{Nat}
\PY{+w}{  }\PY{n+nf}{resT}\PY{+w}{ }\PY{o+ow}{:}\PY{+w}{ }\PY{k+kt}{Type}
\PY{+w}{  }\PY{n+nf}{nsFn}\PY{+w}{ }\PY{o+ow}{:}\PY{+w}{ }resT\PY{+w}{ }\PY{o+ow}{\PYZhy{}\PYZgt{}}\PY{+w}{ }\PY{k+kt}{ATMState}
\PY{+w}{  }\PY{n+nf}{res}\PY{+w}{ }\PY{o+ow}{:}\PY{+w}{ }resT
\PY{+w}{  }\PY{n+nf}{op}\PY{+w}{ }\PY{o+ow}{:}\PY{+w}{ }\PY{k+kt}{ATMOp}\PY{+w}{ }resT\PY{+w}{ }iSt\PY{+w}{ }nsFn
\PY{c+c1}{\PYZhy{}\PYZhy{}\PYZhy{}\PYZhy{}\PYZhy{}\PYZhy{}\PYZhy{}\PYZhy{}\PYZhy{}\PYZhy{}\PYZhy{}\PYZhy{}\PYZhy{}\PYZhy{}\PYZhy{}\PYZhy{}\PYZhy{}\PYZhy{}\PYZhy{}\PYZhy{}\PYZhy{}\PYZhy{}\PYZhy{}\PYZhy{}\PYZhy{}\PYZhy{}\PYZhy{}\PYZhy{}\PYZhy{}\PYZhy{}}
\PY{n+nf}{arbitrary\PYZus{}trace\PYZus{}rhs}\PY{+w}{ }\PY{o+ow}{:}\PY{+w}{ }\PY{k+kt}{Gen}\PY{+w}{ }\PY{o+ow}{(}\PY{k+kt}{ATMTrace}\PY{+w}{ }iSt\PY{+w}{ }\PY{o+ow}{(}\PY{k+kt}{S}\PY{+w}{ }b\PY{o+ow}{))}
  \end{Verbatim}

  Applying \texttt{nsFn} to the generated \texttt{res} gives us the first state
  of our trace:

  \begin{Verbatim}[commandchars=\\\{\}]
arbitrary\PY{+w}{ }\PY{o+ow}{\PYZob{}}depth\PY{o+ow}{=}\PY{o+ow}{(}\PY{k+kt}{S}\PY{+w}{ }b\PY{o+ow}{)\PYZcb{}}\PY{+w}{ }\PY{o+ow}{=}\PY{+w}{ }\PY{k+kr}{do}
\PY{+w}{ }\PY{+w}{ }\PY{o+ow}{(}resT\PY{+w}{ }\PY{o+ow}{**}\PY{+w}{ }nsFn\PY{+w}{ }\PY{o+ow}{**}\PY{+w}{ }\PY{o+ow}{(}\PY{k+kt}{MkOpRes}\PY{+w}{ }op\PY{+w}{ }res\PY{o+ow}{))}\PY{+w}{ }\PY{o+ow}{\PYZlt{}\PYZhy{}}\PY{+w}{ }the
\PY{+w}{ }\PY{+w}{ }\PY{+w}{ }\PY{+w}{ }\PY{+w}{ }\PY{+w}{ }\PY{+w}{ }\PY{+w}{ }\PY{+w}{ }\PY{+w}{ }\PY{+w}{ }\PY{+w}{ }\PY{+w}{ }\PY{+w}{ }\PY{+w}{ }\PY{+w}{ }\PY{+w}{ }\PY{+w}{ }\PY{+w}{ }\PY{+w}{ }\PY{+w}{ }\PY{+w}{ }\PY{+w}{ }\PY{+w}{ }\PY{+w}{ }\PY{+w}{ }\PY{+w}{ }\PY{+w}{ }\PY{+w}{ }\PY{+w}{ }\PY{+w}{ }\PY{+w}{ }\PY{+w}{ }\PY{+w}{ }\PY{o+ow}{(}\PY{k+kt}{Gen}\PY{+w}{ }\PY{c+cm}{\PYZob{}\PYZhy{}}\PY{c+cm}{...}\PY{c+cm}{\PYZhy{}\PYZcb{}}\PY{o+ow}{)}
\PY{+w}{ }\PY{+w}{ }\PY{k+kr}{let}\PY{+w}{ }fstTraceSt\PY{+w}{ }\PY{o+ow}{=}\PY{+w}{ }nsFn\PY{+w}{ }res
\PY{+w}{ }\PY{+w}{ }\PY{o+ow}{?}arbitrary\PYZus{}trace\PYZus{}rhs
  \end{Verbatim}

  Which we again confirm by looking at our new supporting information:

  \begin{Verbatim}[commandchars=\\\{\}]
\PY{+w}{  }\PY{n+nf}{iSt}\PY{+w}{ }\PY{o+ow}{:}\PY{+w}{ }\PY{k+kt}{ATMState}
\PY{+w}{  }\PY{n+nf}{b}\PY{+w}{ }\PY{o+ow}{:}\PY{+w}{ }\PY{k+kt}{Nat}
\PY{+w}{  }\PY{n+nf}{resT}\PY{+w}{ }\PY{o+ow}{:}\PY{+w}{ }\PY{k+kt}{Type}
\PY{+w}{  }\PY{n+nf}{nsFn}\PY{+w}{ }\PY{o+ow}{:}\PY{+w}{ }resT\PY{+w}{ }\PY{o+ow}{\PYZhy{}\PYZgt{}}\PY{+w}{ }\PY{k+kt}{ATMState}
\PY{+w}{  }\PY{n+nf}{res}\PY{+w}{ }\PY{o+ow}{:}\PY{+w}{ }resT
\PY{+w}{  }\PY{n+nf}{op}\PY{+w}{ }\PY{o+ow}{:}\PY{+w}{ }\PY{k+kt}{ATMOp}\PY{+w}{ }resT\PY{+w}{ }iSt\PY{+w}{ }nsFn
\PY{+w}{  }\PY{n+nf}{fstTraceSt}\PY{+w}{ }\PY{o+ow}{:}\PY{+w}{ }\PY{k+kt}{ATMState}
\PY{c+c1}{\PYZhy{}\PYZhy{}\PYZhy{}\PYZhy{}\PYZhy{}\PYZhy{}\PYZhy{}\PYZhy{}\PYZhy{}\PYZhy{}\PYZhy{}\PYZhy{}\PYZhy{}\PYZhy{}\PYZhy{}\PYZhy{}\PYZhy{}\PYZhy{}\PYZhy{}\PYZhy{}\PYZhy{}\PYZhy{}\PYZhy{}\PYZhy{}\PYZhy{}\PYZhy{}\PYZhy{}\PYZhy{}\PYZhy{}\PYZhy{}}
\PY{n+nf}{arbitrary\PYZus{}trace\PYZus{}rhs}\PY{+w}{ }\PY{o+ow}{:}\PY{+w}{ }\PY{k+kt}{Gen}\PY{+w}{ }\PY{o+ow}{(}\PY{k+kt}{ATMTrace}\PY{+w}{ }iSt\PY{+w}{ }\PY{o+ow}{(}\PY{k+kt}{S}\PY{+w}{ }b\PY{o+ow}{))}
  \end{Verbatim}

  The first trace state can only have been obtained by following the
  specification's semantics because the function we applied to generate it was
  the exact function specified in the original type! We now construct the trace
  by storing the operation and its resulting state and recursively generating
  the rest of the trace:

  \begin{Verbatim}[commandchars=\\\{\}]
arbitrary\PY{+w}{ }\PY{o+ow}{\PYZob{}}depth\PY{o+ow}{=}\PY{o+ow}{(}\PY{k+kt}{S}\PY{+w}{ }b\PY{o+ow}{)\PYZcb{}}\PY{+w}{ }\PY{o+ow}{=}\PY{+w}{ }\PY{k+kr}{do}
\PY{+w}{ }\PY{+w}{ }\PY{o+ow}{(}resT\PY{+w}{ }\PY{o+ow}{**}\PY{+w}{ }nsFn\PY{+w}{ }\PY{o+ow}{**}\PY{+w}{ }\PY{o+ow}{(}\PY{k+kt}{MkOpRes}\PY{+w}{ }op\PY{+w}{ }res\PY{o+ow}{))}\PY{+w}{ }\PY{o+ow}{\PYZlt{}\PYZhy{}}\PY{+w}{ }the
\PY{+w}{ }\PY{+w}{ }\PY{+w}{ }\PY{+w}{ }\PY{+w}{ }\PY{+w}{ }\PY{+w}{ }\PY{+w}{ }\PY{+w}{ }\PY{+w}{ }\PY{+w}{ }\PY{+w}{ }\PY{+w}{ }\PY{+w}{ }\PY{+w}{ }\PY{+w}{ }\PY{+w}{ }\PY{+w}{ }\PY{+w}{ }\PY{+w}{ }\PY{+w}{ }\PY{+w}{ }\PY{+w}{ }\PY{+w}{ }\PY{+w}{ }\PY{+w}{ }\PY{+w}{ }\PY{+w}{ }\PY{+w}{ }\PY{+w}{ }\PY{+w}{ }\PY{+w}{ }\PY{+w}{ }\PY{+w}{ }\PY{o+ow}{(}\PY{k+kt}{Gen}\PY{+w}{ }\PY{c+cm}{\PYZob{}\PYZhy{}}\PY{c+cm}{...}\PY{c+cm}{\PYZhy{}\PYZcb{}}\PY{o+ow}{)}
\PY{+w}{ }\PY{+w}{ }\PY{k+kr}{let}\PY{+w}{ }fstTraceSt\PY{+w}{ }\PY{o+ow}{=}\PY{+w}{ }nsFn\PY{+w}{ }res
\PY{+w}{ }\PY{+w}{ }\PY{k+kr}{let}\PY{+w}{ }atmTrace\PY{+w}{ }\PY{o+ow}{=}
\PY{+w}{ }\PY{+w}{ }\PY{+w}{ }\PY{+w}{ }\PY{o+ow}{(}\PY{k+kt}{MkTS}\PY{+w}{ }\PY{o+ow}{(}\PY{k+kt}{MkOpRes}\PY{+w}{ }op\PY{+w}{ }res\PY{o+ow}{)}\PY{+w}{ }fstTraceSt\PY{o+ow}{)}\PY{+w}{ }\PY{o+ow}{::}
\PY{+w}{ }\PY{+w}{ }\PY{+w}{ }\PY{+w}{ }\PY{+w}{ }\PY{+w}{ }\PY{o+ow}{!(}trace\PY{+w}{ }b\PY{+w}{ }fstTraceSt\PY{o+ow}{)}\PY{+w}{ }\PY{+w}{ }\PY{+w}{ }\PY{+w}{ }\PY{+w}{ }\PY{+w}{ }\PY{+w}{ }\PY{+w}{ }
\PY{+w}{ }\PY{+w}{ }pure\PY{+w}{ }\PY{o+ow}{\PYZdl{}}\PY{+w}{ }\PY{k+kt}{MkATMTrace}\PY{+w}{ }iSt\PY{+w}{ }atmTrace
  \end{Verbatim}

  Above, we use a helper function, \texttt{trace}, because \texttt{MkATMTrace}
  expects a \texttt{Vect} of exactly \texttt{bound} elements. We could extract
  this from a recursive call to \texttt{arbitrary}, generating a new
  \texttt{ATMTrace} and then pattern-matching on its constructor to extract the
  \texttt{Vect}, however we prefer this solution of generating the \texttt{Vect}
  in-place using a helper function. Its definition is almost verbatim that of
  the \texttt{abitrary} instance:

  \begin{Verbatim}[commandchars=\\\{\}]
\PY{n+nf}{trace}\PY{+w}{ }\PY{o+ow}{:}\PY{+w}{  }\PY{o+ow}{(}steps\PY{+w}{ }\PY{o+ow}{:}\PY{+w}{ }\PY{k+kt}{Nat}\PY{o+ow}{)}
\PY{+w}{ }\PY{+w}{ }\PY{+w}{ }\PY{+w}{ }\PY{+w}{ }\PY{+w}{ }\PY{o+ow}{\PYZhy{}\PYZgt{}}\PY{+w}{ }\PY{o+ow}{(}st\PY{+w}{ }\PY{o+ow}{:}\PY{+w}{ }\PY{k+kt}{ATMState}\PY{o+ow}{)}
\PY{+w}{ }\PY{+w}{ }\PY{+w}{ }\PY{+w}{ }\PY{+w}{ }\PY{+w}{ }\PY{o+ow}{\PYZhy{}\PYZgt{}}\PY{+w}{ }\PY{k+kt}{Gen}\PY{+w}{ }\PY{o+ow}{(}\PY{k+kt}{Vect}\PY{+w}{ }steps\PY{+w}{ }\PY{k+kt}{TraceStep}\PY{o+ow}{)}
trace\PY{+w}{ }\PY{l+m+mi}{0}\PY{+w}{ }\PY{k+kr}{\PYZus{}}\PY{+w}{ }\PY{o+ow}{=}\PY{+w}{ }pure\PY{+w}{ }\PY{o+ow}{[]}
trace\PY{+w}{ }\PY{o+ow}{(}\PY{k+kt}{S}\PY{+w}{ }k\PY{o+ow}{)}\PY{+w}{ }st\PY{+w}{ }\PY{o+ow}{=}\PY{+w}{ }\PY{k+kr}{do}
\PY{+w}{ }\PY{+w}{ }\PY{o+ow}{(}\PY{k+kr}{\PYZus{}}\PY{+w}{ }\PY{o+ow}{**}\PY{+w}{ }nsFn\PY{+w}{ }\PY{o+ow}{**}\PY{+w}{ }opR\PY{o+ow}{@(}\PY{k+kt}{MkOpRes}\PY{+w}{ }\PY{k+kr}{\PYZus{}}\PY{+w}{ }res\PY{o+ow}{))}\PY{+w}{ }\PY{o+ow}{\PYZlt{}\PYZhy{}}\PY{+w}{ }the
\PY{+w}{ }\PY{+w}{ }\PY{+w}{ }\PY{+w}{ }\PY{+w}{ }\PY{+w}{ }\PY{+w}{ }\PY{+w}{ }\PY{+w}{ }\PY{+w}{ }\PY{+w}{ }\PY{+w}{ }\PY{+w}{ }\PY{+w}{ }\PY{+w}{ }\PY{+w}{ }\PY{+w}{ }\PY{+w}{ }\PY{+w}{ }\PY{+w}{ }\PY{+w}{ }\PY{+w}{ }\PY{+w}{ }\PY{+w}{ }\PY{+w}{ }\PY{+w}{ }\PY{+w}{ }\PY{+w}{ }\PY{+w}{ }\PY{+w}{ }\PY{+w}{ }\PY{+w}{ }\PY{+w}{ }\PY{+w}{ }\PY{o+ow}{(}\PY{k+kt}{Gen}\PY{+w}{ }\PY{c+cm}{\PYZob{}\PYZhy{}}\PY{c+cm}{...}\PY{c+cm}{\PYZhy{}\PYZcb{}}\PY{o+ow}{)}
\PY{+w}{ }\PY{+w}{ }\PY{k+kr}{let}\PY{+w}{ }nextState\PY{+w}{ }\PY{o+ow}{=}\PY{+w}{ }nsFn\PY{+w}{ }res
\PY{+w}{ }\PY{+w}{ }pure\PY{+w}{ }\PY{o+ow}{\PYZdl{}}\PY{+w}{ }\PY{o+ow}{(}\PY{k+kt}{MkTS}\PY{+w}{ }opR\PY{+w}{ }nextState\PY{o+ow}{)}\PY{+w}{ }\PY{o+ow}{::}
\PY{+w}{ }\PY{+w}{ }\PY{+w}{ }\PY{+w}{ }\PY{+w}{ }\PY{+w}{ }\PY{+w}{ }\PY{+w}{ }\PY{+w}{ }\PY{+w}{ }\PY{o+ow}{!(}trace\PY{+w}{ }k\PY{+w}{ }nextState\PY{o+ow}{)}
  \end{Verbatim}

  The as-pattern captures the entire \texttt{OpRes}, meaning we can omit the
  operation being bound to a variable as we never use it in the body of the
  function; and the bang-notation is shorthand for extracting the result of a
  monadic
  computation~\cite{bradyResourceDependentAlgebraicEffects2015}.

  \subsection{\QC-ing the type-level \acrshort{atm}}

  We have our specification, modelled as an \acrshort{ism}, with datatypes for
  generating sample execution traces, so we are now in a position to specify
  properties for \QC to verify.  To start, we check that when we are in the
  \textbf{Ready} state, we always end up in \textbf{CardInserted} after a single
  operation.

  \begin{Verbatim}[commandchars=\\\{\}]
\PY{l+m+mi}{0}\PY{+w}{ }\PY{k+kt}{PROP\PYZus{}readyInsert}\PY{+w}{ }\PY{o+ow}{:}\PY{+w}{ }\PY{k+kt}{Fn}\PY{+w}{ }\PY{o+ow}{(}\PY{k+kt}{ATMTrace}\PY{+w}{ }\PY{k+kt}{Ready}\PY{+w}{ }\PY{l+m+mi}{1}\PY{o+ow}{)}\PY{+w}{ }\PY{k+kt}{Bool}
PROP\PYZus{}readyInsert\PY{+w}{ }\PY{o+ow}{=}\PY{+w}{ }\PY{k+kt}{MkFn}
\PY{+w}{ }\PY{+w}{ }\PY{o+ow}{(\PYZbs{}}\PY{k+kr}{case}\PY{+w}{ }\PY{o+ow}{(}\PY{k+kt}{MkATMTrace}\PY{+w}{ }\PY{k+kr}{\PYZus{}}\PY{+w}{ }\PY{o+ow}{(}
\PY{+w}{ }\PY{+w}{ }\PY{+w}{ }\PY{+w}{ }\PY{+w}{ }\PY{+w}{ }\PY{+w}{ }\PY{+w}{ }\PY{+w}{ }\PY{+w}{ }\PY{+w}{ }\PY{o+ow}{(}\PY{k+kt}{MkTS}\PY{+w}{ }\PY{k+kr}{\PYZus{}}\PY{+w}{ }\PY{k+kt}{CardInserted}\PY{o+ow}{)}\PY{+w}{ }\PY{o+ow}{::}\PY{+w}{ }\PY{o+ow}{[]))}
\PY{+w}{ }\PY{+w}{ }\PY{+w}{ }\PY{+w}{ }\PY{+w}{ }\PY{+w}{ }\PY{+w}{ }\PY{+w}{ }\PY{+w}{ }\PY{+w}{ }\PY{+w}{ }\PY{+w}{ }\PY{+w}{ }\PY{+w}{ }\PY{o+ow}{=\PYZgt{}}\PY{+w}{ }\PY{k+kt}{True}
\PY{+w}{ }\PY{+w}{ }\PY{+w}{ }\PY{+w}{ }\PY{+w}{ }\PY{+w}{ }\PY{+w}{ }\PY{+w}{ }\PY{+w}{ }\PY{o+ow}{(}\PY{k+kt}{MkATMTrace}\PY{+w}{ }\PY{k+kr}{\PYZus{}}\PY{+w}{ }\PY{k+kr}{\PYZus{}}\PY{o+ow}{)}\PY{+w}{ }\PY{o+ow}{=\PYZgt{}}\PY{+w}{ }\PY{k+kt}{False}\PY{o+ow}{)}
    \end{Verbatim}

  Notice that our property is given at quantity 0, so that it is compile time
  only. To \QC this, we wrap the default \texttt{quickCheck} function in a
  type-level one, which tests the given property, specifying whether the test
  should be considered passed if \QC exhausted the arguments.

  \begin{Verbatim}[commandchars=\\\{\}]
\PY{l+m+mi}{0}\PY{+w}{ }\PY{k+kt}{QuickCheck}\PY{+w}{ }\PY{o+ow}{:}\PY{+w}{ }\PY{+w}{ }\PY{k+kt}{Testable}\PY{+w}{ }t
\PY{+w}{ }\PY{+w}{ }\PY{+w}{ }\PY{+w}{ }\PY{+w}{ }\PY{+w}{ }\PY{+w}{ }\PY{+w}{ }\PY{+w}{ }\PY{+w}{ }\PY{+w}{ }\PY{+w}{ }\PY{+w}{ }\PY{o+ow}{=\PYZgt{}}\PY{+w}{ }\PY{o+ow}{(}allowExhaust\PY{+w}{ }\PY{o+ow}{:}\PY{+w}{ }\PY{k+kt}{Bool}\PY{o+ow}{)}
\PY{+w}{ }\PY{+w}{ }\PY{+w}{ }\PY{+w}{ }\PY{+w}{ }\PY{+w}{ }\PY{+w}{ }\PY{+w}{ }\PY{+w}{ }\PY{+w}{ }\PY{+w}{ }\PY{+w}{ }\PY{+w}{ }\PY{o+ow}{\PYZhy{}\PYZgt{}}\PY{+w}{ }\PY{o+ow}{(}prop\PY{+w}{ }\PY{o+ow}{:}\PY{+w}{ }t\PY{o+ow}{)}
\PY{+w}{ }\PY{+w}{ }\PY{+w}{ }\PY{+w}{ }\PY{+w}{ }\PY{+w}{ }\PY{+w}{ }\PY{+w}{ }\PY{+w}{ }\PY{+w}{ }\PY{+w}{ }\PY{+w}{ }\PY{+w}{ }\PY{o+ow}{\PYZhy{}\PYZgt{}}\PY{+w}{ }\PY{k+kt}{Bool}
QuickCheck\PY{+w}{ }allowExhaust\PY{+w}{ }prop\PY{+w}{ }\PY{o+ow}{=}
\PY{+w}{ }\PY{+w}{ }\PY{k+kt}{Maybe}\PY{o+ow}{.}fromMaybe\PY{+w}{ }allowExhaust\PY{+w}{ }\PY{o+ow}{\PYZdl{}}
\PY{+w}{ }\PY{+w}{ }\PY{+w}{ }\PY{+w}{ }\PY{o+ow}{(}quickCheck\PY{+w}{ }prop\PY{o+ow}{).}pass
  \end{Verbatim}

  Using the \iidris built-in data type \texttt{Data.So}, which is inhabited if
  and only if its argument evaluates to \texttt{True}, we can now ask the
  compiler to ensure the property holds:

  \begin{Verbatim}[commandchars=\\\{\}]
\PY{l+m+mi}{0}\PY{+w}{ }\PY{k+kt}{RI\PYZus{}OK}\PY{+w}{ }\PY{o+ow}{:}\PY{+w}{ }\PY{k+kt}{So}\PY{+w}{ }\PY{o+ow}{(}\PY{k+kt}{QuickCheck}\PY{+w}{ }\PY{k+kt}{False}\PY{+w}{ }\PY{k+kt}{PROP\PYZus{}readyInsert}\PY{o+ow}{)}
RI\PYZus{}OK\PY{+w}{ }\PY{o+ow}{=}\PY{+w}{ }\PY{k+kt}{Oh}
  \end{Verbatim}

  As the file loads successfully, we can be confident that our model is sound
  with respect to the specified property. Next, we specify a property which we
  hope \QC will find does not hold: that the \acrshort{atm} eventually gets back
  to an available state within reasonable time.

  \begin{Verbatim}[commandchars=\\\{\}]
\PY{l+m+mi}{0}\PY{+w}{ }\PY{k+kt}{PROP\PYZus{}eventuallyReady}\PY{+w}{ }\PY{o+ow}{:}\PY{+w}{ }\PY{k+kt}{Fn}\PY{+w}{ }\PY{o+ow}{(}\PY{k+kt}{ATMTrace}\PY{+w}{ }\PY{k+kt}{Ready}\PY{+w}{ }\PY{l+m+mi}{10}\PY{o+ow}{)}\PY{+w}{ }\PY{k+kt}{Bool}
PROP\PYZus{}eventuallyReady\PY{+w}{ }\PY{o+ow}{=}\PY{+w}{ }\PY{k+kt}{MkFn}
\PY{+w}{ }\PY{+w}{ }\PY{o+ow}{(\PYZbs{}}\PY{k+kr}{case}\PY{+w}{ }\PY{o+ow}{(}\PY{k+kt}{MkATMTrace}\PY{+w}{ }\PY{k+kr}{\PYZus{}}\PY{+w}{ }trace\PY{o+ow}{)}
\PY{+w}{ }\PY{+w}{ }\PY{+w}{ }\PY{+w}{ }\PY{o+ow}{=\PYZgt{}}\PY{+w}{ }elem\PY{+w}{ }\PY{k+kt}{Ready}\PY{+w}{ }\PY{o+ow}{(}map\PY{+w}{ }\PY{o+ow}{(.}resSt\PY{o+ow}{)}\PY{+w}{ }trace\PY{o+ow}{))}

\PY{l+m+mi}{0}\PY{+w}{ }\PY{k+kt}{ER\PYZus{}OK}\PY{+w}{ }\PY{o+ow}{:}\PY{+w}{ }\PY{k+kt}{So}\PY{+w}{ }\PY{o+ow}{(}\PY{k+kt}{QuickCheck}\PY{+w}{ }\PY{k+kt}{False}\PY{+w}{ }\PY{k+kt}{PROP\PYZus{}eventuallyReady}\PY{o+ow}{)}
ER\PYZus{}OK\PY{+w}{ }\PY{o+ow}{=}\PY{+w}{ }\PY{k+kt}{Oh}
  \end{Verbatim}

  Trying to load the file with this property gives:

  \begin{Verbatim}[commandchars=\\\{\}]
\PY{c+c1}{\PYZhy{}\PYZhy{} Error: While processing right hand side of}
\PY{c+c1}{\PYZhy{}\PYZhy{}        EventuallyReady\PYZus{}OK. When unifying:}
\PY{+w}{ }\PY{+w}{ }\PY{+w}{ }\PY{+w}{ }\PY{k+kt}{So}\PY{+w}{ }\PY{k+kt}{True}
\PY{c+c1}{\PYZhy{}\PYZhy{} and:}
\PY{+w}{ }\PY{+w}{ }\PY{+w}{ }\PY{+w}{ }\PY{k+kt}{So}\PY{+w}{ }\PY{o+ow}{(}\PY{k+kt}{QuickCheck}\PY{+w}{ }\PY{k+kt}{False}\PY{+w}{ }\PY{k+kt}{PROP\PYZus{}eventuallyReady}\PY{o+ow}{)}
\PY{c+c1}{\PYZhy{}\PYZhy{} Mismatch between: True and False}
  \end{Verbatim}

  \QC returns \texttt{False}, indicating that our property is failing.
  Inspecting the reason by running \QC on the property at the \iidris REPL
  reveals the cause of the issue:

  \begin{Verbatim}[commandchars=\\\{\}]
\PY{k+kt}{MkQCRes}\PY{+w}{ }\PY{o+ow}{(}\PY{k+kt}{Just}\PY{+w}{ }\PY{k+kt}{False}\PY{o+ow}{)}\PY{+w}{ }\PY{o+ow}{\PYZlt{}}log\PY{o+ow}{\PYZgt{}}\PY{+w}{ }\PY{l+s}{\PYZdq{}}\PY{l+s}{\PYZdq{}}\PY{l+s}{\PYZdq{}}
\PY{l+s}{Falsifiable, after 4 tests:}
\PY{l+s}{Starting @ Ready:}
\PY{l+s}{[ (\PYZlt{}ATMOp \PYZsq{}Insert \PYZti{} ()\PYZsq{}\PYZgt{}, CardInserted)}
\PY{l+s}{, (\PYZlt{}ATMOp \PYZsq{}CheckPIN 0 \PYZti{} Incorrect\PYZsq{}\PYZgt{}, CardInserted)}
\PY{l+s}{, (\PYZlt{}ATMOp \PYZsq{}CheckPIN 0 \PYZti{} Incorrect\PYZsq{}\PYZgt{}, CardInserted)}
\PY{l+s}{, (\PYZlt{}ATMOp \PYZsq{}CheckPIN 0 \PYZti{} Incorrect\PYZsq{}\PYZgt{}, CardInserted)}
\PY{l+s}{, (\PYZlt{}ATMOp \PYZsq{}CheckPIN 0 \PYZti{} Incorrect\PYZsq{}\PYZgt{}, CardInserted)}
\PY{l+s}{\PYZhy{}\PYZhy{} \PYZlt{}etc\PYZgt{}}
\PY{l+s}{]}\PY{l+s}{\PYZdq{}}\PY{l+s}{\PYZdq{}}\PY{l+s}{\PYZdq{}}
  \end{Verbatim}

  This is the loop that was indeed wrong in the initial specification. Our setup
  has constructed sample programs which use the same semantics as our model and
  implementation, and discovered unintended behaviour in the model itself.

  \textbf{Remark:} The first test is technically incorrect: the \acrshort{ism},
  as it is specified, allows for the user to attempt to \textit{Eject} the card
  in the \textbf{Ready} state, a no-op. However, our generator for
  \texttt{OpRes} never includes this option. This is an inherent shortcoming
  with \QC {\textemdash} it is no silver bullet to incomplete data generators.

  \subsection{Fixing the ATM}

  To fix the model, we index the \textbf{CardInserted}-state by the number of
  retries available, and update \texttt{CheckPIN}'s next-state function to take
  this number into account. This limits the number of permitted PIN attempts.

  \begin{Verbatim}[commandchars=\\\{\}]
\PY{k+kr}{data}\PY{+w}{ }\PY{k+kt}{ATMState}\PY{+w}{ }\PY{o+ow}{=}\PY{+w}{ }\PY{k+kt}{Ready}\PY{+w}{ }\PY{o+ow}{|}\PY{+w}{ }\PY{k+kt}{CardInserted}\PY{+w}{ }\PY{k+kt}{Nat}\PY{+w}{ }\PY{o+ow}{|}\PY{+w}{ }\PY{k+kt}{Session}

\PY{n+nf}{ChkPINfn}\PY{+w}{ }\PY{o+ow}{:}\PY{+w}{ }\PY{o+ow}{(}retries\PY{+w}{ }\PY{o+ow}{:}\PY{+w}{ }\PY{k+kt}{Nat}\PY{o+ow}{)}\PY{+w}{ }\PY{o+ow}{\PYZhy{}\PYZgt{}}\PY{+w}{ }\PY{k+kt}{PINok}\PY{+w}{ }\PY{o+ow}{\PYZhy{}\PYZgt{}}\PY{+w}{ }\PY{k+kt}{ATMState}
ChkPINfn\PY{+w}{ }\PY{l+m+mi}{0}\PY{+w}{ }\PY{+w}{ }\PY{+w}{ }\PY{+w}{ }\PY{+w}{ }\PY{k+kt}{Correct}\PY{+w}{ }\PY{+w}{ }\PY{+w}{ }\PY{o+ow}{=}\PY{+w}{ }\PY{k+kt}{Session}
ChkPINfn\PY{+w}{ }\PY{l+m+mi}{0}\PY{+w}{ }\PY{+w}{ }\PY{+w}{ }\PY{+w}{ }\PY{+w}{ }\PY{k+kt}{Incorrect}\PY{+w}{ }\PY{o+ow}{=}\PY{+w}{ }\PY{k+kt}{Ready}
ChkPINfn\PY{+w}{ }\PY{o+ow}{(}\PY{k+kt}{S}\PY{+w}{ }k\PY{o+ow}{)}\PY{+w}{ }\PY{k+kt}{Correct}\PY{+w}{ }\PY{+w}{ }\PY{+w}{ }\PY{o+ow}{=}\PY{+w}{ }\PY{k+kt}{Session}
ChkPINfn\PY{+w}{ }\PY{o+ow}{(}\PY{k+kt}{S}\PY{+w}{ }k\PY{o+ow}{)}\PY{+w}{ }\PY{k+kt}{Incorrect}\PY{+w}{ }\PY{o+ow}{=}\PY{+w}{ }\PY{k+kt}{CardInserted}\PY{+w}{ }k
  \end{Verbatim}

  When we are out of retries, we \emph{must} get the PIN right or the
  \acrshort{atm} resets. If we were to discard the result on zero retries and
  always reset, we could technically permform the \texttt{CheckPIN} operation a
  fourth time, but would have to discard the result even if the PIN was correct,
  because the machine would return us to \textbf{Ready} regardless of the value.
  This felt incorrect, and so we chose to interpret zero retries as ``final
  try'', rather than ``out of tries''.

  \begin{Verbatim}[commandchars=\\\{\}]
\PY{k+kr}{data}\PY{+w}{ }\PY{k+kt}{ATMOp}\PY{+w}{ }\PY{o+ow}{:}\PY{+w}{ }\PY{+w}{ }\PY{o+ow}{(}t\PY{+w}{ }\PY{o+ow}{:}\PY{+w}{ }\PY{k+kt}{Type}\PY{o+ow}{)}\PY{+w}{ }\PY{o+ow}{\PYZhy{}\PYZgt{}}\PY{+w}{ }\PY{k+kt}{ATMState}
\PY{+w}{ }\PY{+w}{ }\PY{+w}{ }\PY{+w}{ }\PY{+w}{ }\PY{+w}{ }\PY{+w}{ }\PY{+w}{ }\PY{+w}{ }\PY{+w}{ }\PY{+w}{ }\PY{o+ow}{\PYZhy{}\PYZgt{}}\PY{+w}{ }\PY{o+ow}{(}t\PY{+w}{ }\PY{o+ow}{\PYZhy{}\PYZgt{}}\PY{+w}{ }\PY{k+kt}{ATMState}\PY{o+ow}{)}\PY{+w}{ }\PY{o+ow}{\PYZhy{}\PYZgt{}}\PY{+w}{ }\PY{k+kt}{Type}\PY{+w}{ }\PY{k+kr}{where}
\PY{+w}{  }\PY{n+nf}{Insert}\PY{+w}{ }\PY{o+ow}{:}\PY{+w}{ }\PY{k+kt}{ATMOp}\PY{+w}{ }\PY{o+ow}{()}\PY{+w}{ }\PY{k+kt}{Ready}\PY{+w}{ }\PY{o+ow}{(}const\PY{+w}{ }\PY{o+ow}{(}\PY{k+kt}{CardInserted}\PY{+w}{ }\PY{l+m+mi}{2}\PY{o+ow}{))}
\PY{+w}{  }\PY{n+nf}{CheckPIN}\PY{+w}{ }\PY{o+ow}{:}\PY{+w}{  }\PY{o+ow}{(}pin\PY{+w}{ }\PY{o+ow}{:}\PY{+w}{ }\PY{k+kt}{Int}\PY{o+ow}{)}
\PY{+w}{ }\PY{+w}{ }\PY{+w}{ }\PY{+w}{ }\PY{+w}{ }\PY{+w}{ }\PY{+w}{ }\PY{+w}{ }\PY{+w}{ }\PY{+w}{ }\PY{+w}{ }\PY{o+ow}{\PYZhy{}\PYZgt{}}\PY{+w}{ }\PY{k+kt}{ATMOp}\PY{+w}{ }\PY{k+kt}{PINok}\PY{+w}{ }\PY{o+ow}{(}\PY{k+kt}{CardInserted}\PY{+w}{ }tries\PY{o+ow}{)}
\PY{+w}{ }\PY{+w}{ }\PY{+w}{ }\PY{+w}{ }\PY{+w}{ }\PY{+w}{ }\PY{+w}{ }\PY{+w}{ }\PY{+w}{ }\PY{+w}{ }\PY{+w}{ }\PY{+w}{ }\PY{+w}{ }\PY{+w}{ }\PY{+w}{ }\PY{+w}{ }\PY{+w}{ }\PY{+w}{ }\PY{+w}{ }\PY{+w}{ }\PY{+w}{ }\PY{+w}{ }\PY{+w}{ }\PY{+w}{ }\PY{+w}{ }\PY{+w}{ }\PY{o+ow}{(}ChkPINfn\PY{+w}{ }tries\PY{o+ow}{)}
\PY{+w}{  }\PY{n+nf}{Dispense}\PY{+w}{ }\PY{o+ow}{:}\PY{+w}{  }\PY{o+ow}{(}amt\PY{+w}{ }\PY{o+ow}{:}\PY{+w}{ }\PY{k+kt}{Nat}\PY{o+ow}{)}
\PY{+w}{ }\PY{+w}{ }\PY{+w}{ }\PY{+w}{ }\PY{+w}{ }\PY{+w}{ }\PY{+w}{ }\PY{+w}{ }\PY{+w}{ }\PY{+w}{ }\PY{+w}{ }\PY{o+ow}{\PYZhy{}\PYZgt{}}\PY{+w}{ }\PY{k+kt}{ATMOp}\PY{+w}{ }\PY{o+ow}{()}\PY{+w}{ }\PY{k+kt}{Session}\PY{+w}{ }\PY{o+ow}{(}const\PY{+w}{ }\PY{k+kt}{Session}\PY{o+ow}{)}
\PY{+w}{  }\PY{n+nf}{Eject}\PY{+w}{ }\PY{o+ow}{:}\PY{+w}{ }\PY{k+kt}{ATMOp}\PY{+w}{ }\PY{o+ow}{()}\PY{+w}{ }st\PY{+w}{ }\PY{o+ow}{(}const\PY{+w}{ }\PY{k+kt}{Ready}\PY{o+ow}{)}
  \end{Verbatim}

  The file reloads successfully, meaning the type-level property test
  \texttt{ER\_OK} passed. And if we retest the property at the \iidris REPL, we
  get:

  \begin{Verbatim}[commandchars=\\\{\}]
\PY{k+kt}{MkQCRes}\PY{+w}{ }\PY{o+ow}{(}\PY{k+kt}{Just}\PY{+w}{ }\PY{k+kt}{True}\PY{o+ow}{)}\PY{+w}{ }\PY{o+ow}{\PYZlt{}}log\PY{o+ow}{\PYZgt{}}\PY{+w}{ }\PY{l+s}{\PYZdq{}}\PY{l+s}{OK, passed 100 tests}\PY{l+s}{\PYZdq{}}
  \end{Verbatim}

  We are now no longer able to introduce a loop in our implementation, as the
  fourth attempt involves 0 remaining retries, which forces us back into
  \textbf{Ready} thanks to the updated \texttt{ChkPINfn}.

  \begin{Verbatim}[commandchars=\\\{\}]
\PY{k+kr}{failing}\PY{+w}{ }\PY{l+s}{\PYZdq{}}\PY{l+s}{Mismatch between: CardInserted ?tries}
\PY{l+s}{         and Ready.}\PY{l+s}{\PYZdq{}}
\PY{+w}{  }\PY{n+nf}{noLoop}\PY{+w}{ }\PY{o+ow}{:}\PY{+w}{ }\PY{k+kt}{ATM}\PY{+w}{ }\PY{o+ow}{()}\PY{+w}{ }\PY{k+kt}{Ready}\PY{+w}{ }\PY{o+ow}{(}const\PY{+w}{ }\PY{k+kt}{Ready}\PY{o+ow}{)}
\PY{+w}{ }\PY{+w}{ }noLoop\PY{+w}{ }\PY{o+ow}{=}\PY{+w}{ }\PY{k+kr}{do}
\PY{+w}{ }\PY{+w}{ }\PY{+w}{ }\PY{+w}{ }\PY{k+kt}{Op}\PY{+w}{ }\PY{k+kt}{Insert}
\PY{+w}{ }\PY{+w}{ }\PY{+w}{ }\PY{+w}{ }\PY{k+kt}{Incorrect}\PY{+w}{ }\PY{o+ow}{\PYZlt{}\PYZhy{}}\PY{+w}{ }\PY{k+kt}{Op}\PY{+w}{ }\PY{o+ow}{\PYZdl{}}\PY{+w}{ }\PY{k+kt}{CheckPIN}\PY{+w}{ }\PY{l+m+mi}{1234}
\PY{+w}{ }\PY{+w}{ }\PY{+w}{ }\PY{+w}{ }\PY{+w}{ }\PY{+w}{ }\PY{o+ow}{|}\PY{+w}{ }\PY{k+kt}{Correct}\PY{+w}{ }\PY{o+ow}{=\PYZgt{}}\PY{+w}{ }\PY{o+ow}{?}noLoop\PYZus{}rhs\PYZus{}1
\PY{+w}{ }\PY{+w}{ }\PY{+w}{ }\PY{+w}{ }\PY{k+kt}{Incorrect}\PY{+w}{ }\PY{o+ow}{\PYZlt{}\PYZhy{}}\PY{+w}{ }\PY{k+kt}{Op}\PY{+w}{ }\PY{o+ow}{\PYZdl{}}\PY{+w}{ }\PY{k+kt}{CheckPIN}\PY{+w}{ }\PY{l+m+mi}{1243}
\PY{+w}{ }\PY{+w}{ }\PY{+w}{ }\PY{+w}{ }\PY{+w}{ }\PY{+w}{ }\PY{o+ow}{|}\PY{+w}{ }\PY{k+kt}{Correct}\PY{+w}{ }\PY{o+ow}{=\PYZgt{}}\PY{+w}{ }\PY{o+ow}{?}noLoop\PYZus{}rhs\PYZus{}2
\PY{+w}{ }\PY{+w}{ }\PY{+w}{ }\PY{+w}{ }\PY{k+kt}{Incorrect}\PY{+w}{ }\PY{o+ow}{\PYZlt{}\PYZhy{}}\PY{+w}{ }\PY{k+kt}{Op}\PY{+w}{ }\PY{o+ow}{\PYZdl{}}\PY{+w}{ }\PY{k+kt}{CheckPIN}\PY{+w}{ }\PY{l+m+mi}{1432}
\PY{+w}{ }\PY{+w}{ }\PY{+w}{ }\PY{+w}{ }\PY{+w}{ }\PY{+w}{ }\PY{o+ow}{|}\PY{+w}{ }\PY{k+kt}{Correct}\PY{+w}{ }\PY{o+ow}{=\PYZgt{}}\PY{+w}{ }\PY{o+ow}{?}noLoop\PYZus{}rhs\PYZus{}3
\PY{+w}{ }\PY{+w}{ }\PY{+w}{ }\PY{+w}{ }\PY{k+kt}{Incorrect}\PY{+w}{ }\PY{o+ow}{\PYZlt{}\PYZhy{}}\PY{+w}{ }\PY{k+kt}{Op}\PY{+w}{ }\PY{o+ow}{\PYZdl{}}\PY{+w}{ }\PY{k+kt}{CheckPIN}\PY{+w}{ }\PY{l+m+mi}{4231}
\PY{+w}{ }\PY{+w}{ }\PY{+w}{ }\PY{+w}{ }\PY{+w}{ }\PY{+w}{ }\PY{o+ow}{|}\PY{+w}{ }\PY{k+kt}{Correct}\PY{+w}{ }\PY{o+ow}{=\PYZgt{}}\PY{+w}{ }\PY{o+ow}{?}noLoop\PYZus{}rhs\PYZus{}4
\PY{+w}{ }\PY{+w}{ }\PY{+w}{ }\PY{+w}{ }\PY{o+ow}{?}noLoop\PYZus{}rhs
  \end{Verbatim}

  This highlights the power of our new approach: an error in the specification
  can be automatically found and, once fixed, the new model is automatically
  threaded through to both the type checker {\textemdash} verifying all
  implementations {\textemdash} and the sample program generation. This greatly
  increases our confidence that the model is well-behaved, meaningfully tested,
  and correctly implemented.

\section{Generalising}\label{sec:generalising}

The \acrshort{atm} example required significant effort to type-check, program,
and property test. If this were required for every state model, the approach
would be tedious to adopt.  Instead, it would be convenient to only have to
specify the model and its transitions, and get the rest ``for free''.

  \subsection{Generic operations and programs}\label{ssec:generic-op-prog}

  To generalise the data types from the previous section, we need to extract the
  common factor and index over it. As briefly discussed in
  \cref{ssec:atm-op-prog}, the programming part of our approach is largely
  already generalised. To reuse the code with a different system, we need to
  define the new states and transitions (or operations): This gives us our
  indices: the state {\textemdash} \texttt{st : Type} {\textemdash} and the type
  of the operations {\textemdash} \texttt{op}:

  \begin{Verbatim}[commandchars=\\\{\}]
\PY{n+nf}{op}\PY{+w}{ }\PY{o+ow}{:}\PY{+w}{ }\PY{k+kr}{forall}\PY{+w}{ }st\PY{+w}{ }\PY{o+ow}{.}\PY{+w}{ }\PY{o+ow}{(}t\PYZsq{}\PY{+w}{ }\PY{o+ow}{:}\PY{+w}{ }\PY{k+kt}{Type}\PY{o+ow}{)}\PY{+w}{ }\PY{o+ow}{\PYZhy{}\PYZgt{}}\PY{+w}{ }st\PY{+w}{ }\PY{o+ow}{\PYZhy{}\PYZgt{}}\PY{+w}{ }\PY{o+ow}{(}t\PYZsq{}\PY{+w}{ }\PY{o+ow}{\PYZhy{}\PYZgt{}}\PY{+w}{ }st\PY{o+ow}{)}\PY{+w}{ }\PY{o+ow}{\PYZhy{}\PYZgt{}}\PY{+w}{ }\PY{k+kt}{Type}
  \end{Verbatim}

  In order to make the \texttt{Prog} type generic, we index it over the type of
  valid operations for the states:

  \begin{Verbatim}[commandchars=\\\{\}]
\PY{k+kr}{data}\PY{+w}{ }\PY{k+kt}{Prog}\PY{+w}{ }\PY{o+ow}{:}\PY{+w}{ }\PY{+w}{ }\PY{o+ow}{\PYZob{}}\PY{l+m+mi}{0}\PY{+w}{ }stT\PY{+w}{ }\PY{o+ow}{:}\PY{+w}{ }\PY{k+kr}{\PYZus{}}\PY{o+ow}{\PYZcb{}}
\PY{+w}{ }\PY{+w}{ }\PY{+w}{ }\PY{+w}{ }\PY{+w}{ }\PY{+w}{ }\PY{+w}{ }\PY{+w}{ }\PY{+w}{ }\PY{+w}{ }\PY{o+ow}{\PYZhy{}\PYZgt{}}\PY{+w}{ }\PY{o+ow}{(}opT\PY{+w}{ }\PY{o+ow}{:}\PY{+w}{ }\PY{+w}{ }\PY{o+ow}{(}t\PYZsq{}\PY{+w}{ }\PY{o+ow}{:}\PY{+w}{ }\PY{k+kr}{\PYZus{}}\PY{o+ow}{)}\PY{+w}{ }\PY{o+ow}{\PYZhy{}\PYZgt{}}\PY{+w}{ }stT\PY{+w}{ }\PY{o+ow}{\PYZhy{}\PYZgt{}}\PY{+w}{ }\PY{o+ow}{(}t\PYZsq{}\PY{+w}{ }\PY{o+ow}{\PYZhy{}\PYZgt{}}\PY{+w}{ }stT\PY{o+ow}{)}
\PY{+w}{ }\PY{+w}{ }\PY{+w}{ }\PY{+w}{ }\PY{+w}{ }\PY{+w}{ }\PY{+w}{ }\PY{+w}{ }\PY{+w}{ }\PY{+w}{ }\PY{+w}{ }\PY{+w}{ }\PY{+w}{ }\PY{+w}{ }\PY{+w}{ }\PY{+w}{ }\PY{+w}{ }\PY{+w}{ }\PY{o+ow}{\PYZhy{}\PYZgt{}}\PY{+w}{ }\PY{k+kt}{Type}\PY{o+ow}{)}
\PY{+w}{ }\PY{+w}{ }\PY{+w}{ }\PY{+w}{ }\PY{+w}{ }\PY{+w}{ }\PY{+w}{ }\PY{+w}{ }\PY{+w}{ }\PY{+w}{ }\PY{o+ow}{\PYZhy{}\PYZgt{}}\PY{+w}{ }\PY{o+ow}{(}t\PY{+w}{ }\PY{o+ow}{:}\PY{+w}{ }\PY{k+kt}{Type}\PY{o+ow}{)}\PY{+w}{ }\PY{o+ow}{\PYZhy{}\PYZgt{}}\PY{+w}{ }\PY{o+ow}{(}from\PY{+w}{ }\PY{o+ow}{:}\PY{+w}{ }stT\PY{o+ow}{)}
\PY{+w}{ }\PY{+w}{ }\PY{+w}{ }\PY{+w}{ }\PY{+w}{ }\PY{+w}{ }\PY{+w}{ }\PY{+w}{ }\PY{+w}{ }\PY{+w}{ }\PY{o+ow}{\PYZhy{}\PYZgt{}}\PY{+w}{ }\PY{o+ow}{(}to\PY{+w}{ }\PY{o+ow}{:}\PY{+w}{ }t\PY{+w}{ }\PY{o+ow}{\PYZhy{}\PYZgt{}}\PY{+w}{ }stT\PY{o+ow}{)}\PY{+w}{ }\PY{o+ow}{\PYZhy{}\PYZgt{}}\PY{+w}{ }\PY{k+kt}{Type}\PY{+w}{ }\PY{k+kr}{where}
\PY{+w}{    }\PY{n+nf}{Pure}\PY{+w}{ }\PY{o+ow}{:}\PY{+w}{ }\PY{o+ow}{(}x\PY{+w}{ }\PY{o+ow}{:}\PY{+w}{ }t\PY{o+ow}{)}\PY{+w}{ }\PY{o+ow}{\PYZhy{}\PYZgt{}}\PY{+w}{ }\PY{k+kt}{Prog}\PY{+w}{ }opT\PY{+w}{ }t\PY{+w}{ }\PY{o+ow}{(}stFn\PY{+w}{ }x\PY{o+ow}{)}\PY{+w}{ }stFn
\PY{+w}{    }\PY{n+nf}{Op}\PY{+w}{ }\PY{o+ow}{:}\PY{+w}{  }\PY{o+ow}{\PYZob{}}\PY{l+m+mi}{0}\PY{+w}{ }opT\PY{+w}{ }\PY{o+ow}{:}\PY{+w}{ }\PY{+w}{ }\PY{o+ow}{(}t\PYZsq{}\PY{+w}{ }\PY{o+ow}{:}\PY{+w}{ }\PY{k+kr}{\PYZus{}}\PY{o+ow}{)}\PY{+w}{ }\PY{o+ow}{\PYZhy{}\PYZgt{}}\PY{+w}{ }stT\PY{+w}{ }\PY{o+ow}{\PYZhy{}\PYZgt{}}\PY{+w}{ }\PY{o+ow}{(}t\PYZsq{}\PY{+w}{ }\PY{o+ow}{\PYZhy{}\PYZgt{}}\PY{+w}{ }stT\PY{o+ow}{)}
\PY{+w}{ }\PY{+w}{ }\PY{+w}{ }\PY{+w}{ }\PY{+w}{ }\PY{+w}{ }\PY{+w}{ }\PY{+w}{ }\PY{+w}{ }\PY{+w}{ }\PY{+w}{ }\PY{+w}{ }\PY{+w}{ }\PY{+w}{ }\PY{+w}{ }\PY{+w}{ }\PY{+w}{ }\PY{o+ow}{\PYZhy{}\PYZgt{}}\PY{+w}{ }\PY{k+kt}{Type}\PY{o+ow}{\PYZcb{}}
\PY{+w}{ }\PY{+w}{ }\PY{+w}{ }\PY{+w}{ }\PY{+w}{ }\PY{+w}{ }\PY{+w}{ }\PY{o+ow}{\PYZhy{}\PYZgt{}}\PY{+w}{ }opT\PY{+w}{ }t\PY{+w}{ }st\PY{+w}{ }stFn\PY{+w}{ }\PY{o+ow}{\PYZhy{}\PYZgt{}}\PY{+w}{ }\PY{k+kt}{Prog}\PY{+w}{ }opT\PY{+w}{ }t\PY{+w}{ }st\PY{+w}{ }stFn
\PY{+w}{ }\PY{+w}{ }\PY{+w}{ }\PY{+w}{ }\PY{o+ow}{(\PYZgt{}\PYZgt{}=)}\PY{+w}{ }\PY{o+ow}{:}\PY{+w}{ }\PY{k+kt}{Prog}\PY{+w}{ }opT\PY{+w}{ }resT1\PY{+w}{ }st1\PY{+w}{ }stFn1
\PY{+w}{ }\PY{+w}{ }\PY{+w}{ }\PY{+w}{ }\PY{+w}{ }\PY{+w}{ }\PY{+w}{ }\PY{+w}{ }\PY{+w}{ }\PY{+w}{ }\PY{o+ow}{\PYZhy{}\PYZgt{}}\PY{+w}{ }\PY{o+ow}{((}x\PY{+w}{ }\PY{o+ow}{:}\PY{+w}{ }resT1\PY{o+ow}{)}
\PY{+w}{ }\PY{+w}{ }\PY{+w}{ }\PY{+w}{ }\PY{+w}{ }\PY{+w}{ }\PY{+w}{ }\PY{+w}{ }\PY{+w}{ }\PY{+w}{ }\PY{+w}{ }\PY{+w}{ }\PY{+w}{ }\PY{+w}{ }\PY{+w}{ }\PY{o+ow}{\PYZhy{}\PYZgt{}}\PY{+w}{ }\PY{k+kt}{Prog}\PY{+w}{ }opT\PY{+w}{ }resT2\PY{+w}{ }\PY{o+ow}{(}stFn1\PY{+w}{ }x\PY{o+ow}{)}\PY{+w}{ }stFn2\PY{o+ow}{)}
\PY{+w}{ }\PY{+w}{ }\PY{+w}{ }\PY{+w}{ }\PY{+w}{ }\PY{+w}{ }\PY{+w}{ }\PY{+w}{ }\PY{+w}{ }\PY{+w}{ }\PY{o+ow}{\PYZhy{}\PYZgt{}}\PY{+w}{ }\PY{k+kt}{Prog}\PY{+w}{ }opT\PY{+w}{ }resT2\PY{+w}{ }st1\PY{+w}{ }stFn2
  \end{Verbatim}

  This gives us a generic way to describe a program producing a result of some
  type, starting in a given state, and ending in a state depending on the
  result. Note that the program's return type and the operations' return types
  may differ; each operation can return different things, which may be different
  from the return type of the whole program. Using this generalised version,
  anything described in the shape of the \texttt{op}-type automatically gains
  support for \texttt{do}-notation as well as the type checker verifying that
  the program only changes states in accordance with the specification.

  \subsection{Generic traces}\label{ssec:generic-opres-trace}

  Taking the same approach as with programs, we can index the infrastructure
  required for the trace generation by the type of operations to make it
  generic. The first part is \texttt{OpRes} -- capturing the type of an
  operation and the type of result it produced, along with the state it happened
  in and the function describing how to process the result to change state.
  We also need a \texttt{Show} instance, to show counterexamples:

  \begin{Verbatim}[commandchars=\\\{\}]
\PY{k+kr}{record}\PY{+w}{ }\PY{k+kt}{OpRes}\PY{+w}{ }\PY{o+ow}{\PYZob{}}\PY{l+m+mi}{0}\PY{+w}{ }stT\PY{+w}{ }\PY{o+ow}{:}\PY{+w}{ }\PY{k+kr}{\PYZus{}}\PY{o+ow}{\PYZcb{}}
\PY{+w}{ }\PY{+w}{ }\PY{+w}{ }\PY{+w}{ }\PY{+w}{ }\PY{+w}{ }\PY{+w}{ }\PY{+w}{ }\PY{+w}{ }\PY{+w}{ }\PY{+w}{ }\PY{+w}{ }\PY{+w}{ }\PY{o+ow}{(}opT\PY{+w}{ }\PY{o+ow}{:}\PY{+w}{ }\PY{+w}{ }\PY{o+ow}{(}t\PYZsq{}\PY{+w}{ }\PY{o+ow}{:}\PY{+w}{ }\PY{k+kr}{\PYZus{}}\PY{o+ow}{)}\PY{+w}{ }\PY{o+ow}{\PYZhy{}\PYZgt{}}\PY{+w}{ }stT\PY{+w}{ }\PY{o+ow}{\PYZhy{}\PYZgt{}}\PY{+w}{ }\PY{o+ow}{(}t\PYZsq{}\PY{+w}{ }\PY{o+ow}{\PYZhy{}\PYZgt{}}\PY{+w}{ }stT\PY{o+ow}{)}
\PY{+w}{ }\PY{+w}{ }\PY{+w}{ }\PY{+w}{ }\PY{+w}{ }\PY{+w}{ }\PY{+w}{ }\PY{+w}{ }\PY{+w}{ }\PY{+w}{ }\PY{+w}{ }\PY{+w}{ }\PY{+w}{ }\PY{+w}{ }\PY{+w}{ }\PY{+w}{ }\PY{+w}{ }\PY{+w}{ }\PY{o+ow}{\PYZhy{}\PYZgt{}}\PY{+w}{ }\PY{k+kt}{Type}\PY{o+ow}{)}
\PY{+w}{ }\PY{+w}{ }\PY{+w}{ }\PY{+w}{ }\PY{+w}{ }\PY{+w}{ }\PY{+w}{ }\PY{+w}{ }\PY{+w}{ }\PY{+w}{ }\PY{+w}{ }\PY{+w}{ }\PY{+w}{ }\PY{o+ow}{(}resT\PY{+w}{ }\PY{o+ow}{:}\PY{+w}{ }\PY{k+kt}{Type}\PY{o+ow}{)}\PY{+w}{ }\PY{o+ow}{(}currSt\PY{+w}{ }\PY{o+ow}{:}\PY{+w}{ }stT\PY{o+ow}{)}
\PY{+w}{ }\PY{+w}{ }\PY{+w}{ }\PY{+w}{ }\PY{+w}{ }\PY{+w}{ }\PY{+w}{ }\PY{+w}{ }\PY{+w}{ }\PY{+w}{ }\PY{+w}{ }\PY{+w}{ }\PY{+w}{ }\PY{o+ow}{(}\PY{l+m+mi}{0}\PY{+w}{ }nsFn\PY{+w}{ }\PY{o+ow}{:}\PY{+w}{ }resT\PY{+w}{ }\PY{o+ow}{\PYZhy{}\PYZgt{}}\PY{+w}{ }stT\PY{o+ow}{)}\PY{+w}{ }\PY{k+kr}{where}
\PY{+w}{ }\PY{+w}{ }\PY{k+kr}{constructor}\PY{+w}{ }\PY{k+kt}{MkOpRes}
\PY{+w}{  }\PY{n+nf}{op}\PY{+w}{ }\PY{o+ow}{:}\PY{+w}{ }opT\PY{+w}{ }resT\PY{+w}{ }currSt\PY{+w}{ }nsFn
\PY{+w}{  }\PY{n+nf}{res}\PY{+w}{ }\PY{o+ow}{:}\PY{+w}{ }resT
\PY{+w}{ }\PY{+w}{ }\PY{o+ow}{\PYZob{}}\PY{k+kr}{auto}\PY{+w}{ }opShow\PY{+w}{ }\PY{o+ow}{:}\PY{+w}{ }\PY{k+kt}{Show}\PY{+w}{ }\PY{o+ow}{(}opT\PY{+w}{ }resT\PY{+w}{ }currSt\PY{+w}{ }nsFn\PY{o+ow}{)\PYZcb{}}
\PY{+w}{ }\PY{+w}{ }\PY{o+ow}{\PYZob{}}\PY{k+kr}{auto}\PY{+w}{ }rShow\PY{+w}{ }\PY{o+ow}{:}\PY{+w}{ }\PY{k+kt}{Show}\PY{+w}{ }resT\PY{o+ow}{\PYZcb{}}
  \end{Verbatim}

  Both \texttt{TraceStep} and \texttt{Trace} follow the same pattern:

  \begin{Verbatim}[commandchars=\\\{\}]
\PY{k+kr}{record}\PY{+w}{ }\PY{k+kt}{TraceStep}\PY{+w}{ }\PY{o+ow}{(}opT\PY{+w}{ }\PY{o+ow}{:}\PY{+w}{ }\PY{+w}{ }\PY{o+ow}{(}t\PYZsq{}\PY{+w}{ }\PY{o+ow}{:}\PY{+w}{ }\PY{k+kr}{\PYZus{}}\PY{o+ow}{)}\PY{+w}{ }\PY{o+ow}{\PYZhy{}\PYZgt{}}\PY{+w}{ }stT
\PY{+w}{ }\PY{+w}{ }\PY{+w}{ }\PY{+w}{ }\PY{+w}{ }\PY{+w}{ }\PY{+w}{ }\PY{+w}{ }\PY{+w}{ }\PY{+w}{ }\PY{+w}{ }\PY{+w}{ }\PY{+w}{ }\PY{+w}{ }\PY{+w}{ }\PY{+w}{ }\PY{+w}{ }\PY{+w}{ }\PY{+w}{ }\PY{+w}{ }\PY{o+ow}{\PYZhy{}\PYZgt{}}\PY{+w}{ }\PY{o+ow}{(}t\PYZsq{}\PY{+w}{ }\PY{o+ow}{\PYZhy{}\PYZgt{}}\PY{+w}{ }stT\PY{o+ow}{)}\PY{+w}{ }\PY{o+ow}{\PYZhy{}\PYZgt{}}\PY{+w}{ }\PY{k+kt}{Type}\PY{o+ow}{)}\PY{+w}{ }\PY{k+kr}{where}
\PY{+w}{ }\PY{+w}{ }\PY{k+kr}{constructor}\PY{+w}{ }\PY{k+kt}{MkTS}
\PY{+w}{ }\PY{+w}{ }\PY{o+ow}{\PYZob{}}\PY{l+m+mi}{0}\PY{+w}{ }stepRT\PY{+w}{ }\PY{o+ow}{:}\PY{+w}{ }\PY{k+kr}{\PYZus{}}\PY{o+ow}{\PYZcb{}}
\PY{+w}{ }\PY{+w}{ }\PY{o+ow}{\PYZob{}}\PY{l+m+mi}{0}\PY{+w}{ }stepSt\PY{+w}{ }\PY{o+ow}{:}\PY{+w}{ }stT\PY{o+ow}{\PYZcb{}}
\PY{+w}{ }\PY{+w}{ }\PY{o+ow}{\PYZob{}}\PY{l+m+mi}{0}\PY{+w}{ }stepFn\PY{+w}{ }\PY{o+ow}{:}\PY{+w}{ }stepRT\PY{+w}{ }\PY{o+ow}{\PYZhy{}\PYZgt{}}\PY{+w}{ }stT\PY{o+ow}{\PYZcb{}}
\PY{+w}{  }\PY{n+nf}{opRes}\PY{+w}{ }\PY{o+ow}{:}\PY{+w}{ }\PY{k+kt}{OpRes}\PY{+w}{ }opT\PY{+w}{ }stepRT\PY{+w}{ }stepSt\PY{+w}{ }stepFn
\PY{+w}{  }\PY{n+nf}{resSt}\PY{+w}{ }\PY{o+ow}{:}\PY{+w}{ }stT
\PY{+w}{ }\PY{+w}{ }\PY{o+ow}{\PYZob{}}\PY{k+kr}{auto}\PY{+w}{ }showStT\PY{+w}{ }\PY{o+ow}{:}\PY{+w}{ }\PY{k+kt}{Show}\PY{+w}{ }stT\PY{o+ow}{\PYZcb{}}

\PY{k+kr}{data}\PY{+w}{ }\PY{k+kt}{Trace}\PY{+w}{ }\PY{o+ow}{:}\PY{+w}{ }\PY{+w}{ }\PY{o+ow}{(}opT\PY{+w}{ }\PY{o+ow}{:}\PY{+w}{ }\PY{+w}{ }\PY{o+ow}{(}t\PYZsq{}\PY{+w}{ }\PY{o+ow}{:}\PY{+w}{ }\PY{k+kr}{\PYZus{}}\PY{o+ow}{)}\PY{+w}{ }\PY{o+ow}{\PYZhy{}\PYZgt{}}\PY{+w}{ }stT
\PY{+w}{ }\PY{+w}{ }\PY{+w}{ }\PY{+w}{ }\PY{+w}{ }\PY{+w}{ }\PY{+w}{ }\PY{+w}{ }\PY{+w}{ }\PY{+w}{ }\PY{+w}{ }\PY{+w}{ }\PY{+w}{ }\PY{+w}{ }\PY{+w}{ }\PY{+w}{ }\PY{+w}{ }\PY{+w}{ }\PY{+w}{ }\PY{o+ow}{\PYZhy{}\PYZgt{}}\PY{+w}{ }\PY{o+ow}{(}t\PYZsq{}\PY{+w}{ }\PY{o+ow}{\PYZhy{}\PYZgt{}}\PY{+w}{ }stT\PY{o+ow}{)}\PY{+w}{ }\PY{o+ow}{\PYZhy{}\PYZgt{}}\PY{+w}{ }\PY{k+kt}{Type}\PY{o+ow}{)}
\PY{+w}{ }\PY{+w}{ }\PY{+w}{ }\PY{+w}{ }\PY{+w}{ }\PY{+w}{ }\PY{+w}{ }\PY{+w}{ }\PY{+w}{ }\PY{+w}{ }\PY{+w}{ }\PY{o+ow}{\PYZhy{}\PYZgt{}}\PY{+w}{ }stT\PY{+w}{ }\PY{o+ow}{\PYZhy{}\PYZgt{}}\PY{+w}{ }\PY{k+kt}{Nat}\PY{+w}{ }\PY{o+ow}{\PYZhy{}\PYZgt{}}\PY{+w}{ }\PY{k+kt}{Type}\PY{+w}{ }\PY{k+kr}{where}
\PY{+w}{  }\PY{n+nf}{MkTrace}\PY{+w}{ }\PY{o+ow}{:}\PY{+w}{ }\PY{k+kt}{Show}\PY{+w}{ }stT\PY{+w}{ }\PY{o+ow}{=\PYZgt{}}\PY{+w}{ }\PY{o+ow}{(}initSt\PY{+w}{ }\PY{o+ow}{:}\PY{+w}{ }stT\PY{o+ow}{)}\PY{+w}{ }\PY{o+ow}{\PYZhy{}\PYZgt{}}\PY{+w}{ }\PY{o+ow}{\PYZob{}}bound\PY{+w}{ }\PY{o+ow}{:}\PY{+w}{ }\PY{k+kt}{Nat}\PY{o+ow}{\PYZcb{}}
\PY{+w}{ }\PY{+w}{ }\PY{+w}{ }\PY{+w}{ }\PY{+w}{ }\PY{+w}{ }\PY{+w}{ }\PY{+w}{ }\PY{+w}{ }\PY{+w}{ }\PY{o+ow}{\PYZhy{}\PYZgt{}}\PY{+w}{ }\PY{o+ow}{(}trace\PY{+w}{ }\PY{o+ow}{:}\PY{+w}{ }\PY{k+kt}{Vect}\PY{+w}{ }bound\PY{+w}{ }\PY{o+ow}{(}\PY{k+kt}{TraceStep}\PY{+w}{ }opT\PY{o+ow}{))}
\PY{+w}{ }\PY{+w}{ }\PY{+w}{ }\PY{+w}{ }\PY{+w}{ }\PY{+w}{ }\PY{+w}{ }\PY{+w}{ }\PY{+w}{ }\PY{+w}{ }\PY{o+ow}{\PYZhy{}\PYZgt{}}\PY{+w}{ }\PY{k+kt}{Trace}\PY{+w}{ }opT\PY{+w}{ }initSt\PY{+w}{ }bound
  \end{Verbatim}

  \subsection{The \texttt{Traceable} interface}\label{ssec:traceable}

  To generate traces, we need to know which
  operations are valid given a current state. We could define this as an
  instance of \texttt{Arbitrary}, however the type declaration is repetitive and
  not idiomatic \iidris. The declaration for the \texttt{ATMOp} and
  \texttt{OpRes} from \cref{ssec:atm-op-prog,ssec:atm-plumbing}, for example,
  would be:

  \begin{Verbatim}[commandchars=\\\{\}]
\PY{o+ow}{\PYZob{}}st\PY{+w}{ }\PY{o+ow}{:}\PY{+w}{ }\PY{k+kt}{ATMState}\PY{o+ow}{\PYZcb{}}\PY{+w}{ }\PY{o+ow}{\PYZhy{}\PYZgt{}}
\PY{k+kt}{Arbitrary}\PY{+w}{ }\PY{o+ow}{(}resT\PY{+w}{ }\PY{o+ow}{:}\PY{+w}{ }\PY{k+kt}{Type}\PY{+w}{ }\PY{o+ow}{**}\PY{+w}{ }nsFn\PY{+w}{ }\PY{o+ow}{:}\PY{+w}{ }resT\PY{+w}{ }\PY{o+ow}{\PYZhy{}\PYZgt{}}\PY{+w}{ }\PY{k+kt}{ATMState}
\PY{+w}{ }\PY{+w}{ }\PY{+w}{ }\PY{+w}{ }\PY{+w}{ }\PY{+w}{ }\PY{+w}{ }\PY{+w}{ }\PY{+w}{ }\PY{+w}{ }\PY{+w}{ }\PY{o+ow}{**}\PY{+w}{ }\PY{k+kt}{OpRes}\PY{+w}{ }\PY{k+kt}{ATMOp}\PY{+w}{ }resT\PY{+w}{ }st\PY{+w}{ }nsFn\PY{o+ow}{)}\PY{+w}{ }\PY{k+kr}{where}
\PY{+w}{ }\PY{+w}{ }arbitrary\PY{+w}{ }\PY{o+ow}{\PYZob{}}st\PY{o+ow}{\PYZcb{}}\PY{+w}{ }\PY{o+ow}{=}\PY{+w}{ }\PY{o+ow}{?}arbitrary\PYZus{}rhs
  \end{Verbatim}

  While we could define this by pattern matching on the implicit `\texttt{st}'
  argument, requiring an implicit argument to define an interface is uncommon,
  as is using pattern matching in a function which does not take any explicit
  arguments. Furthermore, when defining it for a different \acrshort{ism}, we
  would only change the \texttt{opT} and \texttt{stT}, leaving everything else
  the same, which suggests there is a pattern to factor out. We introduce the
  \texttt{Traceable} interface as shorthand for these longer definitions,
  capturing their similarities. An operation is \emph{traceable} if for some
  given current state, we can return a generator producing valid transitions
  away from that state.

  \begin{Verbatim}[commandchars=\\\{\}]
\PY{k+kr}{interface}\PY{+w}{ }\PY{k+kt}{Traceable}\PY{+w}{ }\PY{o+ow}{(}\PY{l+m+mi}{0}\PY{+w}{ }opT\PY{+w}{ }\PY{o+ow}{:}\PY{+w}{ }\PY{o+ow}{(}t\PYZsq{}\PY{+w}{ }\PY{o+ow}{:}\PY{+w}{ }\PY{k+kr}{\PYZus{}}\PY{o+ow}{)}\PY{+w}{ }\PY{o+ow}{\PYZhy{}\PYZgt{}}\PY{+w}{ }stT
\PY{+w}{ }\PY{+w}{ }\PY{+w}{ }\PY{+w}{ }\PY{+w}{ }\PY{+w}{ }\PY{+w}{ }\PY{+w}{ }\PY{+w}{ }\PY{+w}{ }\PY{+w}{ }\PY{+w}{ }\PY{+w}{ }\PY{+w}{ }\PY{+w}{ }\PY{+w}{ }\PY{+w}{ }\PY{+w}{ }\PY{+w}{ }\PY{+w}{ }\PY{+w}{ }\PY{+w}{ }\PY{o+ow}{\PYZhy{}\PYZgt{}}\PY{+w}{ }\PY{o+ow}{(}t\PYZsq{}\PY{+w}{ }\PY{o+ow}{\PYZhy{}\PYZgt{}}\PY{+w}{ }stT\PY{o+ow}{)}\PY{+w}{ }\PY{o+ow}{\PYZhy{}\PYZgt{}}\PY{+w}{ }\PY{k+kt}{Type}\PY{o+ow}{)}\PY{+w}{ }\PY{k+kr}{where}
\PY{+w}{  }\PY{n+nf}{options}\PY{+w}{ }\PY{o+ow}{:}\PY{+w}{  }\PY{o+ow}{(}st\PY{+w}{ }\PY{o+ow}{:}\PY{+w}{ }stT\PY{o+ow}{)}
\PY{+w}{ }\PY{+w}{ }\PY{+w}{ }\PY{+w}{ }\PY{+w}{ }\PY{+w}{ }\PY{+w}{ }\PY{+w}{ }\PY{+w}{ }\PY{+w}{ }\PY{o+ow}{\PYZhy{}\PYZgt{}}\PY{+w}{ }\PY{k+kt}{Gen}\PY{+w}{ }\PY{o+ow}{(}resT\PY{+w}{ }\PY{o+ow}{:}\PY{+w}{ }\PY{k+kt}{Type}\PY{+w}{ }\PY{o+ow}{**}\PY{+w}{ }nsFn\PY{+w}{ }\PY{o+ow}{:}\PY{+w}{ }resT\PY{+w}{ }\PY{o+ow}{\PYZhy{}\PYZgt{}}\PY{+w}{ }stT
\PY{+w}{ }\PY{+w}{ }\PY{+w}{ }\PY{+w}{ }\PY{+w}{ }\PY{+w}{ }\PY{+w}{ }\PY{+w}{ }\PY{+w}{ }\PY{+w}{ }\PY{+w}{ }\PY{+w}{ }\PY{+w}{ }\PY{+w}{ }\PY{+w}{ }\PY{+w}{ }\PY{+w}{ }\PY{+w}{ }\PY{o+ow}{**}\PY{+w}{ }\PY{k+kt}{OpRes}\PY{+w}{ }opT\PY{+w}{ }resT\PY{+w}{ }st\PY{+w}{ }nsFn\PY{o+ow}{)}
  \end{Verbatim}

  When giving an instance of \texttt{Traceable}, the type checker can
  immediately propagate and infer the values for \texttt{opT} and \texttt{stT}
  respectively, saving us the trouble of writing out the lengthy declaration.
  Our framework is still operating inside the \texttt{Gen} monad, so all of
  \QC's combinators, along with \texttt{do} notation, can be used to construct
  more complex generators.
  This shows the strength of our approach: both
  complicated models and test generators can be implemented in the same file!

  \subsection{\texttt{Arbitrary} for generic \acrshortpl{ism}}\label{ssec:generic-arb}

  With the supporting data structures and records generalised, we can implement
  a version of \texttt{Arbitrary} which will work for \emph{any} \acrshort{ism}
  that implements \texttt{Traceable}. The approach is the same as used in
  \cref{ssec:arb-opres,ssec:arb-atm-trace}, except with everything
  indexed by the type of the permitted operations. An implementation never has
  to worry about the implicit state argument to \texttt{arbitrary} as this is
  required via the more straightforward \texttt{Traceable}.

  \begin{Verbatim}[commandchars=\\\{\}]
\PY{o+ow}{\PYZob{}}\PY{l+m+mi}{0}\PY{+w}{ }stT\PY{+w}{ }\PY{o+ow}{:}\PY{+w}{ }\PY{k+kr}{\PYZus{}}\PY{o+ow}{\PYZcb{}}\PY{+w}{ }\PY{o+ow}{\PYZhy{}\PYZgt{}}\PY{+w}{ }\PY{o+ow}{\PYZob{}}\PY{l+m+mi}{0}\PY{+w}{ }opT\PY{+w}{ }\PY{o+ow}{:}\PY{+w}{ }\PY{k+kr}{\PYZus{}}\PY{o+ow}{\PYZcb{}}\PY{+w}{ }\PY{o+ow}{\PYZhy{}\PYZgt{}}\PY{+w}{ }\PY{o+ow}{\PYZob{}}st\PY{+w}{ }\PY{o+ow}{:}\PY{+w}{ }stT\PY{o+ow}{\PYZcb{}}\PY{+w}{ }\PY{o+ow}{\PYZhy{}\PYZgt{}}
\PY{k+kt}{Traceable}\PY{+w}{ }opT\PY{+w}{ }\PY{o+ow}{=\PYZgt{}}
\PY{k+kt}{Arbitrary}\PY{+w}{ }\PY{o+ow}{(}resT\PY{+w}{ }\PY{o+ow}{:}\PY{+w}{ }\PY{k+kt}{Type}\PY{+w}{ }\PY{o+ow}{**}\PY{+w}{ }nsFn\PY{+w}{ }\PY{o+ow}{:}\PY{+w}{ }resT\PY{+w}{ }\PY{o+ow}{\PYZhy{}\PYZgt{}}\PY{+w}{ }stT\PY{+w}{ }\PY{o+ow}{**}
\PY{+w}{ }\PY{+w}{ }\PY{+w}{ }\PY{+w}{ }\PY{+w}{ }\PY{+w}{ }\PY{+w}{ }\PY{+w}{ }\PY{+w}{ }\PY{+w}{ }\PY{+w}{ }\PY{k+kt}{OpRes}\PY{+w}{ }opT\PY{+w}{ }resT\PY{+w}{ }st\PY{+w}{ }nsFn\PY{o+ow}{)}\PY{+w}{ }\PY{k+kr}{where}
\PY{+w}{ }\PY{+w}{ }\PY{k+kr}{where}
\PY{+w}{ }\PY{+w}{ }arbitrary\PY{+w}{ }\PY{o+ow}{\PYZob{}}st\PY{o+ow}{\PYZcb{}}\PY{+w}{ }\PY{o+ow}{=}\PY{+w}{ }options\PY{+w}{ }st

\PY{o+ow}{\PYZob{}}\PY{l+m+mi}{0}\PY{+w}{ }stT\PY{+w}{ }\PY{o+ow}{:}\PY{+w}{ }\PY{k+kr}{\PYZus{}}\PY{o+ow}{\PYZcb{}}\PY{+w}{ }\PY{o+ow}{\PYZhy{}\PYZgt{}}\PY{+w}{ }\PY{o+ow}{\PYZob{}}iSt\PY{+w}{ }\PY{o+ow}{:}\PY{+w}{ }stT\PY{o+ow}{\PYZcb{}}\PY{+w}{ }\PY{o+ow}{\PYZhy{}\PYZgt{}}\PY{+w}{ }\PY{o+ow}{\PYZob{}}bound\PY{+w}{ }\PY{o+ow}{:}\PY{+w}{ }\PY{k+kt}{Nat}\PY{o+ow}{\PYZcb{}}\PY{+w}{ }\PY{o+ow}{\PYZhy{}\PYZgt{}}
\PY{o+ow}{\PYZob{}}opT\PY{+w}{ }\PY{o+ow}{:}\PY{+w}{ }\PY{+w}{ }\PY{o+ow}{(}t\PYZsq{}\PY{+w}{ }\PY{o+ow}{:}\PY{+w}{ }\PY{k+kt}{Type}\PY{o+ow}{)}\PY{+w}{ }\PY{o+ow}{\PYZhy{}\PYZgt{}}\PY{+w}{ }stT\PY{+w}{ }\PY{o+ow}{\PYZhy{}\PYZgt{}}\PY{+w}{ }\PY{o+ow}{(}t\PYZsq{}\PY{+w}{ }\PY{o+ow}{\PYZhy{}\PYZgt{}}\PY{+w}{ }stT\PY{o+ow}{)}\PY{+w}{ }\PY{o+ow}{\PYZhy{}\PYZgt{}}\PY{+w}{ }\PY{k+kt}{Type}\PY{o+ow}{\PYZcb{}}\PY{+w}{ }\PY{o+ow}{\PYZhy{}\PYZgt{}}
\PY{k+kt}{Show}\PY{+w}{ }stT\PY{+w}{ }\PY{o+ow}{=\PYZgt{}}
\PY{k+kt}{Traceable}\PY{+w}{ }opT\PY{+w}{ }\PY{o+ow}{=\PYZgt{}}
\PY{k+kt}{Arbitrary}\PY{+w}{ }\PY{o+ow}{(}resT\PY{+w}{ }\PY{o+ow}{**}\PY{+w}{ }nsFnT\PY{+w}{ }\PY{o+ow}{**}
\PY{+w}{ }\PY{+w}{ }\PY{+w}{ }\PY{+w}{ }\PY{+w}{ }\PY{+w}{ }\PY{+w}{ }\PY{+w}{ }\PY{+w}{ }\PY{+w}{ }\PY{+w}{ }\PY{k+kt}{OpRes}\PY{+w}{ }opT\PY{+w}{ }resT\PY{+w}{ }iSt\PY{+w}{ }nsFnT\PY{o+ow}{)}\PY{+w}{ }\PY{o+ow}{=\PYZgt{}}
\PY{k+kt}{Arbitrary}\PY{+w}{ }\PY{o+ow}{(}\PY{k+kt}{Trace}\PY{+w}{ }opT\PY{+w}{ }iSt\PY{+w}{ }bound\PY{o+ow}{)}
\PY{+w}{ }\PY{+w}{ }\PY{k+kr}{where}
\PY{+w}{ }\PY{+w}{ }arbitrary\PY{+w}{ }\PY{o+ow}{\PYZob{}}bound\PY{+w}{ }\PY{o+ow}{=}\PY{+w}{ }\PY{l+m+mi}{0}\PY{o+ow}{\PYZcb{}}\PY{+w}{ }\PY{o+ow}{=}
\PY{+w}{ }\PY{+w}{ }\PY{+w}{ }\PY{+w}{ }pure\PY{+w}{ }\PY{o+ow}{\PYZdl{}}\PY{+w}{ }\PY{k+kt}{MkTrace}\PY{+w}{ }iSt\PY{+w}{ }\PY{o+ow}{[]}
\PY{+w}{ }\PY{+w}{ }arbitrary\PY{+w}{ }\PY{o+ow}{\PYZob{}}bound\PY{+w}{ }\PY{o+ow}{=}\PY{+w}{ }\PY{o+ow}{(}\PY{k+kt}{S}\PY{+w}{ }k\PY{o+ow}{)\PYZcb{}}\PY{+w}{ }\PY{o+ow}{=}\PY{+w}{ }\PY{k+kr}{do}
\PY{+w}{ }\PY{+w}{ }\PY{+w}{ }\PY{+w}{ }\PY{o+ow}{(}\PY{k+kr}{\PYZus{}}\PY{+w}{ }\PY{o+ow}{**}\PY{+w}{ }nsFn\PY{+w}{ }\PY{o+ow}{**}\PY{+w}{ }opRes\PY{o+ow}{@(}\PY{k+kt}{MkOpRes}\PY{+w}{ }op\PY{+w}{ }res\PY{o+ow}{))}\PY{+w}{ }\PY{o+ow}{\PYZlt{}\PYZhy{}}
\PY{+w}{ }\PY{+w}{ }\PY{+w}{ }\PY{+w}{ }\PY{+w}{ }\PY{+w}{ }the\PY{+w}{ }\PY{o+ow}{(}\PY{k+kt}{Gen}\PY{+w}{ }\PY{o+ow}{(}rT\PY{+w}{ }\PY{o+ow}{**}\PY{+w}{ }fnT\PY{+w}{ }\PY{o+ow}{**}
\PY{+w}{ }\PY{+w}{ }\PY{+w}{ }\PY{+w}{ }\PY{+w}{ }\PY{+w}{ }\PY{+w}{ }\PY{+w}{ }\PY{+w}{ }\PY{+w}{ }\PY{+w}{ }\PY{+w}{ }\PY{+w}{ }\PY{+w}{ }\PY{+w}{ }\PY{+w}{ }\PY{k+kt}{OpRes}\PY{+w}{ }opT\PY{+w}{ }rT\PY{+w}{ }iSt\PY{+w}{ }fnT\PY{o+ow}{))}\PY{+w}{ }arbitrary
\PY{+w}{ }\PY{+w}{ }\PY{+w}{ }\PY{+w}{ }\PY{k+kr}{let}\PY{+w}{ }fstTraceSt\PY{+w}{ }\PY{o+ow}{=}\PY{+w}{ }nsFn\PY{+w}{ }res
\PY{+w}{ }\PY{+w}{ }\PY{+w}{ }\PY{+w}{ }\PY{k+kr}{let}\PY{+w}{ }traceHead\PY{+w}{ }\PY{o+ow}{=}\PY{+w}{ }\PY{k+kt}{MkTS}\PY{+w}{ }opRes\PY{+w}{ }fstTraceSt
\PY{+w}{ }\PY{+w}{ }\PY{+w}{ }\PY{+w}{ }traceTail\PY{+w}{ }\PY{o+ow}{\PYZlt{}\PYZhy{}}\PY{+w}{ }trace\PY{+w}{ }k\PY{+w}{ }fstTraceSt
\PY{+w}{ }\PY{+w}{ }\PY{+w}{ }\PY{+w}{ }pure\PY{+w}{ }\PY{o+ow}{(}\PY{k+kt}{MkTrace}\PY{+w}{ }iSt\PY{+w}{ }\PY{o+ow}{(}traceHead\PY{+w}{ }\PY{o+ow}{::}\PY{+w}{ }traceTail\PY{o+ow}{))}
\PY{+w}{ }\PY{+w}{ }\PY{+w}{ }\PY{+w}{ }\PY{k+kr}{where}
\PY{+w}{      }\PY{n+nf}{trace}\PY{+w}{ }\PY{o+ow}{:}\PY{+w}{  }\PY{o+ow}{(}steps\PY{+w}{ }\PY{o+ow}{:}\PY{+w}{ }\PY{k+kt}{Nat}\PY{o+ow}{)}\PY{+w}{ }\PY{o+ow}{\PYZhy{}\PYZgt{}}\PY{+w}{ }\PY{o+ow}{(}st\PY{+w}{ }\PY{o+ow}{:}\PY{+w}{ }stT\PY{o+ow}{)}
\PY{+w}{ }\PY{+w}{ }\PY{+w}{ }\PY{+w}{ }\PY{+w}{ }\PY{+w}{ }\PY{+w}{ }\PY{+w}{ }\PY{+w}{ }\PY{+w}{ }\PY{+w}{ }\PY{+w}{ }\PY{o+ow}{\PYZhy{}\PYZgt{}}\PY{+w}{ }\PY{k+kt}{Gen}\PY{+w}{ }\PY{o+ow}{(}\PY{k+kt}{Vect}\PY{+w}{ }steps\PY{+w}{ }\PY{o+ow}{(}\PY{k+kt}{TraceStep}\PY{+w}{ }opT\PY{o+ow}{))}
\PY{+w}{ }\PY{+w}{ }\PY{+w}{ }\PY{+w}{ }\PY{+w}{ }\PY{+w}{ }trace\PY{+w}{ }\PY{l+m+mi}{0}\PY{+w}{ }\PY{k+kr}{\PYZus{}}\PY{+w}{ }\PY{o+ow}{=}\PY{+w}{ }pure\PY{+w}{ }\PY{o+ow}{[]}
\PY{+w}{ }\PY{+w}{ }\PY{+w}{ }\PY{+w}{ }\PY{+w}{ }\PY{+w}{ }trace\PY{+w}{ }\PY{o+ow}{(}\PY{k+kt}{S}\PY{+w}{ }j\PY{o+ow}{)}\PY{+w}{ }st\PY{+w}{ }\PY{o+ow}{=}\PY{+w}{ }\PY{k+kr}{do}
\PY{+w}{ }\PY{+w}{ }\PY{+w}{ }\PY{+w}{ }\PY{+w}{ }\PY{+w}{ }\PY{+w}{ }\PY{+w}{ }\PY{o+ow}{(}\PY{k+kr}{\PYZus{}}\PY{+w}{ }\PY{o+ow}{**}\PY{+w}{ }stFn\PY{+w}{ }\PY{o+ow}{**}\PY{+w}{ }opR\PY{o+ow}{@(}\PY{k+kt}{MkOpRes}\PY{+w}{ }op\PY{+w}{ }res\PY{o+ow}{))}\PY{+w}{ }\PY{o+ow}{\PYZlt{}\PYZhy{}}
\PY{+w}{ }\PY{+w}{ }\PY{+w}{ }\PY{+w}{ }\PY{+w}{ }\PY{+w}{ }\PY{+w}{ }\PY{+w}{ }\PY{+w}{ }\PY{+w}{ }the\PY{+w}{ }\PY{o+ow}{(}\PY{k+kt}{Gen}\PY{+w}{ }\PY{o+ow}{(}x\PY{+w}{ }\PY{o+ow}{**}\PY{+w}{ }y\PY{+w}{ }\PY{o+ow}{**}
\PY{+w}{ }\PY{+w}{ }\PY{+w}{ }\PY{+w}{ }\PY{+w}{ }\PY{+w}{ }\PY{+w}{ }\PY{+w}{ }\PY{+w}{ }\PY{+w}{ }\PY{+w}{ }\PY{+w}{ }\PY{+w}{ }\PY{+w}{ }\PY{+w}{ }\PY{+w}{ }\PY{+w}{ }\PY{+w}{ }\PY{+w}{ }\PY{+w}{ }\PY{k+kt}{OpRes}\PY{+w}{ }opT\PY{+w}{ }x\PY{+w}{ }st\PY{+w}{ }y\PY{o+ow}{))}\PY{+w}{ }arbitrary
\PY{+w}{ }\PY{+w}{ }\PY{+w}{ }\PY{+w}{ }\PY{+w}{ }\PY{+w}{ }\PY{+w}{ }\PY{+w}{ }\PY{k+kr}{let}\PY{+w}{ }nextSt\PY{+w}{ }\PY{o+ow}{=}\PY{+w}{ }stFn\PY{+w}{ }res
\PY{+w}{ }\PY{+w}{ }\PY{+w}{ }\PY{+w}{ }\PY{+w}{ }\PY{+w}{ }\PY{+w}{ }\PY{+w}{ }pure\PY{+w}{ }\PY{o+ow}{\PYZdl{}}\PY{+w}{ }\PY{o+ow}{(}\PY{k+kt}{MkTS}\PY{+w}{ }opR\PY{+w}{ }nextSt\PY{o+ow}{)}\PY{+w}{ }\PY{o+ow}{::}
\PY{+w}{ }\PY{+w}{ }\PY{+w}{ }\PY{+w}{ }\PY{+w}{ }\PY{+w}{ }\PY{+w}{ }\PY{+w}{ }\PY{+w}{ }\PY{+w}{ }\PY{+w}{ }\PY{+w}{ }\PY{+w}{ }\PY{+w}{ }\PY{+w}{ }\PY{+w}{ }\PY{o+ow}{!(}trace\PY{+w}{ }j\PY{+w}{ }nextSt\PY{o+ow}{)}
  \end{Verbatim}

  This completes the generalisation, allowing us to model, verify, implement,
  and test any specification as long as the states, transitions, and options
  from each state are given.

  \subsection{Evaluation: The \acrshort{arq} Protocol}\label{ssec:arq}

  We evaluate our generalisation by implementing a different system,
  the \acrfull{arq} protocol. The \acrshort{arq} protocol works by sending a
  single packet containing some data and a packet number, and then waiting for
  an acknowledgement of the packet number before advancing to sending the next
  packet~\cite{linErrorControlCoding1983}.
  We chose \acrshort{arq} because it is simple enough to be
  understandable, while also presenting a some interesting
  challenges: there is an external second party involved, whose behaviour we
  cannot know; and, due to packet numbering, there are potentially infinite
  states.



  \subsubsection{The states of \acrshort{arq}}\label{ssec:arq-states}

  Na{\"i}vely, the protocol only has two states: \textbf{Ready} and
  \textbf{Waiting}. However, the semantics of \acrshort{arq} introduce a third
  state, \textbf{Acked}. When we receive an acknowledgement {\textemdash} an
  \texttt{Ack} {\textemdash} for a certain packet, we need to check that it
  acknowledges the sequence number we sent and retry if the \texttt{Ack} was for
  another packet (potentially due to data corruption on the return trip).
  Checking the acknowledged sequence number can then either require us to
  retransmit the same packet or, if everything is fine, to proceed to sending
  the next packet in the sequence.

  \begin{Verbatim}[commandchars=\\\{\}]
\PY{k+kr}{data}\PY{+w}{ }\PY{k+kt}{ARQState}\PY{+w}{ }\PY{o+ow}{=}\PY{+w}{ }\PY{k+kt}{Ready}\PY{+w}{ }\PY{k+kt}{Nat}\PY{+w}{ }\PY{o+ow}{|}\PY{+w}{ }\PY{k+kt}{Waiting}\PY{+w}{ }\PY{k+kt}{Nat}
\PY{+w}{ }\PY{+w}{ }\PY{+w}{ }\PY{+w}{ }\PY{+w}{ }\PY{+w}{ }\PY{+w}{ }\PY{+w}{ }\PY{+w}{ }\PY{+w}{ }\PY{+w}{ }\PY{+w}{ }\PY{+w}{ }\PY{+w}{ }\PY{o+ow}{|}\PY{+w}{ }\PY{k+kt}{Acked}\PY{+w}{ }\PY{k+kt}{Nat}\PY{+w}{ }\PY{k+kt}{Nat}
  \end{Verbatim}

  Each state takes the current sequence number of the packet being transmitted,
  with \textbf{Acked} additionally taking the acknowledged sequence number so
  that we can verify it.

  We define a simple packet record, along with a data type for capturing the
  possible outcomes of waiting on an acknowledgement.

  \begin{Verbatim}[commandchars=\\\{\}]
\PY{k+kr}{record}\PY{+w}{ }\PY{k+kt}{Pkt}\PY{+w}{ }\PY{k+kr}{where}
\PY{+w}{ }\PY{+w}{ }\PY{k+kr}{constructor}\PY{+w}{ }\PY{k+kt}{MkPkt}
\PY{+w}{  }\PY{n+nf}{pl}\PY{+w}{ }\PY{o+ow}{:}\PY{+w}{ }\PY{k+kt}{Bits8}
\PY{+w}{  }\PY{n+nf}{sn}\PY{+w}{ }\PY{o+ow}{:}\PY{+w}{ }\PY{k+kt}{Nat}

\PY{k+kr}{data}\PY{+w}{ }\PY{k+kt}{WaitRes}\PY{+w}{ }\PY{o+ow}{=}\PY{+w}{ }\PY{k+kt}{Ack}\PY{+w}{ }\PY{k+kt}{Nat}\PY{+w}{ }\PY{o+ow}{|}\PY{+w}{ }\PY{k+kt}{Timeout}
  \end{Verbatim}

  Note that this captures the fact that we cannot know how the other side will
  reply, if at all. We are not trying to simulate timed automata to model the
  exact timeouts required. Instead, we model the possible outcomes and test that
  our protocol is well-behaved under these scenarios.

  \subsubsection{The \texttt{Next} state function,}\label{ssec:arq-nextStFn}

  transitions to \textbf{Acked} if an \texttt{Ack}-reply was received
  {\textemdash} keeping track of both the packet number and the reply number
  {\textemdash} or immediately back to \textbf{Ready} if the reply never came,
  forcing us to retry sending the same packet.

  \begin{Verbatim}[commandchars=\\\{\}]
\PY{n+nf}{Next}\PY{+w}{ }\PY{o+ow}{:}\PY{+w}{ }\PY{o+ow}{(}n\PY{+w}{ }\PY{o+ow}{:}\PY{+w}{ }\PY{k+kt}{Nat}\PY{o+ow}{)}\PY{+w}{ }\PY{o+ow}{\PYZhy{}\PYZgt{}}\PY{+w}{ }\PY{k+kt}{WaitRes}\PY{+w}{ }\PY{o+ow}{\PYZhy{}\PYZgt{}}\PY{+w}{ }\PY{k+kt}{ARQState}
Next\PY{+w}{ }n\PY{+w}{ }\PY{o+ow}{(}\PY{k+kt}{Ack}\PY{+w}{ }a\PY{o+ow}{)}\PY{+w}{ }\PY{o+ow}{=}\PY{+w}{ }\PY{k+kt}{Acked}\PY{+w}{ }n\PY{+w}{ }a
Next\PY{+w}{ }n\PY{+w}{ }\PY{k+kt}{Timeout}\PY{+w}{ }\PY{o+ow}{=}\PY{+w}{ }\PY{k+kt}{Ready}\PY{+w}{ }n
  \end{Verbatim}

  \subsubsection{Moving on to \acrshort{arq} operations,}\label{ssec:arq-ops}

  \textit{Send} takes a packet to send and ensures the state types keeps track
  of the current packet number, and \textit{Wait} proceeds with a wait result.
  The more interesting transitions, \textit{Proceed} and \textit{Retry}, take a
  proof that the acknowledged number and the packet number are equal or that
  they cannot be equal, respectively. This adds some overhead to programming
  with the operations, but we chose to include this as it nicely show how
  dependent types integrate with our new approach:

  \begin{Verbatim}[commandchars=\\\{\}]
\PY{k+kr}{data}\PY{+w}{ }\PY{k+kt}{ARQOp}\PY{+w}{ }\PY{o+ow}{:}\PY{+w}{ }\PY{o+ow}{(}t\PY{+w}{ }\PY{o+ow}{:}\PY{+w}{ }\PY{k+kr}{\PYZus{}}\PY{o+ow}{)}\PY{+w}{ }\PY{o+ow}{\PYZhy{}\PYZgt{}}\PY{+w}{ }\PY{k+kt}{ARQState}\PY{+w}{ }\PY{o+ow}{\PYZhy{}\PYZgt{}}\PY{+w}{ }\PY{o+ow}{(}t\PY{+w}{ }\PY{o+ow}{\PYZhy{}\PYZgt{}}\PY{+w}{ }\PY{k+kt}{ARQState}\PY{o+ow}{)}
\PY{+w}{ }\PY{+w}{ }\PY{+w}{ }\PY{+w}{ }\PY{+w}{ }\PY{+w}{ }\PY{+w}{ }\PY{+w}{ }\PY{+w}{ }\PY{+w}{ }\PY{+w}{ }\PY{o+ow}{\PYZhy{}\PYZgt{}}\PY{+w}{ }\PY{k+kt}{Type}\PY{+w}{ }\PY{k+kr}{where}
\PY{+w}{  }\PY{n+nf}{Send}\PY{+w}{ }\PY{o+ow}{:}\PY{+w}{ }\PY{o+ow}{(}pkt\PY{+w}{ }\PY{o+ow}{:}\PY{+w}{ }\PY{k+kt}{Pkt}\PY{o+ow}{)}\PY{+w}{ }\PY{o+ow}{\PYZhy{}\PYZgt{}}\PY{+w}{ }\PY{k+kt}{ARQOp}\PY{+w}{ }\PY{o+ow}{()}\PY{+w}{ }\PY{o+ow}{(}\PY{k+kt}{Ready}\PY{+w}{ }pkt\PY{o+ow}{.}sn\PY{o+ow}{)}
\PY{+w}{ }\PY{+w}{ }\PY{+w}{ }\PY{+w}{ }\PY{+w}{ }\PY{+w}{ }\PY{+w}{ }\PY{+w}{ }\PY{+w}{ }\PY{+w}{ }\PY{+w}{ }\PY{+w}{ }\PY{+w}{ }\PY{+w}{ }\PY{+w}{ }\PY{+w}{ }\PY{+w}{ }\PY{+w}{ }\PY{+w}{ }\PY{+w}{ }\PY{+w}{ }\PY{+w}{ }\PY{+w}{ }\PY{+w}{ }\PY{o+ow}{(}const\PY{+w}{ }\PY{o+ow}{\PYZdl{}}\PY{+w}{ }\PY{k+kt}{Waiting}\PY{+w}{ }\PY{o+ow}{(}pkt\PY{o+ow}{.}sn\PY{o+ow}{))}
\PY{+w}{  }\PY{n+nf}{Wait}\PY{+w}{ }\PY{o+ow}{:}\PY{+w}{ }\PY{k+kt}{ARQOp}\PY{+w}{ }\PY{k+kt}{WaitRes}\PY{+w}{ }\PY{o+ow}{(}\PY{k+kt}{Waiting}\PY{+w}{ }n\PY{o+ow}{)}\PY{+w}{ }\PY{o+ow}{(}Next\PY{+w}{ }n\PY{o+ow}{)}
\PY{+w}{  }\PY{n+nf}{Proceed}\PY{+w}{ }\PY{o+ow}{:}\PY{+w}{ }\PY{o+ow}{(}ok\PY{+w}{ }\PY{o+ow}{:}\PY{+w}{ }a\PY{+w}{ }\PY{o+ow}{=}\PY{o+ow}{=}\PY{o+ow}{=}\PY{+w}{ }n\PY{o+ow}{)}
\PY{+w}{ }\PY{+w}{ }\PY{+w}{ }\PY{+w}{ }\PY{o+ow}{\PYZhy{}\PYZgt{}}\PY{+w}{ }\PY{k+kt}{ARQOp}\PY{+w}{ }\PY{o+ow}{()}\PY{+w}{ }\PY{o+ow}{(}\PY{k+kt}{Acked}\PY{+w}{ }n\PY{+w}{ }a\PY{o+ow}{)}\PY{+w}{ }\PY{o+ow}{(}const\PY{+w}{ }\PY{o+ow}{\PYZdl{}}\PY{+w}{ }\PY{k+kt}{Ready}\PY{+w}{ }\PY{o+ow}{(}\PY{k+kt}{S}\PY{+w}{ }n\PY{o+ow}{))}
\PY{+w}{  }\PY{n+nf}{Retry}\PY{+w}{ }\PY{o+ow}{:}\PY{+w}{ }\PY{o+ow}{(}\PY{k+kt}{Not}\PY{+w}{ }\PY{o+ow}{(}a\PY{+w}{ }\PY{o+ow}{=}\PY{o+ow}{=}\PY{o+ow}{=}\PY{+w}{ }n\PY{o+ow}{))}
\PY{+w}{ }\PY{+w}{ }\PY{+w}{ }\PY{+w}{ }\PY{o+ow}{\PYZhy{}\PYZgt{}}\PY{+w}{ }\PY{k+kt}{ARQOp}\PY{+w}{ }\PY{o+ow}{()}\PY{+w}{ }\PY{o+ow}{(}\PY{k+kt}{Acked}\PY{+w}{ }n\PY{+w}{ }a\PY{o+ow}{)}\PY{+w}{ }\PY{o+ow}{(}const\PY{+w}{ }\PY{o+ow}{\PYZdl{}}\PY{+w}{ }\PY{k+kt}{Ready}\PY{+w}{ }n\PY{o+ow}{)}
  \end{Verbatim}

  \subsubsection{Finally, we need a \texttt{Traceable} instance}\label{ssec:arq-traceable}

  When in the \textbf{Ready} state, we have access to the sequence number we are
  meant to be sending, and so we can construct an arbitrary packet and
  \textit{Send} it (we use a placeholder payload of 255 rather than
  \texttt{arbitrary} for brevity). Once we have received an \texttt{Ack} and are
  in the \textbf{Acked} state, we need to check whether the two numbers are
  equal. If they do, the only thing we can do is to advance to sending the next
  packet. If they cannot be equal, the only thing we can do is to retry sending
  the packet. This may sound like we have no control over the frequency of
  accepted versus rejected acknowledgements, however we \emph{can} control this
  by simulating an unreliable network from the \textbf{Waiting} state: 20\% of
  the time we do not get a reply, timing out instead; 5\% of the time we get an
  \texttt{arbitrary} acknowledgement; and the remaining 75\% of the time we
  successfully transmit and get a valid acknowledgement back:

  \begin{Verbatim}[commandchars=\\\{\}]
\PY{k+kt}{Traceable}\PY{+w}{ }\PY{k+kt}{ARQOp}\PY{+w}{ }\PY{k+kr}{where}
\PY{+w}{ }\PY{+w}{ }options\PY{+w}{ }\PY{o+ow}{(}\PY{k+kt}{Ready}\PY{+w}{ }k\PY{o+ow}{)}\PY{+w}{ }\PY{o+ow}{=}\PY{+w}{ }pure
\PY{+w}{ }\PY{+w}{ }\PY{+w}{ }\PY{+w}{ }\PY{o+ow}{(}\PY{k+kr}{\PYZus{}}\PY{+w}{ }\PY{o+ow}{**}\PY{+w}{ }\PY{k+kr}{\PYZus{}}\PY{+w}{ }\PY{o+ow}{**}\PY{+w}{ }\PY{k+kt}{MkOpRes}\PY{+w}{ }\PY{o+ow}{(}\PY{k+kt}{Send}\PY{+w}{ }\PY{o+ow}{(}\PY{k+kt}{MkPkt}\PY{+w}{ }\PY{l+m+mi}{255}\PY{+w}{ }k\PY{o+ow}{))}\PY{+w}{ }\PY{o+ow}{())}

\PY{+w}{ }\PY{+w}{ }options\PY{+w}{ }\PY{o+ow}{(}\PY{k+kt}{Waiting}\PY{+w}{ }k\PY{o+ow}{)}\PY{+w}{ }\PY{o+ow}{=}\PY{+w}{ }frequency
\PY{+w}{ }\PY{+w}{ }\PY{+w}{ }\PY{+w}{ }\PY{o+ow}{[}\PY{+w}{ }\PY{o+ow}{(}\PY{l+m+mi}{4},\PY{+w}{ }pure\PY{+w}{ }\PY{o+ow}{(}\PY{k+kr}{\PYZus{}}\PY{+w}{ }\PY{o+ow}{**}\PY{+w}{ }\PY{k+kr}{\PYZus{}}\PY{+w}{ }\PY{o+ow}{**}\PY{+w}{ }\PY{k+kt}{MkOpRes}\PY{+w}{ }\PY{k+kt}{Wait}\PY{+w}{ }\PY{k+kt}{Timeout}\PY{o+ow}{))}
\PY{+w}{ }\PY{+w}{ }\PY{+w}{ }\PY{+w}{ },\PY{+w}{ }\PY{o+ow}{(}\PY{l+m+mi}{1},\PY{+w}{ }\PY{k+kr}{do}\PY{+w}{ }pure
\PY{+w}{ }\PY{+w}{ }\PY{+w}{ }\PY{+w}{ }\PY{+w}{ }\PY{+w}{ }\PY{+w}{ }\PY{+w}{ }\PY{o+ow}{(}\PY{k+kr}{\PYZus{}}\PY{+w}{ }\PY{o+ow}{**}\PY{+w}{ }\PY{k+kr}{\PYZus{}}\PY{+w}{ }\PY{o+ow}{**}\PY{+w}{ }\PY{k+kt}{MkOpRes}\PY{+w}{ }\PY{k+kt}{Wait}\PY{+w}{ }\PY{o+ow}{(}\PY{k+kt}{Ack}\PY{+w}{ }\PY{o+ow}{!}arbitrary\PY{o+ow}{)))}
\PY{+w}{ }\PY{+w}{ }\PY{+w}{ }\PY{+w}{ },\PY{+w}{ }\PY{o+ow}{(}\PY{l+m+mi}{15},\PY{+w}{ }pure\PY{+w}{ }\PY{o+ow}{(}\PY{k+kr}{\PYZus{}}\PY{+w}{ }\PY{o+ow}{**}\PY{+w}{ }\PY{k+kr}{\PYZus{}}\PY{+w}{ }\PY{o+ow}{**}\PY{+w}{ }\PY{k+kt}{MkOpRes}\PY{+w}{ }\PY{k+kt}{Wait}\PY{+w}{ }\PY{o+ow}{(}\PY{k+kt}{Ack}\PY{+w}{ }k\PY{o+ow}{)))}
\PY{+w}{ }\PY{+w}{ }\PY{+w}{ }\PY{+w}{ }\PY{o+ow}{]}

\PY{+w}{ }\PY{+w}{ }options\PY{+w}{ }\PY{o+ow}{(}\PY{k+kt}{Acked}\PY{+w}{ }n\PY{+w}{ }a\PY{o+ow}{)}\PY{+w}{ }\PY{o+ow}{=}\PY{+w}{ }\PY{k+kr}{case}\PY{+w}{ }decEq\PY{+w}{ }a\PY{+w}{ }n\PY{+w}{ }\PY{k+kr}{of}
\PY{+w}{ }\PY{+w}{ }\PY{+w}{ }\PY{+w}{ }\PY{o+ow}{(}\PY{k+kt}{Yes}\PY{+w}{ }prf\PY{o+ow}{)}\PY{+w}{ }\PY{o+ow}{=\PYZgt{}}
\PY{+w}{ }\PY{+w}{ }\PY{+w}{ }\PY{+w}{ }\PY{+w}{ }\PY{+w}{ }pure\PY{+w}{ }\PY{o+ow}{(}\PY{k+kr}{\PYZus{}}\PY{+w}{ }\PY{o+ow}{**}\PY{+w}{ }\PY{k+kr}{\PYZus{}}\PY{+w}{ }\PY{o+ow}{**}\PY{+w}{ }\PY{k+kt}{MkOpRes}\PY{+w}{ }\PY{o+ow}{(}\PY{k+kt}{Proceed}\PY{+w}{ }prf\PY{o+ow}{)}\PY{+w}{ }\PY{o+ow}{())}
\PY{+w}{ }\PY{+w}{ }\PY{+w}{ }\PY{+w}{ }\PY{o+ow}{(}\PY{k+kt}{No}\PY{+w}{ }contra\PY{o+ow}{)}\PY{+w}{ }\PY{o+ow}{=\PYZgt{}}
\PY{+w}{ }\PY{+w}{ }\PY{+w}{ }\PY{+w}{ }\PY{+w}{ }\PY{+w}{ }pure\PY{+w}{ }\PY{o+ow}{(}\PY{k+kr}{\PYZus{}}\PY{+w}{ }\PY{o+ow}{**}\PY{+w}{ }\PY{k+kr}{\PYZus{}}\PY{+w}{ }\PY{o+ow}{**}\PY{+w}{ }\PY{k+kt}{MkOpRes}\PY{+w}{ }\PY{o+ow}{(}\PY{k+kt}{Retry}\PY{+w}{ }contra\PY{o+ow}{)}\PY{+w}{ }\PY{o+ow}{())}
  \end{Verbatim}

  \subsubsection{This is all we need}\label{ssec:arq-all-we-need}

  We have now defined everything the programmer needs to define to use our new
  approach. Thanks to our generalisation, we can now plug our new stateful model
  into \texttt{Prog} and immediately get access to \texttt{do}-notation and
  type-level state transition verification:

  \begin{Verbatim}[commandchars=\\\{\}]
\PY{n+nf}{sendN}\PY{+w}{ }\PY{o+ow}{:}\PY{+w}{ }\PY{o+ow}{(}n\PY{+w}{ }\PY{o+ow}{:}\PY{+w}{ }\PY{k+kt}{Nat}\PY{o+ow}{)}
\PY{+w}{ }\PY{+w}{ }\PY{o+ow}{\PYZhy{}\PYZgt{}}\PY{+w}{ }\PY{k+kt}{Prog}\PY{+w}{ }\PY{k+kt}{ARQOp}\PY{+w}{ }\PY{o+ow}{()}\PY{+w}{ }\PY{o+ow}{(}\PY{k+kt}{Ready}\PY{+w}{ }n\PY{o+ow}{)}\PY{+w}{ }\PY{o+ow}{(}const\PY{+w}{ }\PY{o+ow}{\PYZdl{}}\PY{+w}{ }\PY{k+kt}{Ready}\PY{+w}{ }\PY{o+ow}{(}\PY{k+kt}{S}\PY{+w}{ }n\PY{o+ow}{))}
sendN\PY{+w}{ }n\PY{+w}{ }\PY{o+ow}{=}\PY{+w}{ }\PY{k+kr}{do}
\PY{+w}{ }\PY{+w}{ }\PY{k+kt}{Op}\PY{+w}{ }\PY{o+ow}{\PYZdl{}}\PY{+w}{ }\PY{k+kt}{Send}\PY{+w}{ }\PY{o+ow}{(}\PY{k+kt}{MkPkt}\PY{+w}{ }\PY{l+m+mi}{255}\PY{+w}{ }n\PY{o+ow}{)}
\PY{+w}{ }\PY{+w}{ }\PY{o+ow}{(}\PY{k+kt}{Ack}\PY{+w}{ }a\PY{o+ow}{)}\PY{+w}{ }\PY{o+ow}{\PYZlt{}\PYZhy{}}\PY{+w}{ }\PY{k+kt}{Op}\PY{+w}{ }\PY{k+kt}{Wait}
\PY{+w}{ }\PY{+w}{ }\PY{+w}{ }\PY{+w}{ }\PY{o+ow}{|}\PY{+w}{ }\PY{k+kt}{Timeout}\PY{+w}{ }\PY{o+ow}{=\PYZgt{}}\PY{+w}{ }sendN\PY{+w}{ }n
\PY{+w}{ }\PY{+w}{ }\PY{k+kr}{case}\PY{+w}{ }decEq\PY{+w}{ }a\PY{+w}{ }n\PY{+w}{ }\PY{k+kr}{of}
\PY{+w}{ }\PY{+w}{ }\PY{+w}{ }\PY{+w}{ }\PY{+w}{ }\PY{+w}{ }\PY{+w}{ }\PY{o+ow}{(}\PY{k+kt}{Yes}\PY{+w}{ }prf\PY{o+ow}{)}\PY{+w}{ }\PY{o+ow}{=\PYZgt{}}\PY{+w}{ }\PY{k+kt}{Op}\PY{+w}{ }\PY{o+ow}{\PYZdl{}}\PY{+w}{ }\PY{k+kt}{Proceed}\PY{+w}{ }prf
\PY{+w}{ }\PY{+w}{ }\PY{+w}{ }\PY{+w}{ }\PY{+w}{ }\PY{+w}{ }\PY{+w}{ }\PY{o+ow}{(}\PY{k+kt}{No}\PY{+w}{ }contra\PY{o+ow}{)}\PY{+w}{ }\PY{o+ow}{=\PYZgt{}}\PY{+w}{ }\PY{k+kr}{do}
\PY{+w}{ }\PY{+w}{ }\PY{+w}{ }\PY{+w}{ }\PY{+w}{ }\PY{+w}{ }\PY{+w}{ }\PY{+w}{ }\PY{+w}{ }\PY{k+kt}{Op}\PY{+w}{ }\PY{o+ow}{\PYZdl{}}\PY{+w}{ }\PY{k+kt}{Retry}\PY{+w}{ }contra
\PY{+w}{ }\PY{+w}{ }\PY{+w}{ }\PY{+w}{ }\PY{+w}{ }\PY{+w}{ }\PY{+w}{ }\PY{+w}{ }\PY{+w}{ }sendN\PY{+w}{ }n

\PY{n+nf}{prog}\PY{+w}{ }\PY{o+ow}{:}\PY{+w}{ }\PY{k+kt}{Prog}\PY{+w}{ }\PY{k+kt}{ARQOp}\PY{+w}{ }\PY{o+ow}{()}\PY{+w}{ }\PY{o+ow}{(}\PY{k+kt}{Ready}\PY{+w}{ }\PY{l+m+mi}{0}\PY{o+ow}{)}\PY{+w}{ }\PY{o+ow}{(}const\PY{+w}{ }\PY{o+ow}{\PYZdl{}}\PY{+w}{ }\PY{k+kt}{Ready}\PY{+w}{ }\PY{l+m+mi}{3}\PY{o+ow}{)}
prog\PY{+w}{ }\PY{o+ow}{=}\PY{+w}{ }\PY{k+kr}{do}\PY{+w}{ }sendN\PY{+w}{ }\PY{l+m+mi}{0}\PY{+w}{ };\PY{+w}{ }sendN\PY{+w}{ }\PY{l+m+mi}{1}\PY{+w}{ };\PY{+w}{ }sendN\PY{+w}{ }\PY{l+m+mi}{2}

\PY{k+kr}{failing}\PY{+w}{ }\PY{l+s}{\PYZdq{}}\PY{l+s}{Mismatch between: 1 and 0}\PY{l+s}{\PYZdq{}}
\PY{+w}{  }\PY{n+nf}{bad}\PY{+w}{ }\PY{o+ow}{:}\PY{+w}{ }\PY{k+kt}{Prog}\PY{+w}{ }\PY{k+kt}{ARQOp}\PY{+w}{ }\PY{o+ow}{()}\PY{+w}{ }\PY{o+ow}{(}\PY{k+kt}{Ready}\PY{+w}{ }\PY{l+m+mi}{0}\PY{o+ow}{)}\PY{+w}{ }\PY{o+ow}{(}const\PY{+w}{ }\PY{o+ow}{\PYZdl{}}\PY{+w}{ }\PY{k+kt}{Ready}\PY{+w}{ }\PY{l+m+mi}{2}\PY{o+ow}{)}
\PY{+w}{ }\PY{+w}{ }bad\PY{+w}{ }\PY{o+ow}{=}\PY{+w}{ }\PY{k+kr}{do}\PY{+w}{ }sendN\PY{+w}{ }\PY{l+m+mi}{1}
  \end{Verbatim}

  Additionally, although the program above may run forever, we can increase our
  confidence that it will not. \texttt{Traceable} allows us to use type-level
  \QC, meaning we can write a property and check it at compile-time:

  \begin{Verbatim}[commandchars=\\\{\}]
\PY{l+m+mi}{0}\PY{+w}{ }\PY{k+kt}{PROP\PYZus{}sendThreeOK}\PY{+w}{ }\PY{o+ow}{:}\PY{+w}{ }\PY{k+kt}{Fn}\PY{+w}{ }\PY{o+ow}{(}\PY{k+kt}{Trace}\PY{+w}{ }\PY{k+kt}{ARQOp}\PY{+w}{ }\PY{o+ow}{(}\PY{k+kt}{Ready}\PY{+w}{ }\PY{l+m+mi}{0}\PY{o+ow}{)}\PY{+w}{ }\PY{l+m+mi}{20}\PY{o+ow}{)}
\PY{+w}{ }\PY{+w}{ }\PY{+w}{ }\PY{+w}{ }\PY{+w}{ }\PY{+w}{ }\PY{+w}{ }\PY{+w}{ }\PY{+w}{ }\PY{+w}{ }\PY{+w}{ }\PY{+w}{ }\PY{+w}{ }\PY{+w}{ }\PY{+w}{ }\PY{+w}{ }\PY{+w}{ }\PY{+w}{ }\PY{+w}{ }\PY{+w}{ }\PY{+w}{ }\PY{+w}{ }\PY{+w}{ }\PY{+w}{ }\PY{k+kt}{Bool}
PROP\PYZus{}sendThreeOK\PY{+w}{ }\PY{o+ow}{=}\PY{+w}{ }\PY{k+kt}{MkFn}\PY{+w}{ }\PY{o+ow}{(\PYZbs{}}\PY{k+kr}{case}\PY{+w}{ }\PY{o+ow}{(}\PY{k+kt}{MkTrace}\PY{+w}{ }\PY{k+kr}{\PYZus{}}\PY{+w}{ }trace\PY{o+ow}{)}\PY{+w}{ }\PY{o+ow}{=\PYZgt{}}
\PY{+w}{ }\PY{+w}{ }elem\PY{+w}{ }\PY{o+ow}{(}\PY{k+kt}{Ready}\PY{+w}{ }\PY{l+m+mi}{3}\PY{o+ow}{)}\PY{+w}{ }\PY{o+ow}{\PYZdl{}}\PY{+w}{ }\PY{o+ow}{(.}resSt\PY{o+ow}{)}\PY{+w}{ }\PY{o+ow}{\PYZlt{}\PYZdl{}\PYZgt{}}\PY{+w}{ }trace\PY{o+ow}{)}

\PY{l+m+mi}{0}\PY{+w}{ }\PY{k+kt}{QC\PYZus{}sendThreeOK}\PY{+w}{ }\PY{o+ow}{:}\PY{+w}{ }\PY{k+kt}{So}\PY{+w}{ }\PY{o+ow}{(}\PY{k+kt}{QuickCheck}\PY{+w}{ }\PY{k+kt}{False}
\PY{+w}{ }\PY{+w}{ }\PY{+w}{ }\PY{+w}{ }\PY{+w}{ }\PY{+w}{ }\PY{+w}{ }\PY{+w}{ }\PY{+w}{ }\PY{+w}{ }\PY{+w}{ }\PY{+w}{ }\PY{+w}{ }\PY{+w}{ }\PY{+w}{ }\PY{+w}{ }\PY{+w}{ }\PY{+w}{ }\PY{+w}{ }\PY{+w}{ }\PY{+w}{ }\PY{+w}{ }\PY{+w}{ }\PY{k+kt}{PROP\PYZus{}sendThreeOK}\PY{o+ow}{)}
QC\PYZus{}sendThreeOK\PY{+w}{ }\PY{o+ow}{=}\PY{+w}{ }\PY{k+kt}{Oh}
  \end{Verbatim}

  The trace is to a depth of 20 because it takes at least 3 transitions to
  reliably send a single packet. Since there are no reported mismatches between
  \texttt{True} and \texttt{False} on file loading, we know that the property
  holds. While we have not \emph{proven} that our program is guaranteed to
  terminate, we have increased our confidence that it does, without
  having to leave the language or modelling framework we are already using,
  and with a guarantee that the types, program, and test all use the same
  model and rules.

\section{Evaluation \& Further Work}\label{sec:evaln-further}

The types, state transitions, and \texttt{Traceable} implementation for
\acrshort{arq} come to around 30 lines of code, in contrast to the nearly 80 we
had to write just for the supports for the \acrshort{atm} in
\cref{ssec:atm-plumbing}. These are tricky lines of code {\textemdash} weaving
the state in the types and making sure the trace generation follows the correct
sequence {\textemdash} so having them in a generalised form saves us from the
risk of incorrectly reimplementing them, in addition to also saving us a lot of
tedious work.

It could be argued that most \acrshortpl{dsl} are simple enough to manually
reason about. However, as we have seen, seemingly trivial \acrshortpl{dsl} like
the one for the \acrshort{atm} are easy to get wrong and this mistake can remain
undetected for years. For more involved use cases like network protocols,
resource management, or concurrency, where the host type system is leveraged to
provide certain desirable guarantees for the target
domain~\cite{czarneckiDSLImplementationMetaOCaml2004,%
      bhattiDomainSpecificLanguages2009,%
      bradyCorrectbyConstructionConcurrencyUsing2010,%
      bradyResourceSafeSystemsProgramming2012,%
      florPiwareHardwareDescription2018,%
      castro-perezZooidDSLCertified2021%
},
the risk of the \acrshort{dsl} accidentally introducing subtle inconsistencies
and unintended behaviour, is likely to be much higher, weakening the goal of
eliminating certain bugs by using a stricter host language. Combined with the
ubiquity of stateful systems, we therefore believe that our approach is
worthwhile and intend to model and test more advanced protocols in the future.

Throughout the paper, all the traces have had seemingly magic numbers as their
bound.  The numbers were determined partially on reasoning: it is possible to
get a good estimate of the depth needed by taking the number of transitions in
the \acrshort{ism}, deciding on an upper bound for when the system should be in
the desired state, and multiplying this by the number of transitions (or the
number of states) in the system. Should the properties fail, they can be
examined to either reveal a valid error case, possibly a fault in the
generators, or a false positive caused by the model (as happened
in~\cite{hughesExperiencesQuickCheckTesting2016}).
For true positives the depth can be doubled until they pass, at which point a
binary search can be used to find the smallest valid bound. We believe this to
be part of the confidence building: the programmers can increase the bound until
they are confident in their typed models, or they can decide that the current
bound is sufficient. In our experience, this is not a hindrance, as the
properties fail quickly, thereby quickly finding the initial upper bound.

The type-level \acrshort{pbt} for the \acrshort{arq} model takes around 3.5
seconds on a reasonably modern
laptop\footnote{x86\_64 Intel Core i7-8750H @ 2.20GHz, boosting to
\textasciitilde 4.08GHz, with 32GiB of SODIMM DDR4 RAM @ 2667MT/s}.
While this may seem slow, it is worth remembering that the type checker is doing
a lot of work. It is solving interface constraints, unifying values, running a
\acrshort{prng}, and doing equality checks for non-trivial types at least 100
times. The \iidris type checker is currently the main bottleneck to our approach
and presents many interesting research questions in terms of how to speed up the
elaboration process, when to expand and inline certain functions and datatypes,
and how much information to keep around in the type checker and elaborator.

In the future, we plan to examine and implement increasingly more advanced
systems. ARQ with Sliding
Window~\cite{linErrorControlCoding1983}
would be a nice extension to the \acrshort{arq} example, as it improves the
throughput of the protocol and presents some new challenges for the state
function: how do we best model the packet window's movements? Pick and Place
machines used for automatic placement of surface mounted
components~\cite{arOpenSourceAutomated2018},
file
systems~\cite{chenPropertyBasedTestingClimbing2022},
and protocols with crash-stop
failures~\cite{barwellDesigningAsynchronousMultiparty2023},
are all stateful systems which will present interesting modelling challenges as
well as provide us with more performance and usability data.
Additionally, it will be interesting to explore what kind of properties we can
check. \acrfull{ltl} and \acrfull{ctl} are used extensively in model
checking~\cite{clarkeModelCheckingMy2008,clarkeModelCheckingState2012},
and there is recent work by O'Connor and Wickstr{\"o}m on combining
\acrshort{ltl} with \acrshort{pbt} for use in testing
\acrfullpl{gui}~\cite{oconnorQuickstromPropertyBased2022}.
As such, it will be interesting to see how big the overlap might be between our
approach and what model checkers can express and verify.

All the code used in this paper is publicly available at:\\
\url{https://github.com/CodingCellist/tyde-24-code}

\section{Conclusion}\label{sec:conclusion}

We successfully implemented \QC in \iidris and demonstrated how it can be
extended to generate arbitrary instances of dependent types. We then highlighted
how dependent types allow us to model stateful systems, that these models are
tricky to get right, and how we can use \QC at the type-level to automatically
detect this and help us fix it. Finally, we generalised the type-level framework
to support any stateful system, and demonstrated its usefulness by modelling,
implementing, and testing a network protocol. Our approach is not a proof
system, however it should help prototype specifications and models faster,
gaining confidence that the current system is sound, before potentially choosing
to model check or to formalise and prove it. We believe that our generalisation
is low-friction enough for wider adoption and are excited to see what this may
lead to.

\begin{acks}
  We are grateful to the anonymous reviewers who took the time to carefully read
  through the paper and write detailed and insightful feedback, including ideas
  for how to improve the code, many of which were incorporated into this final
  version. We are also grateful for the support of EPSRC grant EP/T007265/1.
\end{acks}

\bibliographystyle{ACM-Reference-Format}
\bibliography{00-tyde24}

\end{document}